\documentclass[12pt,psf,epsf]{article}
\usepackage[centertags]{amsmath}
\usepackage{amssymb}
\usepackage{graphicx}
\usepackage{epsfig}
\usepackage{ulem}
\usepackage[english]{babel}
\usepackage{array}
\usepackage{amsthm}
\usepackage{latexsym}
\usepackage[mathcal]{euscript}
\pdfoutput=1
\usepackage{epsfig}
 \usepackage{jheppub}
 \usepackage{hyperref}

\newcommand{\nn}{\nonumber}
\newcommand{\be}{\begin{equation}}
\newcommand{\ee}{\end{equation}}
\newcommand{\bea}{\begin{eqnarray}}
\newcommand{\eea}{\end{eqnarray}}

\newcommand{\e}{\mathrm{e}}

\newcommand{\tr}{\text{tr}}
\newcommand{\Tr}{\text{Tr}}
\newcommand{\ZZ}{\mathbb{Z}}

\newcommand{\HH}{\mathbb{H}}
\newcommand{\RR}{\mathbb{R}}

\DeclareMathOperator\arctanh{arctanh}
\DeclareMathOperator\arcsinh{arcsinh}

\newtheorem{Proposition}{Proposition}

\title{Thermalization after holographic bilocal quench}

\author{Irina Ya. Aref'eva,}
\author{Mikhail A. Khramtsov,	}
\author{Maria D. Tikhanovskaya}

\affiliation{Steklov Mathematical Institute, Russian Academy of Sciences,\\Gubkina str. 8, 119991, Moscow, Russia}

\emailAdd{arefeva@mi.ras.ru}
\emailAdd{khramtsov@mi.ras.ru}
\emailAdd{tikhanovskaya@mi.ras.ru}

\abstract{We study thermalization in the holographic $(1+1)$-dimensional CFT after simultaneous generation of two high-energy excitations in the antipodal points on the circle. The holographic picture of such quantum quench is the creation of BTZ black hole from a collision of two massless particles. We perform holographic computation of entanglement entropy and mutual information in the boundary theory and analyze their evolution with time. We show that equilibration of the entanglement in the regions which contained one of the initial excitations is generally similar to that in other holographic quench models, but with some important distinctions. We observe that entanglement propagates along a sharp effective light cone from the points of initial excitations on the boundary. The characteristics of entanglement propagation in the global quench models such as entanglement velocity and the light cone velocity also have a meaning in the bilocal quench scenario. We also observe the loss of memory about the initial state during the equilibration process. We find that the memory loss reflects on the time behavior of the entanglement similarly to the global quench case, and it is related to the universal linear growth of entanglement, which comes from the interior of the forming black hole. We also analyze general two-point correlation functions in the framework of the geodesic approximation, focusing on the study of the late time behavior.

}

\begin{document}
\maketitle

\section{Introduction}

The description of thermalization and equilibration in closed quantum systems has been a long-standing theoretical problem. An interesting setting to study non-equilibrium physics in quantum systems is a quantum quench, when one prepares the ground state of a given Hamiltonian $H_1$ and then evolves it unitarily under the action of a different Hamiltonian $H_2 \neq H_1$. The system is going to evolve to some pure state $|\psi(t) \rangle$ after time $t > 0$. The system is said to thermalize if for any subsystem $A$ the density matrix $\rho_A = \Tr_{\bar{A}} |\psi(t) \rangle \langle \psi(t)|$ becomes equal to the thermal density matrix with a certain non-zero temperature. The entanglement entropy of the subsystem $S(A) = - \Tr_A\  \rho_A \log \rho_A$ and related quantities are very useful tools to probe subsystems which relax to thermal equilibrium as the full system evolves in time after the quantum quench. 

The AdS/CFT-correpsondence and holography \cite{Malda,Aharony1999} have given new tools for studying physics out of equilibrium in strongly coupled QFT. According to the holographic dictionary, thermal state in the boundary theory corresponds to a black hole in the bulk. The problem of studying the behavior of observables during thermalization after a quench in strongly-coupled quantum field theory can be treated in the leading order of the semiclassical approximation using non-stationary asymptotically AdS classical gravity solutions which describe formation of a black hole in the bulk. In holographic context, the entanglement entropy then can be calculated according to the Ryu-Takayanagi prescription \cite{RT} generalized to the general time-dependent case by Hubeny, Rangamani and Takayanagi \cite{HRT}, as area of the minimal surface anchored on the boundary region where the subsystem under the study is located. The most well studied holographic model of thermalization in CFT is the Vaidya dust shell collapse \cite{Abajo10,Bal10,Bal2012,Liu13,Liu2013,Li13,Hartman13,Hubeny2013,Ageev2017,Ageev17,Mezei2016,Mezei16,Casini15,Anous16,Ziogas}. This gravity dual models the global quench in CFT \cite{Calabrese05,Calabrese16}, when in the initial time moment a spatially uniform distribution of energy is injected into the system. Another known holographic model of global quantum quench is the end of world brane model \cite{Hartman13}. The entanglement dynamics  of the global quench in CFT were found to share many similarities between these two different models \cite{Hartman13,Mezei16}, hinting at the possibility of some universality of entanglement spreading at least in global quench situations. 

Another type of quantum quenches is the local quench \cite{Calabrese16,Asplund14,Nozaki13,Caputa141,Caputa142,Caputa15,Asplund13,David16,Asplund15,Rangamani15,Rozali17,Erdmenger17}. Such quenches have been studied holographically as perturbations of the zero temperature vacuum \cite{Asplund14,Nozaki13} as well as of thermal equilibrium state \cite{Caputa142,Caputa15,Rangamani15,David16,Rozali17,Erdmenger17,Bai14}. In the present paper we study the thermalization in holographic $(1+1)$-dimensional compact CFT after a particular variation of a local quantum quench. This quench protocol, which we call the bilocal quench, is realized by simultaneously creating two high-energy excitations in the antipodal points on the cylinder. In the bulk the dynamics after this quench are described by the collision of two massless particles in the AdS$_3$ spacetime which leads to the formation of a static massive particle or a BTZ black hole \cite{Matschull}. We focus on the black hole formation case, which describes thermalization in the boundary theory after the quench. 

The main feature of this model, which is not prominent in the Vaidya global quench model of thermalization, is the fact that the AdS$_3$ spacetime with two colliding particles which create a black hole is explicitly described as a topological quotient space of AdS$_3$ by a certain topological identification \cite{Matschull,Bal99,Jevicki02,AA}. This simplifies dealing with boundary physical quantities which are expressed geometrically in the bulk as e.g. geodesic lengths by relating all geodesics in the quotient to certain auxiliary geodesics in the global AdS$_3$ spacetime. The bilocal quench setup is also interesting because the thermalization of a closed system after introduction of two high-energy local excitations is an attractive toy model for description of thermalization in systems such as quark-gluon plasma after a collision of two heavy ions \cite{ArefevaQGP}.

Our main object of study in the present paper is behavior of the entanglement entropy of subsystems in the boundary CFT in the non-equilibrium regime after the quench described above. We perform the holographic computation of the entanglement entropy and mutual information in different subsystems after the bilocal quench, and we analyze the time dependence and spreading of the entanglement. We make a direct comparison to the global quench thermalization models, in particular the model based on the null dust collapse in the Vaidya-AdS spacetime. We find that because of lack of translational invariance in the initial state, the equilibration picture globally is substantially different from the picture given by the global quench. Specifically, subsystems which do not contain one of the initial excitations inside, exhibit thermal behavior of the holographic entanglement entropy right from the beginning of the time evolution. For subsystems, which do contain one of the excitations, however, the entanglement entropy demonstrates non-trivial non-equilibrium dynamics in many ways similar to the global quench situation, but with some substantial differences. Since the bulk spacetime is explicitly represented as a locally AdS$_3$ space with a topological identification, construction of HRT geodesics which calculate entanglement entropy becomes a purely geometrical problem. We discuss it in detail and make some observations about the loss of memory about the initial state upon equilibration of subsystems. We also study the leading behavior of two-point correlation functions in the framework of the geodesic approximation \cite{Bal99}, including the long-time behavior. 

The paper is organized as follows. In the section \ref{sectionSetup} we first set up our conventions and introduce the basic objects which are necessary for description of the bulk holographic dual to the thermalization after the bilocal quench. Then we describe the geometry of the AdS$_3$ spacetime with two colliding massless point particles which create a BTZ black hole and explain how it works as a bulk holographic dual. In the section \ref{sectionGeodesics} we study the boundary-to-boundary geodesics which are necessary for holographic computations in this bulk spacetime. We classify them, calculate the geodesic lengths and prove several statements about their behavior with respect to topological identifications generated by colliding particles. In the section \ref{sectionHEE} we use the results of the section \ref{sectionGeodesics} applied to the bulk spacetime described in the section \ref{sectionBHcreation} in BTZ coordinate patch to perform the holographic calculation of the entanglement entropy and mutual information and study the time dependence of entanglement in detail. In the section \ref{sectionCorr} we continue the holographic study of thermalization by analyzing the two-point correlation functions in the framework of the geodesic approximation. In the section \ref{sectionDiscussion} we recollect the results of the work and discuss their implications and future directions. 

\section{Holographic setup} \label{sectionSetup}

\subsection{Geometry of AdS$_3$ and global defects \label{secGeometry}}

\subsubsection{The AdS$_3$ spacetime}

We start the discussion by establishing the conventions and describing the basic objects which will help us then construct the holographic dual for bilocal quench in the boundary. On the gravity side, we deal with the pure AdS$_3$ spacetime, as well as with asymptotically locally AdS$_3$ solutions of $3$D Einstein equations. Because $3$D gravity is topological, solutions of gravitational equations with negative cosmological constant are global defects in AdS$_3$. More precisely \cite{Maloney07}, they have the general form of AdS'$_3$/$\Gamma$, where $\Gamma$ - a discrete subgroup of the isometry group $SL(2, \mathbb{R})^2$, and AdS'$_3$ is the subset of AdS$_3$ where $\Gamma$ acts discretely. The objects in AdS$_3$ in which we are interested, namely point particles and black holes, are particular examples of such solutions. The AdS$_3$ spacetime has a simple geometry, which allows to use a unified framework to describe global defects in AdS$_3$. 

We begin with the description of pure AdS$_3$ space as a hypersurface in the $4$-dimensional flat spacetime $\mathbb{R}^{2,2}$. It is given by the quadratic equation (we set the AdS radius to $1$):
\be
\label{hyperboloid}
-x_0^2-x_3^2+x_1^2+x_2^2=-1.
\ee
This quadric surface can be parametrized by coordinates, to which we will refer as global coordinates on AdS$_3$:
\bea\label{barrel}
x_0 &=& \cosh\chi\sin \tau\,,\\
x_1 &=& \sinh\chi\cos \phi\,,\nn\\
 x_2 &=& \sinh\chi\sin\phi\,,\nn\\
 x_3 &=& \cosh\chi\cos \tau\,.\nn
\eea
The induced metric on the AdS$_3$ is given by
\begin{equation}
 ds^2=-{\cosh}^2\chi\,d\tau^2+d{\chi}^2+{\sinh}^2\chi\,d\phi^2. \label{barrelmetric}
\end{equation}
Here $\chi \in [0,\ +\infty)$ is the holographic coordinate, and other coordinates have ranges $\phi \in [0, 2\pi)$ and $\tau \in [-\pi, \pi]$. The conformal boundary of the AdS$_3$ spacetime is located at $\chi \to \infty$. The spacetime can be visually represented as a cylinder together with its interior, where $\chi$ plays the role of a radial coordinate, $\tau$ is a coordinate along the vertical axis of the cylinder and $\phi$ is the angular coordinate. 
The global coordinates are most suitable for description of global defects in the bulk, since they keep the complete information about the topological identification associated with the given defect. 

Another coordinate system which we will use is obtained by parametrization:
\bea
x_0&=&-\frac{r}{R}\cosh R\ \varphi\,,\nn\\
x_1&=&\sqrt{\frac{r^2}{R^2}-1}\cosh R\ t\,,\nn\\
x_2&=&\frac{r}{R}\sinh R\ \varphi\,,\nn\\
x_3&=&\sqrt{\frac{r^2}{R^2}-1}\sinh R\ t \label{Sch-coor}\,;
\eea
where $t$, $\varphi \in \mathbb{R}$ are the coordinates on the boundary, and $r \in (R, +\infty)$ is the radial coordinate. There is a coordinate singularity at $r = R > 0$. The metric in these coordinates has the form
\be
ds^2 = - (r^2-R^2)\ d t^2 + \frac{d r^2}{r^2-R^2}+r^2 d\varphi^2\,,\label{schw}
\ee
We will refer to these coordinates as BTZ coordinates. In the AdS$_3$ spacetime, this patch covers only a part of the global AdS$_3$. Note that the choice of the BTZ coordinate system is ambiguous. This ambiguity is the choice of the part of the global AdS$_3$ spacetime to cover with a BTZ patch. Namely, different choices of the patch can be implemented by changing the signs in front of the square root in the parametrization formulas. 

As we will see, the BTZ coordinates are most natural for the holographic description of thermalization in CFT on a cylinder. However, description of topological defects in AdS$_3$ is more convenient in the global coordinates. Hence we will need the transformation formulas from the global coordinates to BTZ patch. They are obtained using the embedding coordinate parametrizations (\ref{barrel}) and (\ref{Sch-coor}):
\bea
x_0 = \cosh \chi \sin \tau &=&-\frac{r}{R}\cosh  R\ \varphi\nn\\
x_2 = \sinh \chi \sin \phi &=&\frac{r}{R}\sinh R\ \varphi\nn\\
x_3 = \cosh \chi \cos \tau &=&\sqrt{\frac{r^2}{R^2}-1}\sinh R\ t\nn\\
x_1 = \sinh \chi \cos \phi &=&\sqrt{\frac{r^2}{R^2}-1}\cosh R\ t\label{Sch-transf}\,,
\eea

To deal with classical solutions, which are quotients of AdS$_3$, it is most convenient to use the algebraic representation of AdS$_3$. The AdS$_3$ spacetime can be described as the $SL(2, \mathbb{R})$ group manifold. We can treat points in AdS$_3$ as matrices:
\begin{equation}\label{SL2}
 X=x_3\mathbf{1}+\sum_{\mu=0,1,2}\gamma_{\mu}x^\mu=
 \begin{pmatrix}
  x_3+x_2 & x_0+x_1 \\
  x_1-x_0 & x_3-x_2\\
 \end{pmatrix}\,;
\end{equation}
where the matrix basis is introduced
\begin{align}
&\mathbf{1}=\begin{pmatrix} 1 & 0 \\ 0 & 1 \end{pmatrix},\qquad
 \gamma_0=\begin{pmatrix} \phantom{-}0 & 1 \\  -1 & 0 \end{pmatrix},\qquad
 \gamma_1=\begin{pmatrix} 0 & 1 \\ 1 & 0 \end{pmatrix},\qquad
 \gamma_2=\begin{pmatrix} 1 & \phantom{-}0 \\ 0 & -1 \end{pmatrix},
\label{eq2.4}\
\end{align}
In this notation the condition $\det X=1$ then gives the hypersurface equation ~\eqref{hyperboloid}. 

The physical quantities on the boundary which we are interested in are calculated from geodesics in the bulk. To study  geodesics on quotients of AdS$_3$, it is most convenient to work in terms of matrix notations. The geodesics in AdS$_3$ embedded into $\RR^{(2,2)}$ can be described as solutions of the Lagrangian  \cite{Bengtsson,Arefeva09}:
\be
\mathcal{L} = \frac{1}{2} \dot{x}^2 + \lambda (x^2 + 1)\,,
\ee
where $\lambda$ is a Lagrange multiplier.
The geodesic length in AdS$_3$ can be expressed in terms of the scalar product in the embedding spacetime $\RR^{(2, 2)}$. Suppose that $x$ and $y$ are two points in the embedding space, and we denote their respective matrices defined according to (\ref{SL2}) as $X$ and $Y$. Then if points $x$ and $y$ belong to the AdS$_3$ hyperboloid, i. e. $\det X = \det Y = 1$, then it is true that 
\be
\eta_{MN} x^{M} y^N = -\frac12 \tr\  X Y^{-1} =  -\frac12 \tr\  X^{-1} Y\,.
\ee
The length of a spacelike geodesic between points $x$ and $y$ is expressed by formula 
\be
\cosh \mathcal{L}_{\text{spacelike}} (x,\ y) = \frac12 \tr\  X Y^{-1}\,; \label{Lspacelike}
\ee
and length of a timelike geodesic is given by the formula
\be
\cos \mathcal{L}_{\text{timelike}} (x,\ y) = \frac12 \tr\  X Y^{-1}\,. \label{Ltimelike}
\ee

The isometry group of AdS$_3$ is the group $SO(2, 2) \simeq SL(2, \mathbb{R}) \times SL(2, \mathbb{R})$, which acts on matrix $X$ as follows: 
\be
X \to {\bf g} X {\bf h}^{-1}\,, \quad {\bf g},\ {\bf h} \in SL(2, \mathbb{R})\,;
\ee
This group has an $PSL(2, \mathbb{R}) = SL(2, \mathbb{R})/\ZZ_2$ subgroup which corresponds to isometries which leave the origin of AdS$_3$ (which is represented by the unit matrix) fixed. It is realized by choosing ${\bf u} = {\bf g}^{-1} = {\bf h}$ as an element of the $SL(2, \RR)$ group up to an overall sign. Then it can represent an isometry of AdS$_3$ which preserves the origin by acting on $X$ via conjugation:
\be
X \to X^* = \textbf{u}^{-1} X \textbf{u}\,,\label{conjugation}
\ee
Point-like objects in AdS$_3$ such as particles and black holes are obtained from empty AdS$_3$ by taking a topological quotient. The identification is defined by the isometry ${\bf u}$ acting on the $SL(2, \RR)$ group manifold via conjugation (\ref{conjugation}). We will refer to the identification isometry ${\bf u}$ as the holonomy of the topological defect, in agreement with the discussion in \cite{Matschull}. Let us now proceed to concrete discussion of topological defects which we deal with in the present investigation. 

\subsubsection{Massless point particles in AdS$_3$}
Our main ingredient for constructing the bulk spacetime is a couple of massless particles. A point particle in $(2+1)$-dimensional gravity produces a defect, which holnomy is determined by the momentum vector of the particle \cite{Matschull97}. The most general form of a holonomy of a point particle with momentum $p^\mu$ is given by 
\be
{\bf u} = u\ \mathbf{1} + p^\mu \gamma_\mu\,;
\ee
The condition $\det {\bf u} = 1$ then means that 
\be
u^2 - p_\mu p^\mu = 1\,;
\ee
\begin{figure}[t]
\centering
\includegraphics[width=6cm]{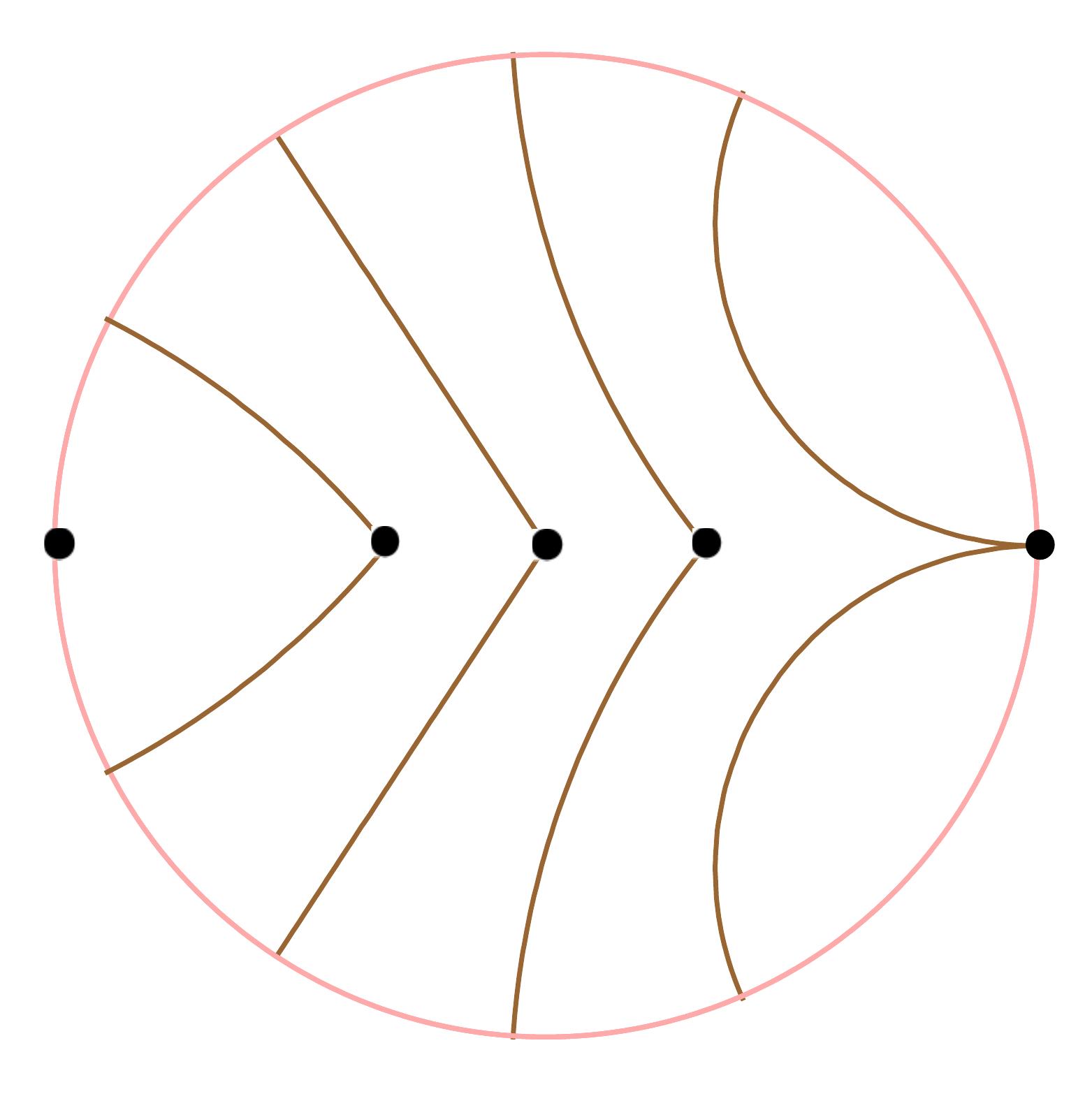}A.
\includegraphics[width=5cm]{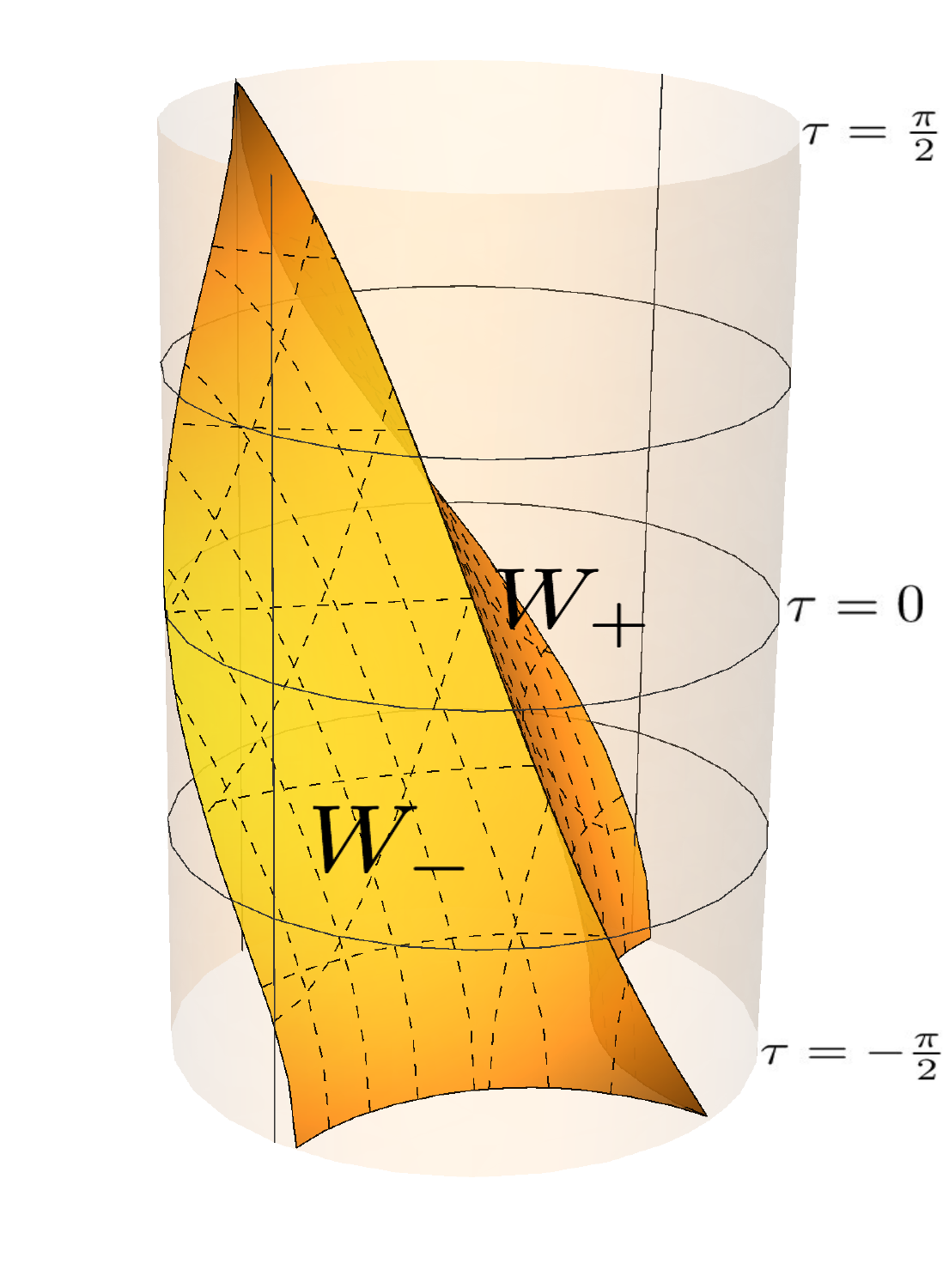}B.
\caption{\textbf{A}. Cartoon of propagation of a massless particle in the AdS$_3$ spacetime projected onto synchronous time slices of the AdS$_3$ cylinder in different moments of time. The particle moves from left (right) to right (left), and the spacetime is obtained by cutting out the wedge behind (in front of) the particle between the surfaces $W_\pm$. \textbf{B}. $3$D plot of massless particle moving through the AdS$_3$ spacetime in global coordinates. The intersection of surfaces $W_\pm$ is the worldline of the particle.}
\label{masslessPic}
\end{figure}
In the present work we focus on the case of massless particles, which means that we have to set $p^2 = 0$. Then from the equation above we have\footnote{The ambiguity of the sign of $u$ here is the ambiguity of the overall sign of the holonomy, and thus can be fixed arbitrarily.} $u = 1$. Thus, the holonomy of a massless particle is given by
\be
{\bf u}_{\text{massless}} = \mathbf{1} + p^\mu \gamma_\mu\,;  \label{Umassless}
\ee
It produces an identification in AdS$_3$, which glues together two surfaces $W_-$ and $W_+$: 
\be
W_+ = {\bf u}^{-1} W_- {\bf u}\,;
\ee
and these surfaces intersect along the worldline of the particle, see Fig.\ref{masslessPic}. The surfaces $W_\pm$ intersect time slices of AdS$_3$ along the equal-time geodesics which we denote as $w_\pm$ (see Fig.\ref{masslessPic}A). Thus in the AdS$_3$ spacetime the massless particle cuts out the wedge between surfaces $W_-$ and $W_+$. One can cut out the wedge either in front of the particle worldline, or behind the worldline. Or, equivalently, one can think that on the Fig.\ref{masslessPic} the particle moves either from left to right, or from right to left. Sometimes we will call the region space which is cut out by the identification, i. e. the complement of the fundamental domain to the global AdS$_3$, as the dead zone, and we call the boundary of the fundamental domain as living space. The holonomy of a massless particle belongs to the parabolic conjugacy class, since $|\tr\ {\bf u}| = 2$ in this case. The fixed point of a parabolic holonomy is on the boundary of the $\HH_2$ \cite{Spectral}. That means that a massless particle can actually reach the boundary of the AdS$_3$ spacetime, and the turning points of its worldline at $\tau = \frac{\pi}{2}+\pi n$ are located there. The motion of the particle is periodic with return points located at the boundary. 

\subsubsection{Maximally extended BTZ black hole}

The bulk dual for the thermalization process in the CFT$_2$ is the creation of the BTZ black hole in the bulk. More specifically, in the present work we consider formation of the static BTZ black hole from point particle collisions. The black hole formed from matter in a dynamical process of some kind is dual to a pure state on the boundary, in contrast to the eternal (maximally extended) black hole, which is dual to the mixed thermal state on the boundary \cite{Maldacena01}. However, we will use the eternal black hole geometry as a reference point for description of the black hole formed from particle collisions. 

\begin{figure}[t]
\centering
\includegraphics[width=5.5cm]{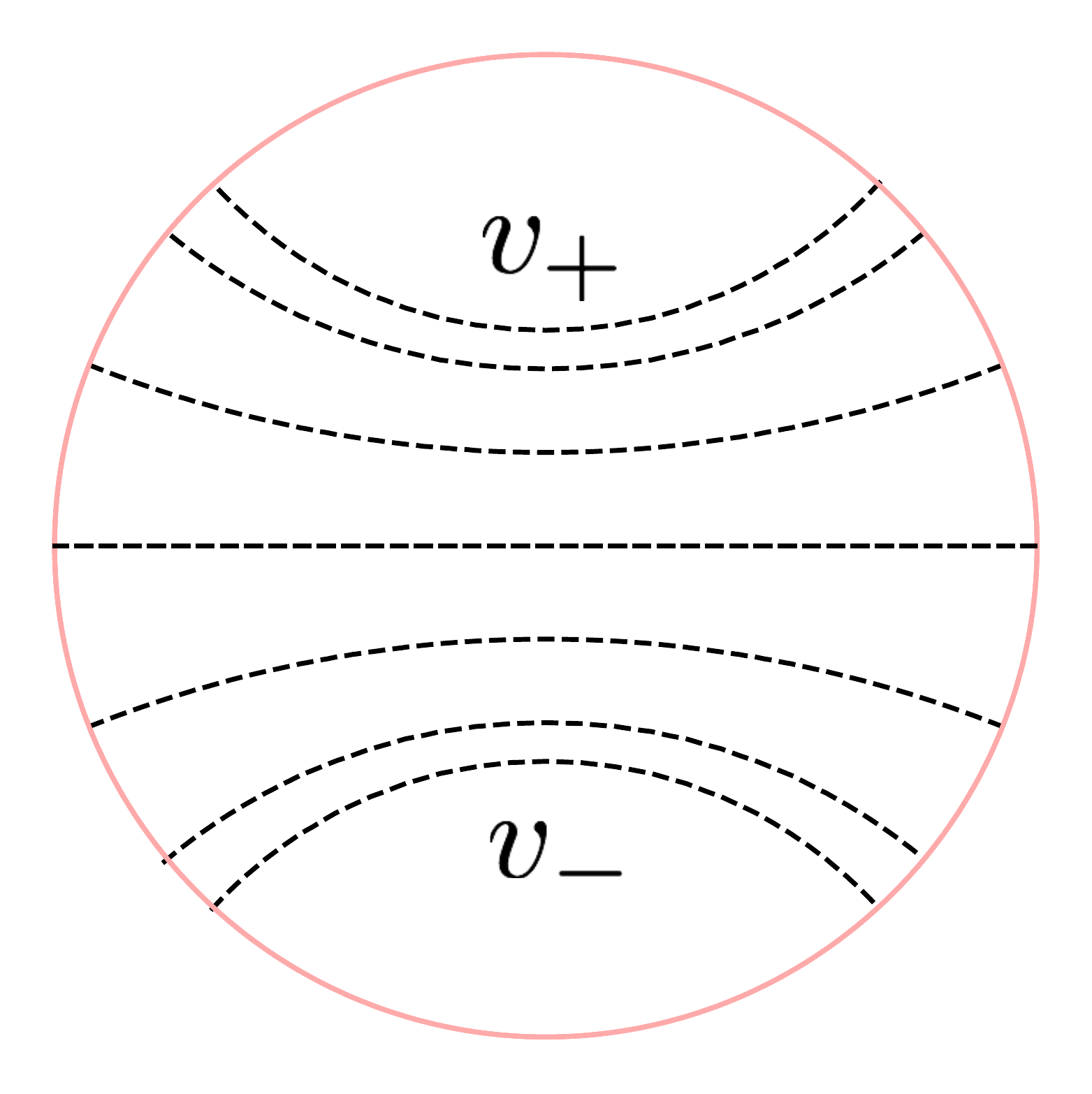}A
\includegraphics[width=7cm]{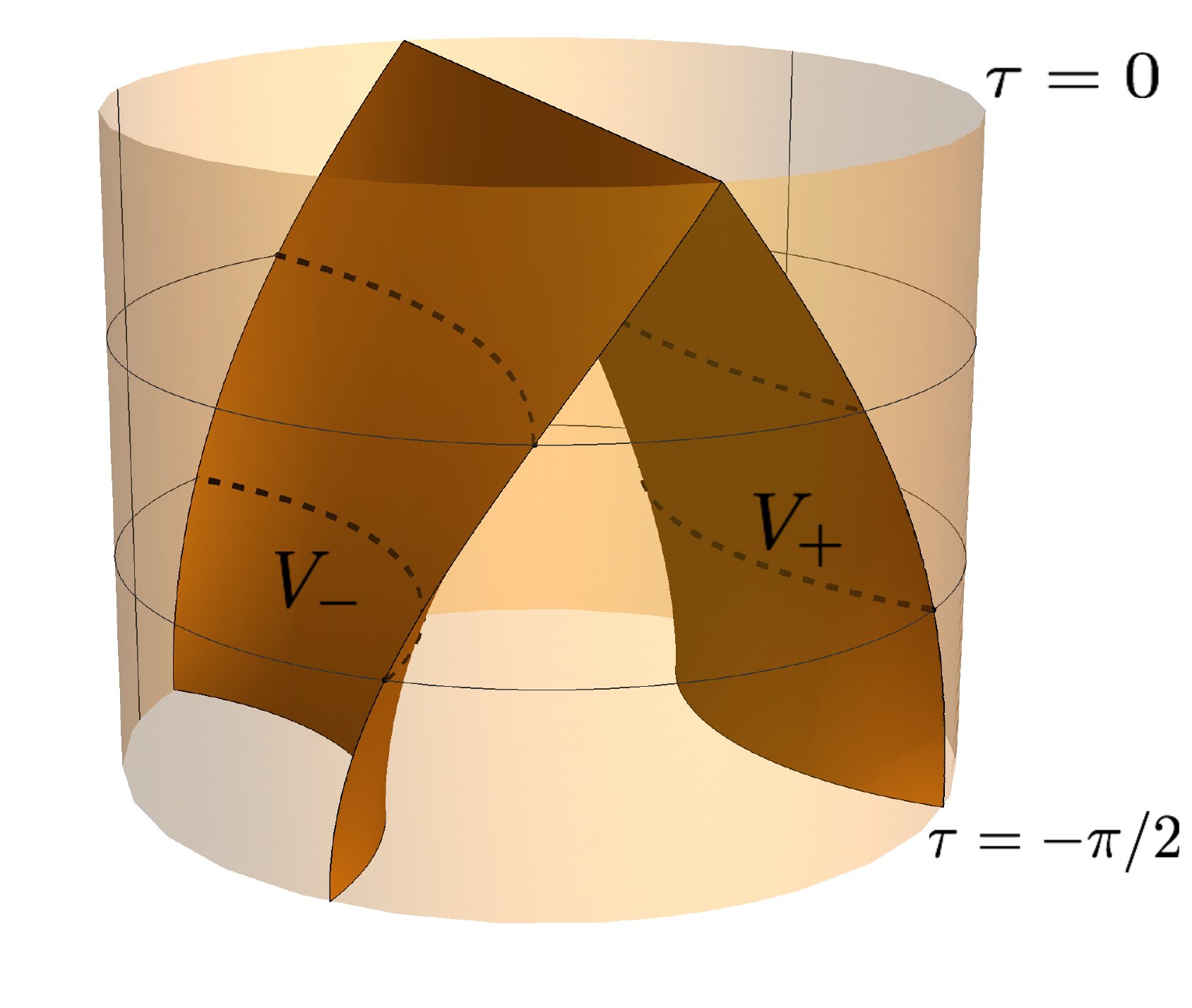}B
\caption{Maximally extended BTZ black hole in global coordinates. \textbf{A}. Projection of surfaces $V_\pm$ onto equal time slices in different moments of global time. The dashed lines are identified. \textbf{B}. $3$D plot of the BTZ black hole identification surfaces $V_\pm$ between $\tau = -\frac{\pi}{2}$ and $\tau = 0$. 
}
\label{bh0-m}
\end{figure}

The maximally extended BTZ black hole in global coordinates is described as an AdS$_3$ space quotient by a hyperbolic $SL(2, \mathbb{R})$ element. We focus on the static case. The corresponding holonomy has the following form \cite{Matschull}:
\begin{equation}\label{btz-hol} 
    {\bf u}_{\text{BTZ}} = - e^{-\mu\,\gamma_1} = - \cosh \mu \, {\bf 1} + \sinh \mu \, \gamma_1 = \begin{pmatrix}
    -\cosh \mu & \sinh \mu \\
    \sinh \mu & -\cosh \mu
    \end{pmatrix}\,;
\end{equation}
Here $\mu > 0$ is a parameter related to the mass of the black hole. This holonomy generates an identification which identifies two surfaces $V_\pm$:
\be
V_+ = {\bf u}_{\text{BTZ}}^{-1} V_- {\bf u}_{\text{BTZ}}\,; 
\ee
In global coordinates these surfaces are defined by equations \cite{Matschull}:
\begin{equation}\label{btz-faces}
   V_\pm: \quad   \tanh \chi \, \sin \phi 
           = \mp \sin t \, \tanh\mu .   
\end{equation}
These surfaces intersect AdS$_3$ time slices along equal-time geodesics $v_\pm$, as shown in Fig.\ref{bh0-m}. The maximally extended BTZ spacetime is defined as the region of the AdS$_3$ spacetime between the surfaces $V_\pm$. The part outside of this region is the dead zone which is cut out from the spacetime. The surface $V_+$ and $V_-$ do not intersect, except for $\tau=n\pi$, $n\in \ZZ$, where they intersect along the horizontal diameter of the time slice disc. The spacetime is singular in these moments of time. 
The spacetime manifold as the region between the two surfaces has two boundaries. Holographically this means that the maximally extended BTZ black hole is dual to the thermofield double state on the boundary. The spacetime has an event horizon, which consists of two surfaces described by the equation
\be\label{hor-eq} 
  \cos \phi \, \tanh \chi = \cos \tau\,. 
\ee
Spacetime splits into four regions, and for $\tau \in [-\pi,\ 0]$ has the same global causal
structure as the maximally extended AdS-Schwarzschild black hole, i.e. we have
two external regions, to the left and to the right of both horizons,
which are causally completely disconnected. At $\tau = -\pi$ we have the past singularity of the spacetime, and at $\tau = 0$ we have the future singularity. Each of these two regions can be covered by a BTZ coordinate patch, with metric given by (\ref{schw}) with horizon located at $R = \frac{\mu}{\pi}$. The action of the holonomy (\ref{btz-hol}) results in the identification $\varphi \sim \varphi + 2 \pi$. The length of the horizon equals $ 2\mu = 2\pi R$. The horizon radius is related to the mass of the black hole by the relation
\be
M = \frac{R^2}{8 G} = \frac{\mu^2}{8 \pi^2 G}\,, \label{BTZmass}
\ee
where $G$ is three-dimensional Newton constant. The Hawking temperature of the black hole which equals the temperature in the dual theory is given by 
\be
T = \frac{R}{2\pi}\,. \label{BTZtemp}
\ee
\subsection{Black hole creation from particle collisions in AdS$_3$} \label{sectionBHcreation}

Now let us discuss the picture of massless point particle collisions. We begin by setting up the stage in global AdS$_3$ as presented by Matschull in \cite{Matschull}. After that, we will make a transition to the BTZ coordinate patch in a similar way to \cite{Jevicki02}, which gives the natural dual description of the thermalizing CFT on a cylinder. We will need both pictures for our analysis, the first one containing all the data we need for our holographic computations, and the second one for straightforward definition of holographic observables and temporal evolution in the boundary theory after the quench. 

\subsubsection{Black hole creation in global coordinates}

An AdS$_3$ quotient spacetime contains a black hole if its total defect holonomy belongs to the hyperbolic conjugacy class. i.e. that it coincides with the BTZ holonomy (\ref{btz-hol}) up to a coordinate transformation. The total holonomy of two colliding massless particles is a product of two holonomies of each particle. This product holonomy is not neccessarily hyperbolic, but it depends on the energies of the particles. The black hole creation threshold is thus can be expressed \cite{Matschull} as the condition:
\be
\frac{1}{2} \tr\ {\bf u}_{\text{total}} =: - \cosh \mu < 1\,; \label{EnergyBound}
\ee
This translates into a lower bound of energy for colliding particles, which in itself produces a lower bound for the energy of excitations in the bilocal quench which would thermalize the boundary CFT. Since we are interested in thernalization after the quench, we consider only those particle collisions which create black holes. Thus the topological identification in the spacetime is constructed in such a way that we have two singularities with holonomies ${\bf u}_1$ and ${\bf u}_2$ corresponding to massless particles, and the total holonomy when circling around both particles must equal the BTZ black hole holonomy (\ref{btz-hol}). The resulting spacetime can be obtained by making additional cutting and gluing in the maximally extended BTZ black hole spacetime. 

More specifically that means that we have to choose two holonomies for particles ${\bf u}_1$ and ${\bf u}_2$ such that their product would equal ${\bf u}_{\text{BTZ}}$. The choice of holonomy of a massless particle is dictated by the choice of its momentum vector, according to (\ref{Umassless}). 
Suppose that two massless particles start from points $\phi=\theta$ and $\phi = 0$ from the boundary at $\tau = -\frac{\pi}{2}$. they move along the radial worldlines given by the equation 
\be
\tanh \frac{\chi}{2} = - \tan \frac{\tau}{2}\,; \label{Globalworldline}
\ee
At the moment of global time $\tau = 0$, particles meet each other at the origin $\chi = 0$, and the collision happens. 

We can choose one holonomy, say ${\bf u}_1$, freely, and let the product constraint determine the other one. Since we can multiply in two different orders, there will be two possible choices for the holonomy of the second particle\footnote{Note that in \cite{Matschull} the numeration of particles is reversed.}: 
\be
{\bf u}_{2+} = {\bf u}_1^{-1} {\bf u}_{\text{BTZ}}\,,\qquad {\bf u}_{2-} = {\bf u}_{\text{BTZ}}\ {\bf u}_1^{-1}\,; \label{U_2}
\ee
From these equations, one can find the parameters of the particles (see \cite{Matschull} for more details). We choose the momentum of the reference particle $1$ such that it has the energy $\tan \epsilon_1$ and it moves along the radial direction, starting from the point $\phi = 0$. 
The corresponding holonomy, from (\ref{Umassless}), reads
\be
{\bf u}_1 = {\bf 1} + \tan \epsilon_1 (\gamma_0 - \gamma_1)\,; \label{U1}
\ee
The second particle which starts from $\phi = \theta$ will have the energy $\tan \epsilon_2$. The equation (\ref{U_2}) then dictates that the particle $2$ moves with along the radial geodesic with angle $\phi = \pm \theta$, where $\sin \theta = \tanh \mu$. It has the energy $\tan \epsilon_2 = \cosh \mu \coth \frac{\mu}{2}$, and the first particle moves along the geodesic has the energy  $\tan \epsilon_1 = \coth \frac{\mu}{2}$. The last expression appears in many formulas in this work, so we introduce the notation:
\be
\mathcal{E} := \tan \epsilon_1 = \coth \frac{\mu}{2}\,. \label{E}
\ee
The resulting holonomy of the second particle reads
\be
{\bf u}_{2\pm} = {\bf 1} + \tan \epsilon_2 (\gamma_0 -\cos \theta\ \gamma_1 \mp \sin \theta\ \gamma_2)\,; \label{U2}
\ee
Henceforth all parameters of the infalling particles are determined through the holonomy from the black hole mass parameter $\mu$. Let us now recollect the kinematic data of the particles in the BTZ black hole creation process in global coordinates:
\begin{itemize}
\item Particle $1$: energy $\tan \epsilon_1 = \coth \frac{\mu}{2}$, angle $\phi = 0$\,;
\item Particle $2$: energy $\tan \epsilon_2 = \cosh \mu \coth \frac{\mu}{2}$, angle $\sin \phi = \pm \tanh \mu$. 
\end{itemize}
Having defined the holonomy of the particle $2$ as a product of the other particle inverse holonomy with the black hole holonomy, we now can try to represent the geometry of the identification by this holonomy through the identifications corresponding to ${\bf u}_1$ and ${\bf u}_{\text{BTZ}}$. 
\begin{figure}[t]
\centering
\includegraphics[width=5cm]{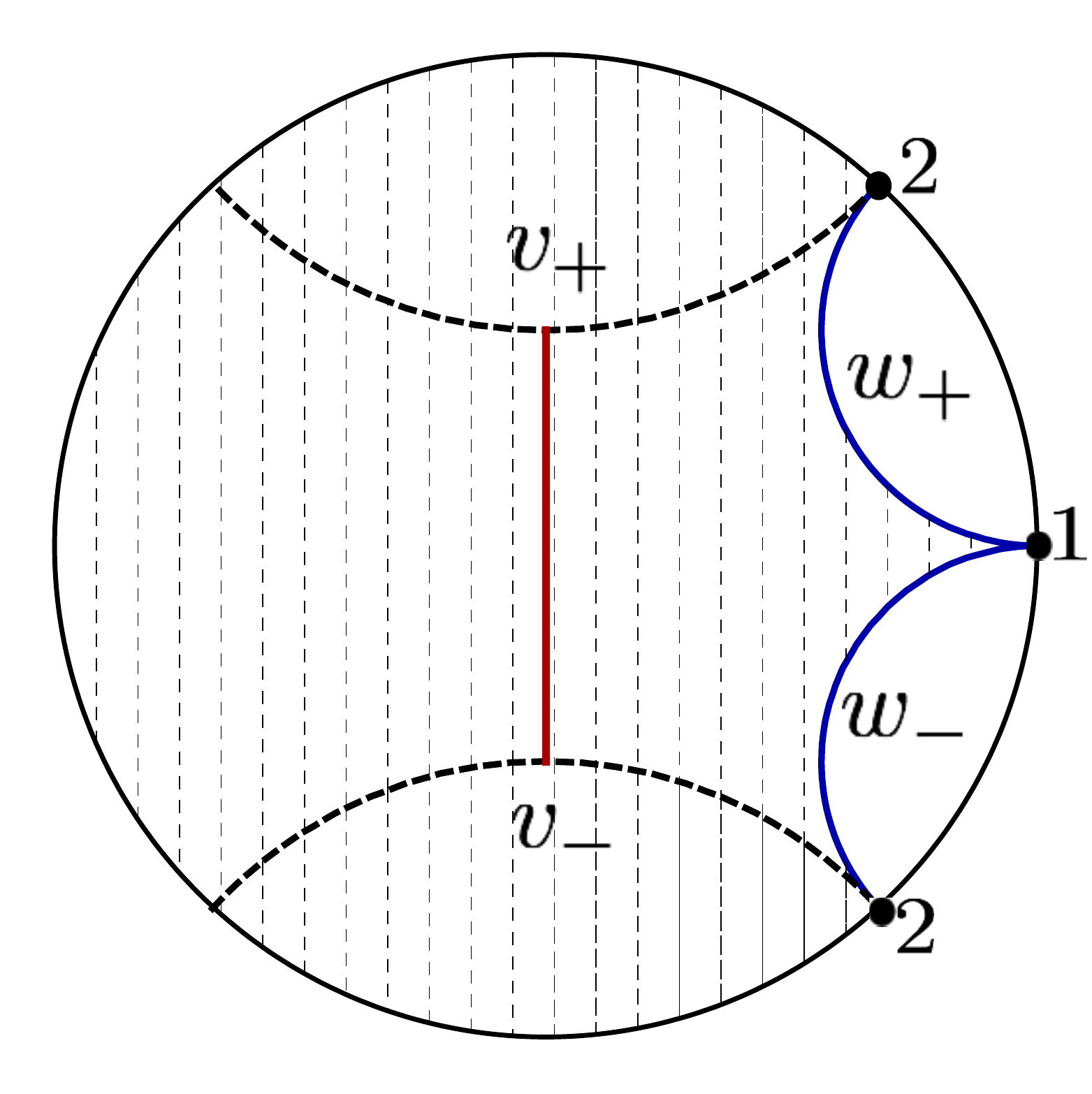}A.
\includegraphics[width=5cm]{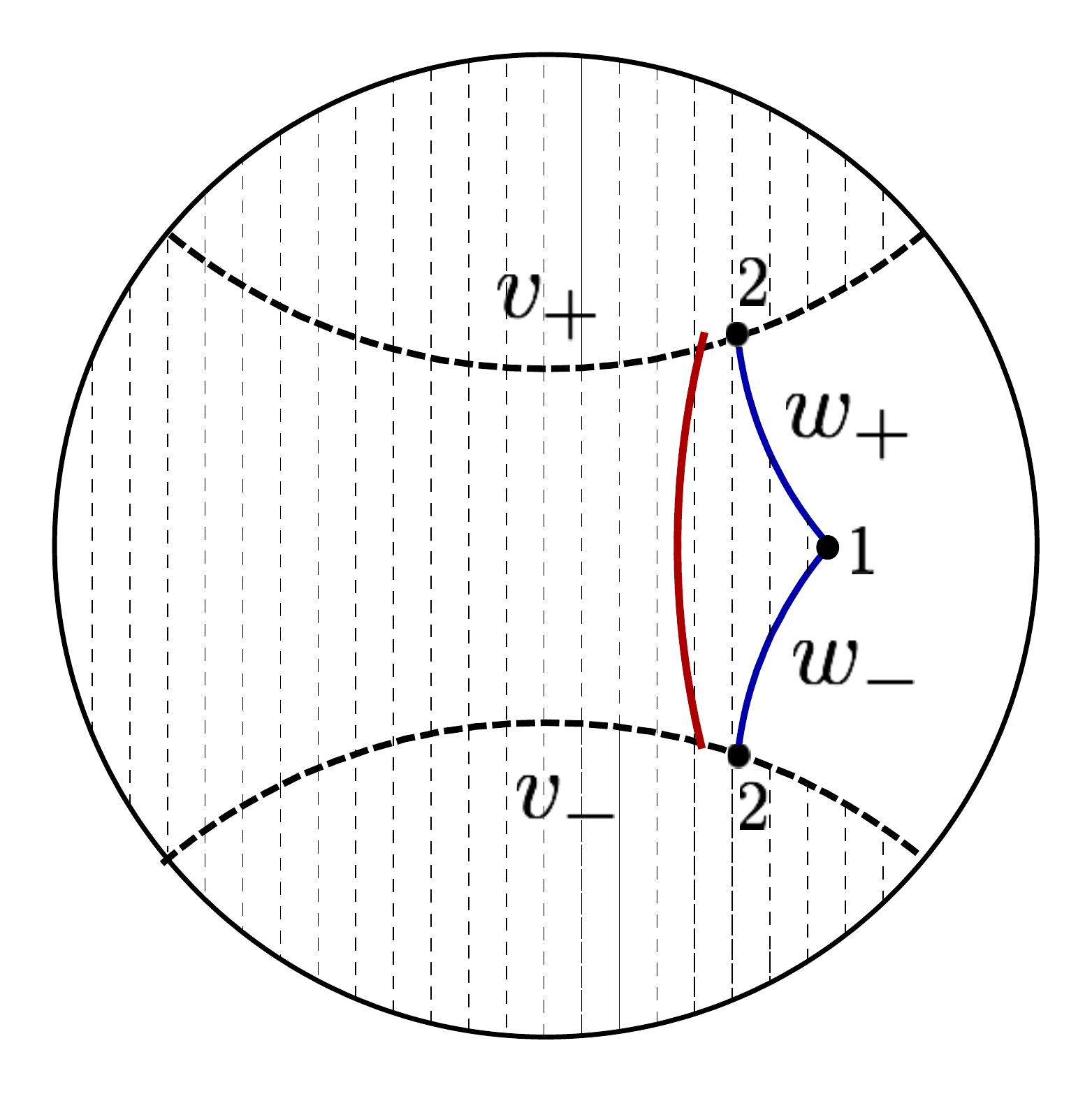}B.\\
\includegraphics[width=5cm]{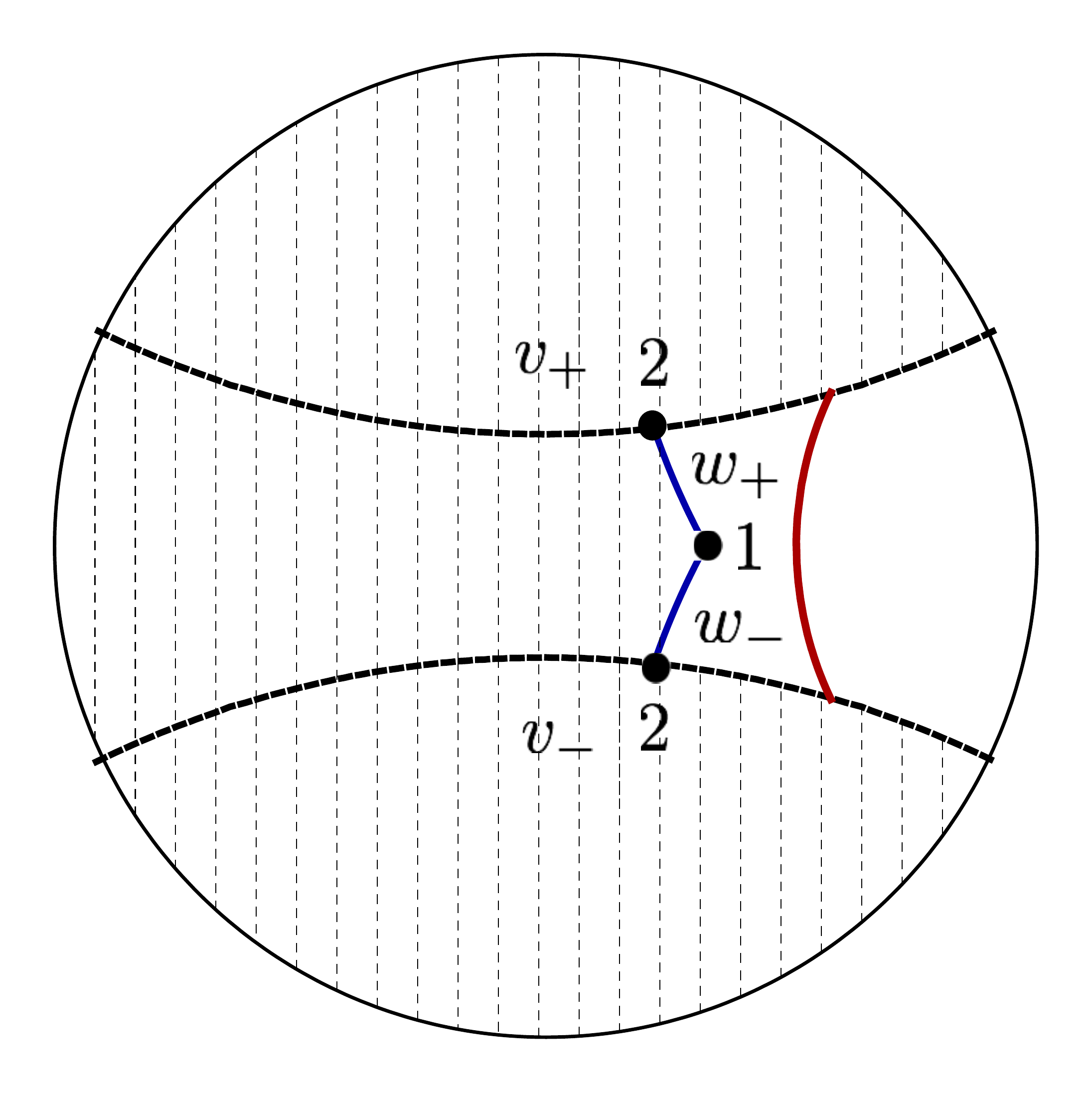}C.
\includegraphics[width=5cm]{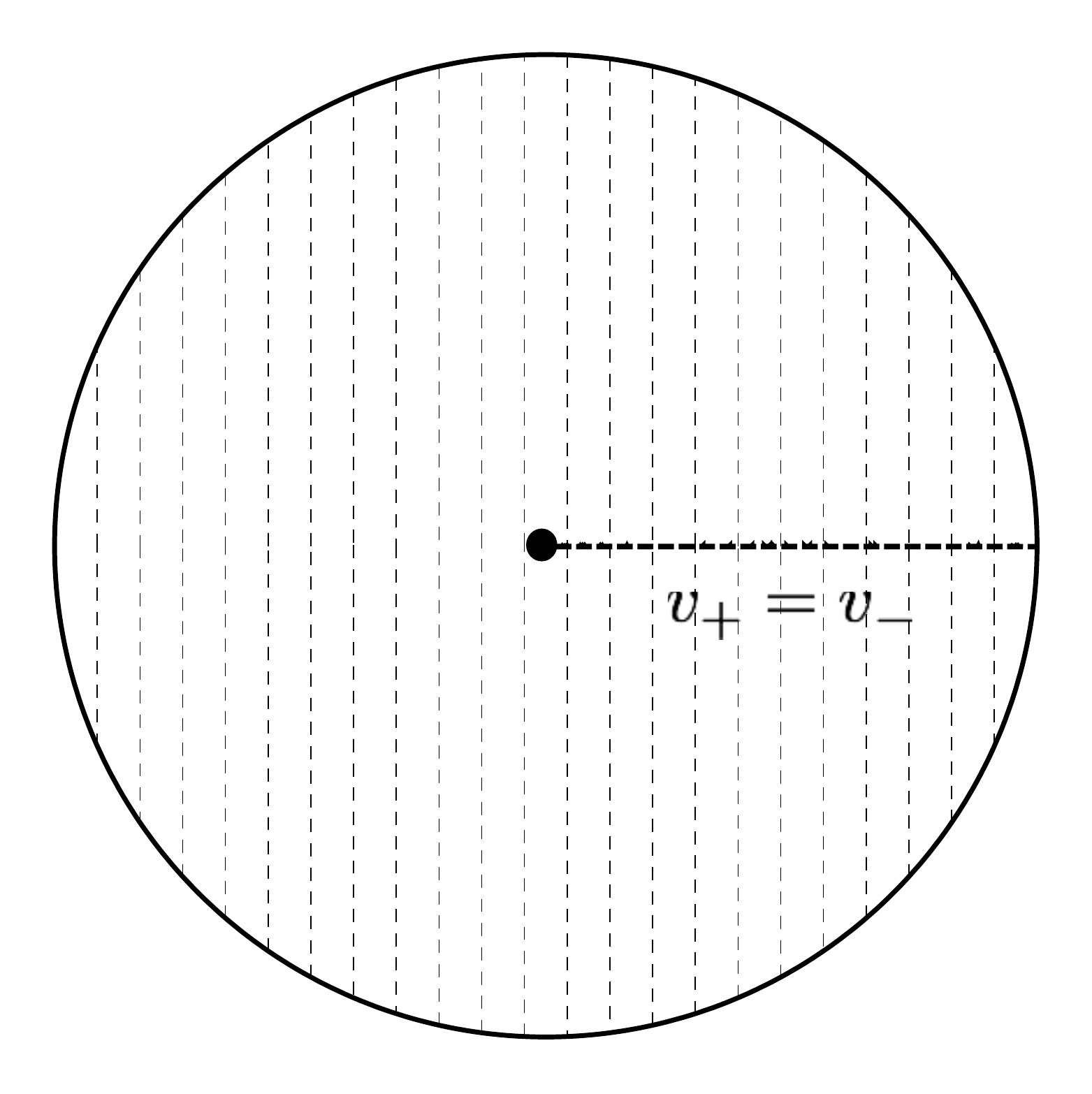}D.
\caption{Collision of particles in the BTZ rest frame. The dark red curve represents the horizon of the black hole which is about to form. The dark red curve is the apparent horizon.}
\label{falling_particles}
\end{figure}

One can show \cite{Matschull} that the second particle sits precisely on the intersection of a wedge face of the particle $1$ with an identification geodesic of the BTZ black hole. The choice of the sign corresponds to the choice of the copy of the second particle with respect to the isometry of the first particle, plus corresponding to the copy located on the $W_+$ face, and the minus sign corresponding to the copy located on the $W_-$ wedge face. The holonomy of a defect can be thought of as an action of the identification which one encounters when moving along the closed contour with the defect inside. For example, when considering a time slice of AdS with a single particle, the identification cuts out a wedge with faces $W_{\pm}$ (see Fig.\ref{masslessPic}), which are identified by the action of the holonomy:
\be
{\bf u}_1\ :\quad W_- \to W_+\,;
\ee
The surfaces $W_\pm$ are given by equation \cite{Matschull}:
\be
\tan \chi \sin(\epsilon_1 \pm \phi) = -\sin \epsilon_1 \sin \tau\,, \qquad \tan \epsilon_1 = \coth \frac{\mu}{2}\,; \label{WedgeFace}
\ee
The forming BTZ black hole is represented by another holonomy, which identifies two surfaces $V_\pm$, which are described by the equation (\ref{btz-faces}): 
\be
{\bf u}_{\text{BTZ}}\ :\quad V_- \to V_+\,;
\ee
Using these identifications, we can represent the action of ${\bf u}_{2\pm}$ defined as composition of the upper two holonomies by (\ref{U_2}). That means that once we circle around the particle $2$, we have to go through the identification $V_- \to V_+$ and through $W_+ \to W_-$ (note that the ${\bf u}_2$ enters in (\ref{U_2}) as inverse) for any closed contour which lies inside a time slice and contains \textit{only} the second particle. 

\begin{figure}[t]
\centering
\includegraphics[width=7cm]{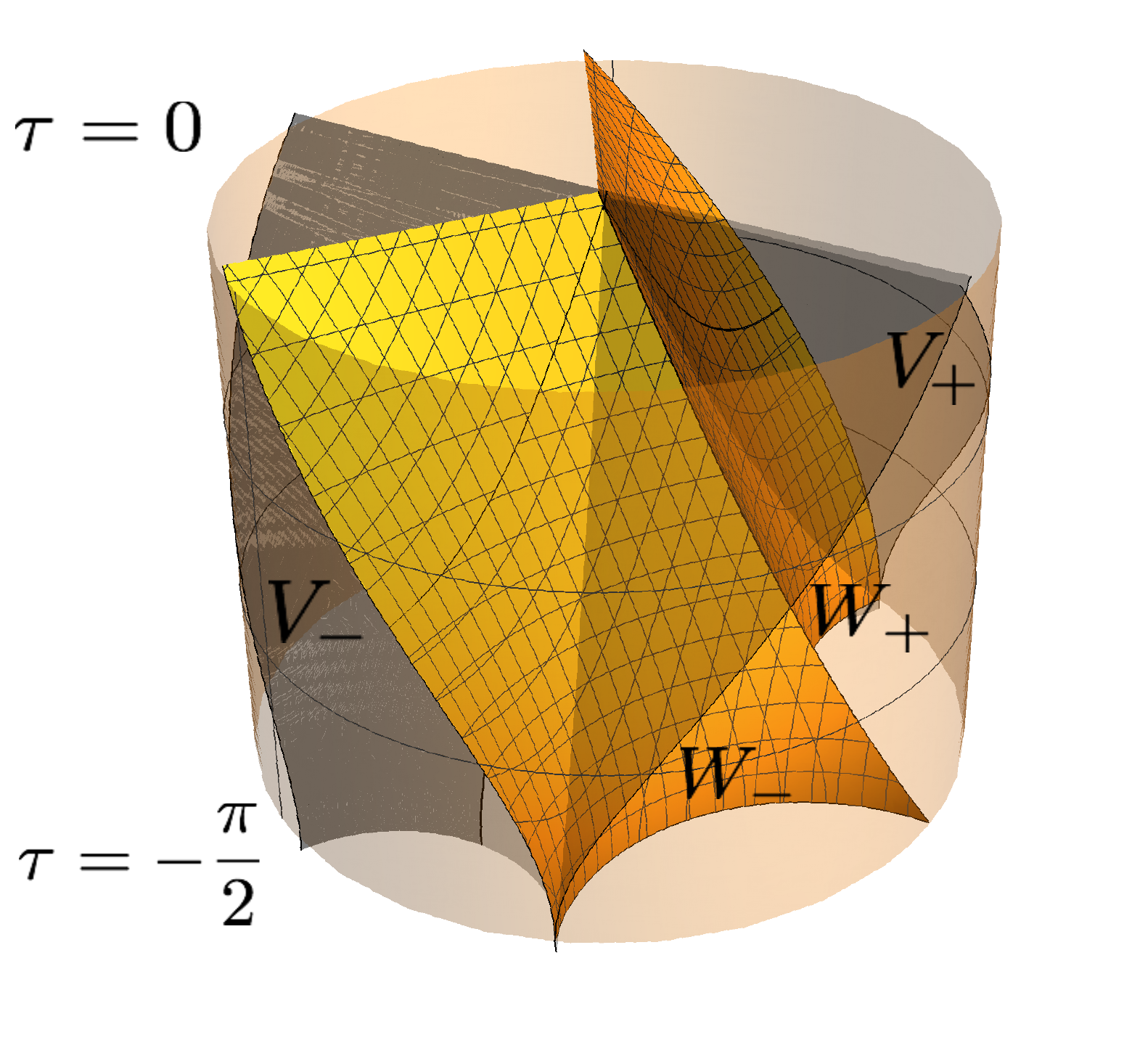}
\caption{$3$D picture of identification surfaces in AdS$_3$ which correspond to the BTZ black hole creation in the rest frame in global coordinates.}
\label{Matschull3D}
\end{figure}

For the spacetime with two particles colliding into a black hole, we have to impose two more analogous requirements. So, if we circle along a contour containing both particles, we have to pass through the BTZ identification $V_- \to V_+$. If we circle along the contour around the particle $1$, we have to pass only through the identification $W_- \to W_+$. Combining these requirements, one arrives to the conclusion that the geometry of identifications in the black hole rest frame looks as illustrated on Figs. \ref{falling_particles} and \ref{Matschull3D}. 

Having described the collision picture in global coordinates, we now have to make some important remarks. First, the black hole which is formed in the collision is not an eternal one, it has only one external region with respect to the apparent horizon. The boundary of this spacetime has only one connected component, which holographically means that this black hole is dual to a pure state in the boundary, as expected in our quench scenario. This situation is very similar to black hole formation from the cloud of collapsing dust in the AdS-Vaidya metric. However, while in the latter case the pure state black hole is usually illustrated through a Penrose diagram, we have the precise picture on Fig.\ref{falling_particles} of the full spacetime in global coordinates similar to the Penrose diagram of a pure state black hole, but with more detail because we have no spherical symmetry. In particular, the future singularity in a Penrose diagram would correspond to the moment of the collision of particles $\tau = 0$, when the spacetime in global coordinates shrinks into a singularity, see Fig.\ref{falling_particles}D. However, we emphasize that there is much more information contained this picture than in Penrose diagram, because our spacetime is not just a cut out piece of AdS$_3$, but a topological quotient. This simplifies the holographic calculation procedure, and yields some interesting details. For example, while the second external region of the BTZ black hole never becomes a part of the spacetime in the collision process and remains inside of the identification dead zone, it actually influences the behavior of holographic observables. This phenomenon will be pointed out precisely when we will discuss HRT geodesics which govern the behavior of holographic entanglement entropy. 
Another important point is that from the bulk point of view it is most intuitive to perform the diagnostic of black hole formation in the center of mass reference frame \cite{Matschull}, where particles start from the opposite sides of the AdS$_3$ cylinder and move towards each other head-on. Unlike the black hole rest frame, the center of mass frame picture also covers the case of low energies, when a static massive particle is created instead of the black hole. However, the resulting holonomy in that picture (in the high energy case) is not equal to a holonomy of a static BTZ black hole, but it is related to ${\bf u}_\text{BTZ}$ by a conjugation, which corresponds to the coordinate transformation from the black hole rest frame to the center of mass frame. However, while intuitively attractive, the center of mass picture is not a natural gravitational dual to the CFT$_2$ on a cylinder, because the living space is changing with time as wedges move, and it cannot be mapped straightforwardly to a cylinder by a simple coordinate transoformation, unlike the black hole rest frame picture. Nevertheless, the bulk spacetimes with defects which make the living space time-dependent were also studied in the context of holography, e.g. in case of a single moving particle \cite{Bal99,AAT,AKT,AB}, colliding massless particles in center of mass frame \cite{Bal99,AA}, moving particles which orbit around the origin of AdS$_3$ \cite{Arefeva2015}. 

\subsubsection{Colliding particles in BTZ coordinates} \label{222}

We are finally ready to discuss the direct holographic dual to the CFT on a cylinder which equilibrates after the bilocal quench. We make transformation from the global coordianates to BTZ coordinates introduced in section \ref{secGeometry}. The collision of particles in BTZ coordinates in AdS$_3$ was discussed previously in holographic context in \cite{Bal99} and in context of near-horizon dynamics of black holes in \cite{Jevicki02}. 

\begin{figure}[t]
\centering
\includegraphics[width=7cm]{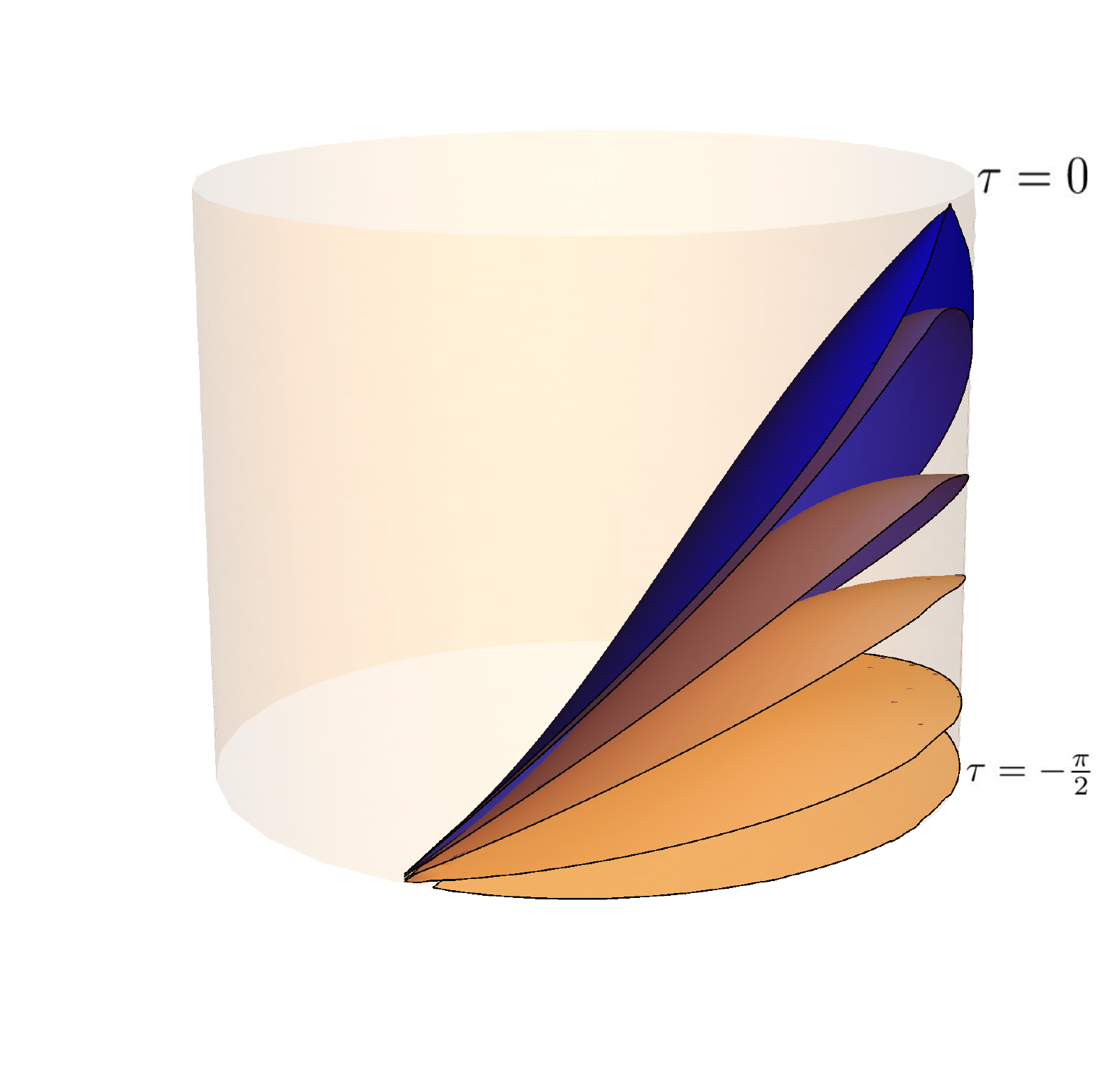}
\caption{Embedding of BTZ coordinate time slices into AdS$_3$ cylinder. The blue surface is the horizon, which coincides with the $t= \infty$ time slice.}
\label{schw-3D}
\end{figure}

The transformation formulas from global coordinates to BTZ coordinates are given by equations (\ref{Sch-transf}), and the metric is given by (\ref{schw}). We set the radius of the coordinate horizon $R = \frac{\mu}{\pi}$ and we will express all quantities appearing from this point in terms of $R$, since it is proportional to the temperature. In this case one can show that the surface $r=R$ coincides with the part of the surface of the horizon of maximally extended BTZ black hole given by equation (\ref{hor-eq}) which bounds the patch covered by our parametrization in BTZ coordinates. Hence we  see that the horizon of the black hole which is about to be formed will be located at $r=R$. The initial time slice in global coordinates $\tau = -\frac{\pi}{2}$, when particles start from the boundary, is mapped into the $t = 0$ time slice in the BTZ coordinates. Likewise, the $t = \infty$ time slice in the BTZ coordinates coincides with the horizon surface given by eq. (\ref{hor-eq}). The embedding of finite-time BTZ time slices in the global AdS$_3$ cylinder is shown on Fig.\ref{schw-3D}. Thus, the BTZ coordinates cover a patch which is outside of the horizon of the black hole, or in our case is outside of the apparent horizon of the colliding particles.

\begin{figure}[t]
\centering
\includegraphics[width=7cm]{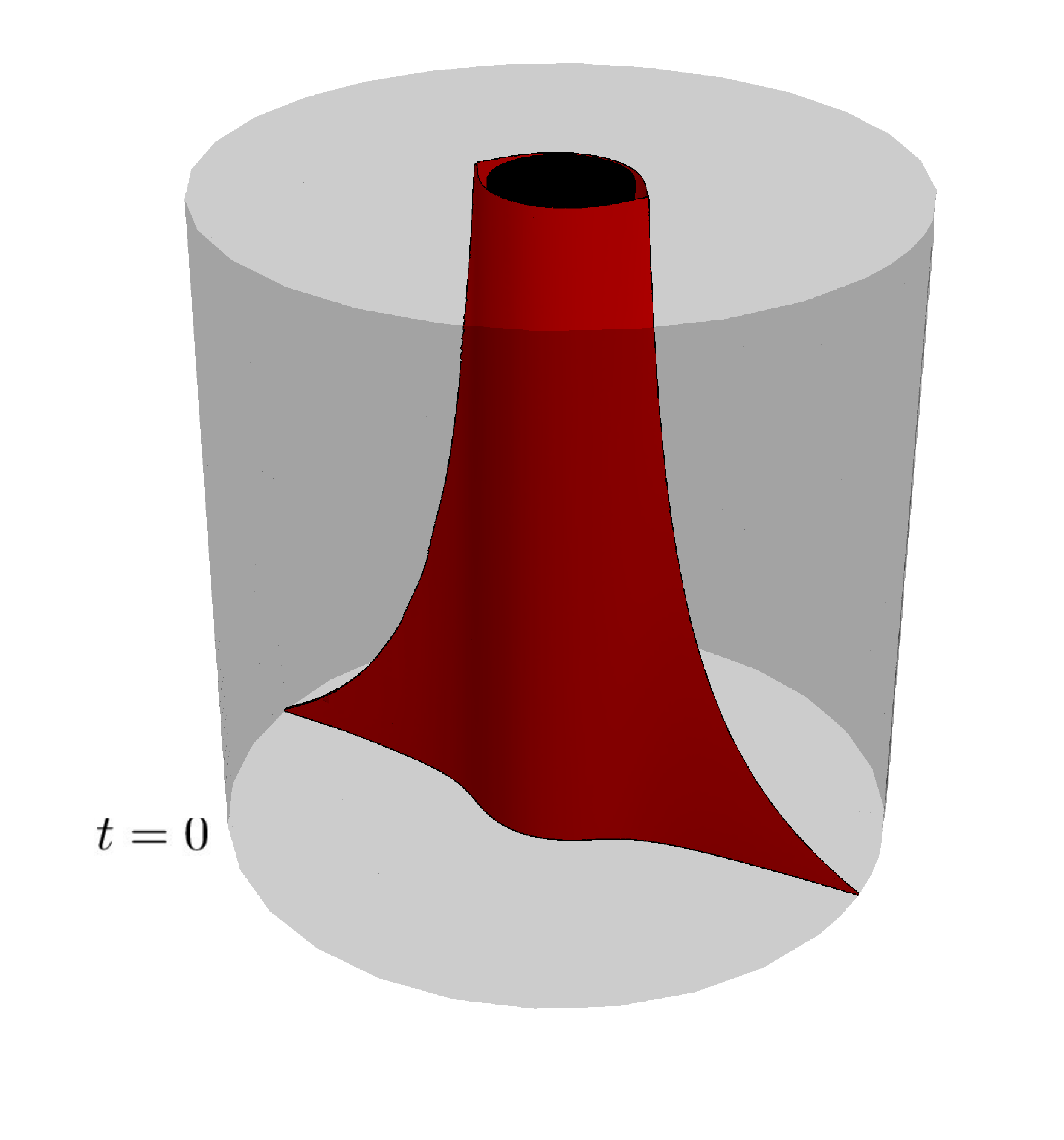}
\caption{Creation of the black hole by colliding particles in BTZ coordinates. Red surfaces are the faces of identification $W_\pm$ introduced by colliding particles. As particles move towards one another, they asymptotically approach the black hole horizon, showed as the black cylinder (which is obscured behind the identification surfaces) in the center.}
\label{schw-collision-3D}
\end{figure}

\begin{figure}[t]
\centering
\includegraphics[width=4cm]{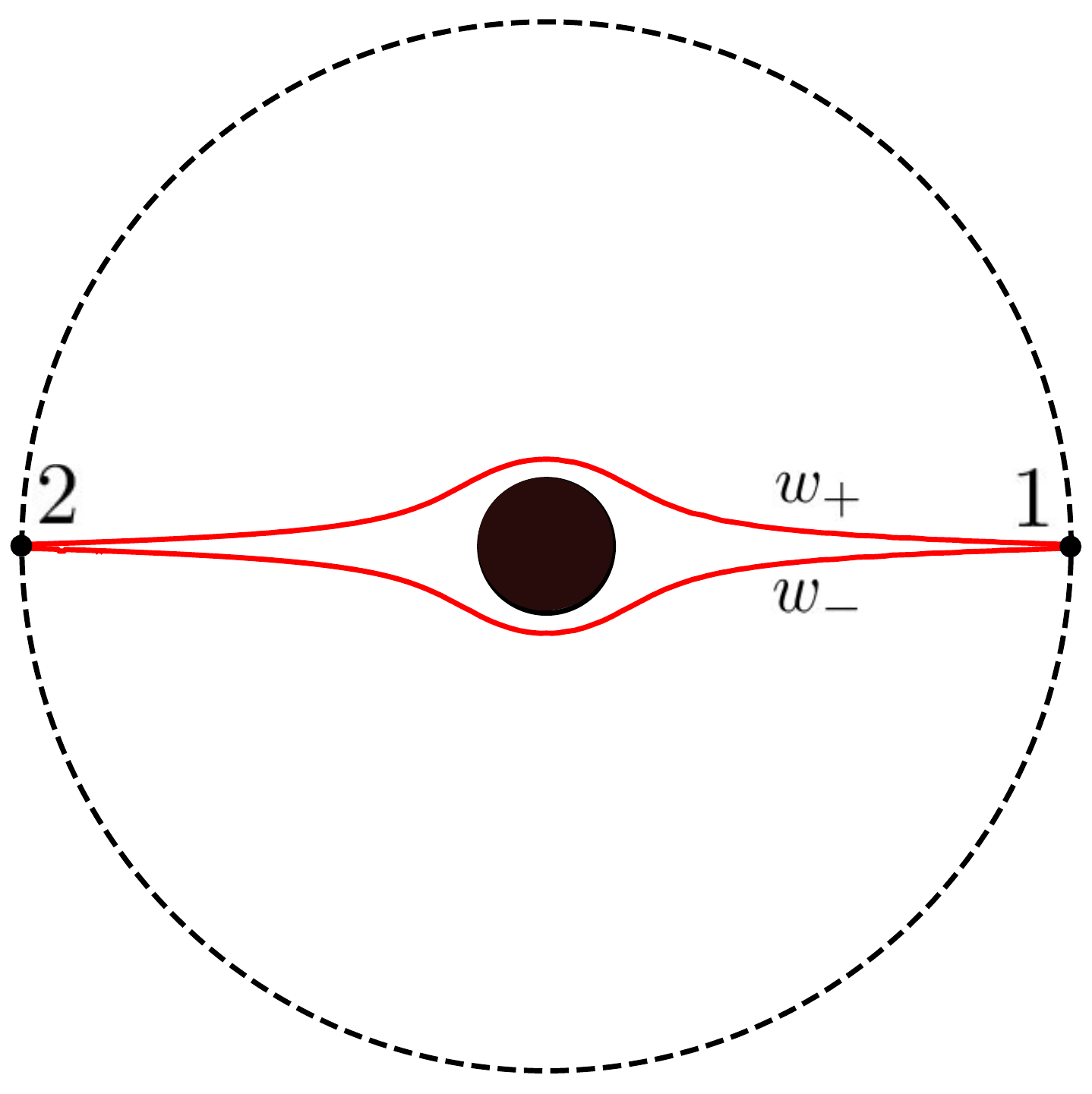}A.
\includegraphics[width=4cm]{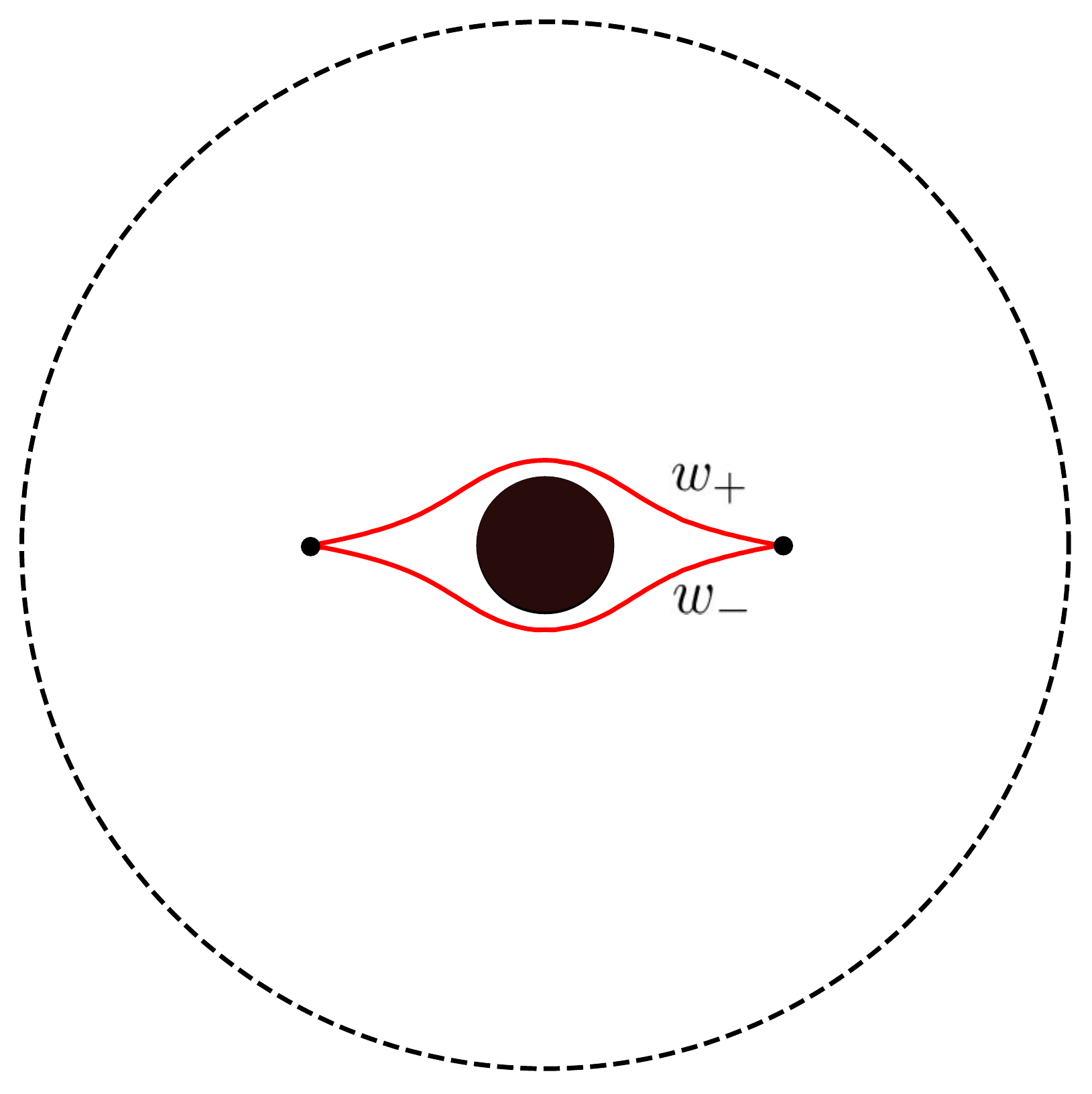}B.
\includegraphics[width=4cm]{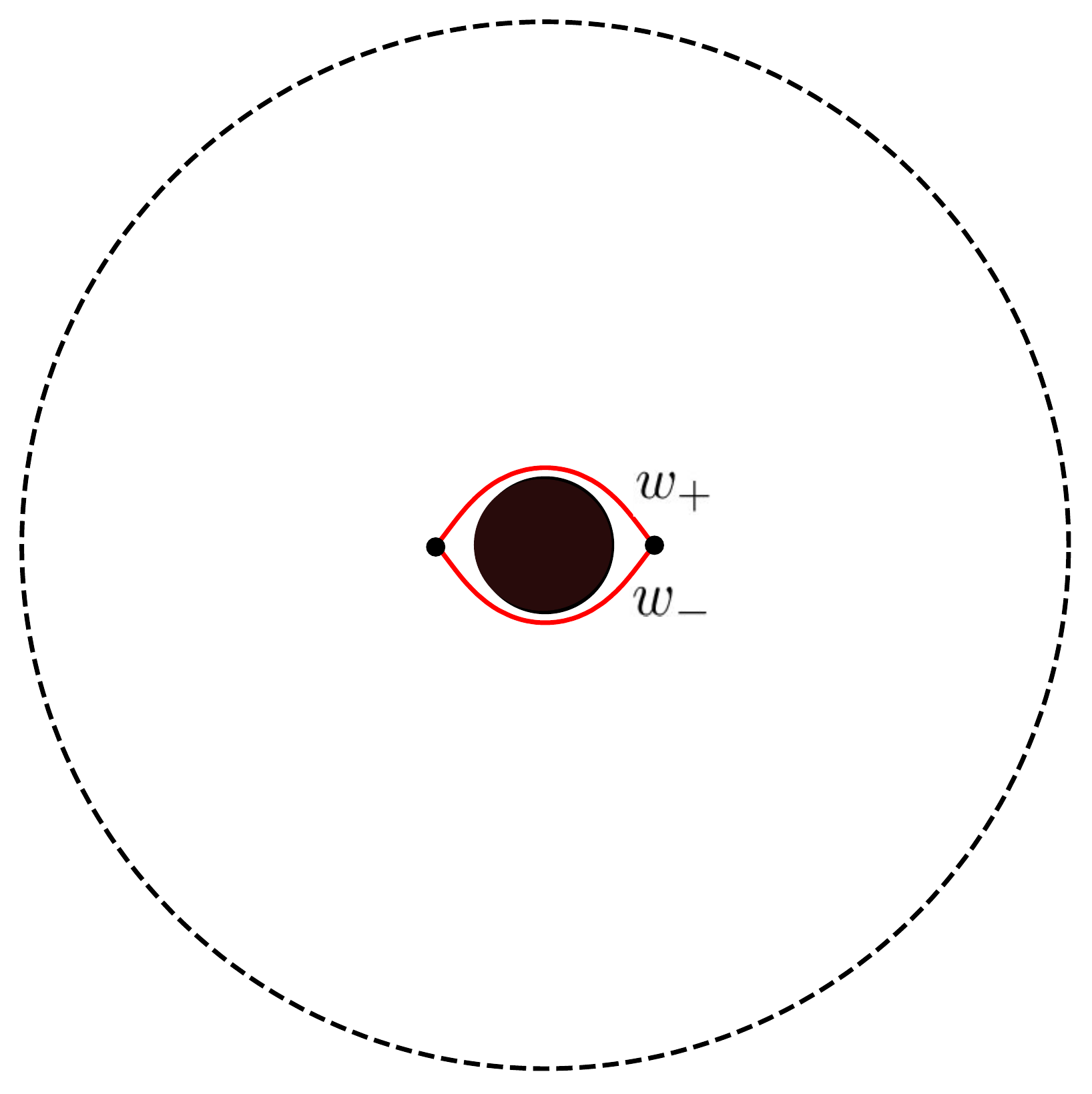}C.
\caption{Cartoon of the black hole creation in the BTZ coordinates. The space between the two red curves is the dead zone cut out by the identification. The black circle is the horizon. \textbf{A}: particles start from the boundary at $t = 0$. \textbf{B}: Particles move through the bulk towards each other. \textbf{C}: Particles asymptotically approach the horizon.}
\label{schw-cartoon}
\end{figure}

Now have to answer the question: how all the cutting and gluing on global AdS$_3$ performed in the previous section is reflected in the BTZ coordinate patch? In the global coordinates we have two sets of identification surfaces: BTZ identification $V_- \to V_+$ defined by equations (\ref{btz-faces}) and the particle $1$ wedge $W_- \to W_+$. The BTZ black hole identification in BTZ coordinates leads to the periodic condition for the angular coordinate:
\be
V_- \sim  V_+\ \Leftrightarrow\  \varphi \sim \varphi + 2\pi\,.
\ee
This identification does not depend on the BTZ coordinate time $t$, and thus the living space in these coordinates is a cylinder, which is exactly what we want. We use transformations (\ref{Sch-transf}) to get the initial data for particles in the BTZ coordinates. Dividing the first equation by the second equation in (\ref{Sch-transf}), one gets 
\be
\tanh \chi \frac{\sin \phi}{\sin \tau} = - \tanh R \varphi\,; \label{Schw1}
\ee
Taking the boundary limit of this formula and substituting the above angle values for both particles in the global coordinates, one gets $\varphi_1 = 0$ for the particle $1$ and $\varphi_2 = \pm \pi$ for the particle $2$. We choose the fundamental domain in the BTZ coordinates as $\varphi \in [-\pi, \pi]$. In this case the identification in global coordinates $\phi =  -\theta \sim \phi = \theta$ translates precisely into the identification $\varphi = -\pi \sim \varphi = \pi$. 
Thus we arrive at a picture where two particles move towards each other head-on from the antipodal points of the cylinder, particle $1$ moving along the $\varphi = 0$ and particle $2$ moving along the  $\varphi=\pi\sim -\pi$ worldline \cite{Jevicki02}, see Fig.\ref{schw-collision-3D}. 
What is left is to describe the wedge cut out by the particle $1$. To derive the equations for its faces, we transform the equation (\ref{WedgeFace}) in global coordinates to BTZ coordinates, which is the following:
\be
\tanh \chi \sin(\arctan (\coth \frac{\mu}{2}) \pm \phi) = - \sin \tau \sin (\arctan(\coth\frac{\mu}{2}))\,;
\ee
We can expand the sine of sum into two terms and use the formula (\ref{Schw1}) for one term and an analogous formula
\be
\tanh \chi \frac{\cos \phi}{\sin \tau} = -\sqrt{1-\frac{R^2}{r^2}} \frac{\cosh R\ t}{\cosh R \ \varphi}\,,
\ee
for the second term. We arrive at the following expression for the $W_\pm$ wedge faces in the BTZ coordinate patch (our choice of coordinate differs from that of Jevicki and Thaler \cite{Jevicki02} by a rescaling):
\be
\sqrt{1-\frac{R^2}{r^2}} \cosh R\ t = \cosh R \varphi \mp \tanh \frac{\mu}{2} \sinh R\ \varphi\,; \label{W+-Schw}
\ee
These identification surfaces are anchored onto worldlines of particles given by equation 
\be
r(t) = R \coth\ R\ t\,; \label{BTZworldline}
\ee
and the horizon is located inside the dead zone between $W_-$ and $W_+$. The $3D$ picture of particles moving towards each other in BTZ coordinates is shown in Fig.\ref{schw-collision-3D}, and the cartoon of time evolution is shown on Fig.\ref{schw-cartoon}. Note that from the equation (\ref{BTZworldline}) it follows that particles cannot reach the horizon in finite time. This agrees with earlier observation that we will not see the emergence of the horizon in the BTZ coordinate picture in any finite time. Holographically, this means that the state in the dual theory will always remain pure. 

We conclude this section with a brief discussion of another property of thermalization which is prominent in our holographic description. A common feature of thermalization in a closed quantum system after unitary time evolution of a certain pure state is that at late times the system loses memory about any particular details of the initial state, keeping only information about the extensive characteristics of the initial state, such as total energy, total conserved charge, etc. In our case, these "details" are the initial locations and energies of the individual particles. From the shape of the identification wedge shown on Fig.\ref{schw-collision-3D} we see that at late times the shape of the identification wedge gradually approaches the cylindrical shape, and cusps at the particle worldlines become smoother and smoother with time. One could say that as the time goes by, the bulk spacetime geometry gradually forgets about the parameters of the particles themselves. The thermal state is recovered in the limit of the infinite time, when the wedge completely falls onto the horizon, and complete rotational symmetry is restored. This kind of memory loss is not so evident in the global quench holographic duals, because by definition in the global quench scenario one deals with a translationally invariant initial state. That means that the details of the initial state are already smeared over the entire boundary time slice from the beginning (however the memory loss still absolutely can be observed in the time evolution of physical quantities such as HEE \cite{Liu13,Liu2013,Ageev17}, which we will discuss later). We are certain that one could find the same property for situations where more than two particles create a black hole, or even more complex scenarios of thermalization with a non-homogeneous initial state.

\section{Geodesics in AdS$_3$ with colliding particles} \label{sectionGeodesics}

To proceed with investigation of dynamics of entanglement and two-point correlation functions in the boundary dual of the AdS$_3$ spacetime with colliding particles, we need to study the geodesics in this spacetime with the endpoints located on the boundary. Generally speaking, we have a (locally) asymptotically AdS$_3$ spacetime with two conical singularities, moving along the lightlike worldlines. Geodesics can go from boundary to boundary directly, or they can wind around one defect, or around both of them. To avoid possible confusions, we will refer to the geodesics of the first kind as \textit{direct} geodesics, to the second kind as \textit{crossing} geodesics (the meaning of the name will be clarified in the further discussion), and to the third kind as \textit{winding} geodesics. The main task which we address in this section is to find all geodesics between two given boundary points in BTZ coordinates of colliding particles background, to calculate their lengths and to analyze what happens to geodesics when we evolve the system, that is move the boundary points along the time direction. 

To calculate the lengths of geodesics, it is most convenient to use the $SL(2, \RR)$ group formula (\ref{Lspacelike}). We are interested in geodesics between boundary points $a$ and $b$. These points are parametrized by $SL(2, \RR)$ matrices according to (\ref{SL2}), where points in embedding space are parametrized by BTZ coordinates using (\ref{Sch-coor}):
\bea
&&A = \begin{pmatrix}
  \frac{r_a}{R} \sinh R \varphi_a + \sqrt{\frac{r_a^2}{R^2}-1} \sinh R t_a & -\frac{r_a}{R} \cosh R \varphi_a + \sqrt{\frac{r_a^2}{R^2}-1} \cosh R t_a \\
  \sqrt{\frac{r_a^2}{R^2}-1} \cosh R t_a + \frac{r_a}{R} \cosh R \varphi_a & \sqrt{\frac{r_a^2}{R^2}-1} \sinh R t_a - \frac{r_a}{R} \sinh R \varphi_a\\
 \end{pmatrix}\,;\label{AandB}\\&&
B = \begin{pmatrix}
  \frac{r_b}{R} \sinh R \varphi_b + \sqrt{\frac{r_b^2}{R^2}-1} \sinh R t_b & -\frac{r_b}{R} \cosh R \varphi_b + \sqrt{\frac{r_b^2}{R^2}-1} \cosh R t_b \\
  \sqrt{\frac{r_b^2}{R^2}-1} \cosh R t_b + \frac{r_b}{R} \cosh R \varphi_b & \sqrt{\frac{r_b^2}{R^2}-1} \sinh R t_b - \frac{r_b}{R} \sinh R \varphi_b\\
 \end{pmatrix}\,;\nn
\eea
We set $r_a = r_b = r_0 >> R$ as radial cut-off near the boundary. We'll also introduce the auxiliary notation which we will use throughout the rest of the paper: 
\bea
&&t_0 = \frac12(t_a + t_b)\,, \qquad \Delta t = t_b - t_a\,; \label{t0dt}\\
&& \varphi_0 = \frac{1}{2}(\varphi_a + \varphi_b) \,, \qquad \Delta \varphi = \varphi_b - \varphi_a\,;
\eea

Using the formula (\ref{Lspacelike}), one can now find the length of a direct geodesic between spacelike-separated points $a$ and $b$. In the limit $r_0 \gg R$, it will have the form
\be 
\mathcal{L}_{\text{dir}}(a,\ b) = \log \tr A^{-1} B\,. \label{LdirTr}
\ee
All holographic quantities which we consider in this paper are expressed through lengths of specific geodesics. However, in the length formula itself (\ref{LdirTr}) there is no account for actual existence of the geodesic in the spacetime, since this formula is native to pure AdS$_3$ and not to a specific topological quotient which we consider in this paper. In order to make any holographic calculations correct in such spacetimes, one has to add to the length formulas the data about the interaction of geodesics with topological identifications. In the rest of this section, we are focusing on this issue in the case of AdS$_3$ spacetime with colliding particles described in the previous section. We will be using the parametrizations of geodesics in different coordinate systems, as well as isometry formulas from Appendix \ref{AppIsometries} and \ref{BTZgeodesics}. 

\subsection{Direct geodesics} 
\label{DirGeod}

It is known that between two given points at the boundary in the BTZ coordinates of the BTZ black hole spacetime one can construct one direct geodesic and an infinite number of geodesics which wind around the horizon (see e.g. \cite{Hubeny2013,Hubeny13,Bal14}). Once we introduce the infalling particle topological identification described by the holonomy ${\bf u}_1$, some of those geodesics will cross the identification wedge in some manner. The subject of this subsection is to explain which points on the boundary can be connected by direct geodesics in the geometry  described in sec. \ref{222}. In BTZ coordinates the identification wedge bisects the initial time slice, hence the geodesics between endpoints located on the same side to the collision line will behave differently compared to geodesics between the endpoints located to different sides of the collision line. Since the topological identification is realized by an isometry and the identification surfaces intersect the time slices along the pieces of boundary-to-boundary geodesics themselves, we can prove some statements about the behavior of the geodesics. 

\begin{Proposition} \label{DirectSide}
Suppose that $a$ and $b$ are spacelike-separated points on the boundary such that either $\varphi_a,\ \varphi_b \in (-\pi, 0)$, or $\varphi_a,\ \varphi_b \in (0, \pi)$. Then there always exists a direct geodesic between these two points at any given moment of time and any time separation.
\end{Proposition}

\begin{figure}[t]
\centering
\includegraphics[width=6cm]{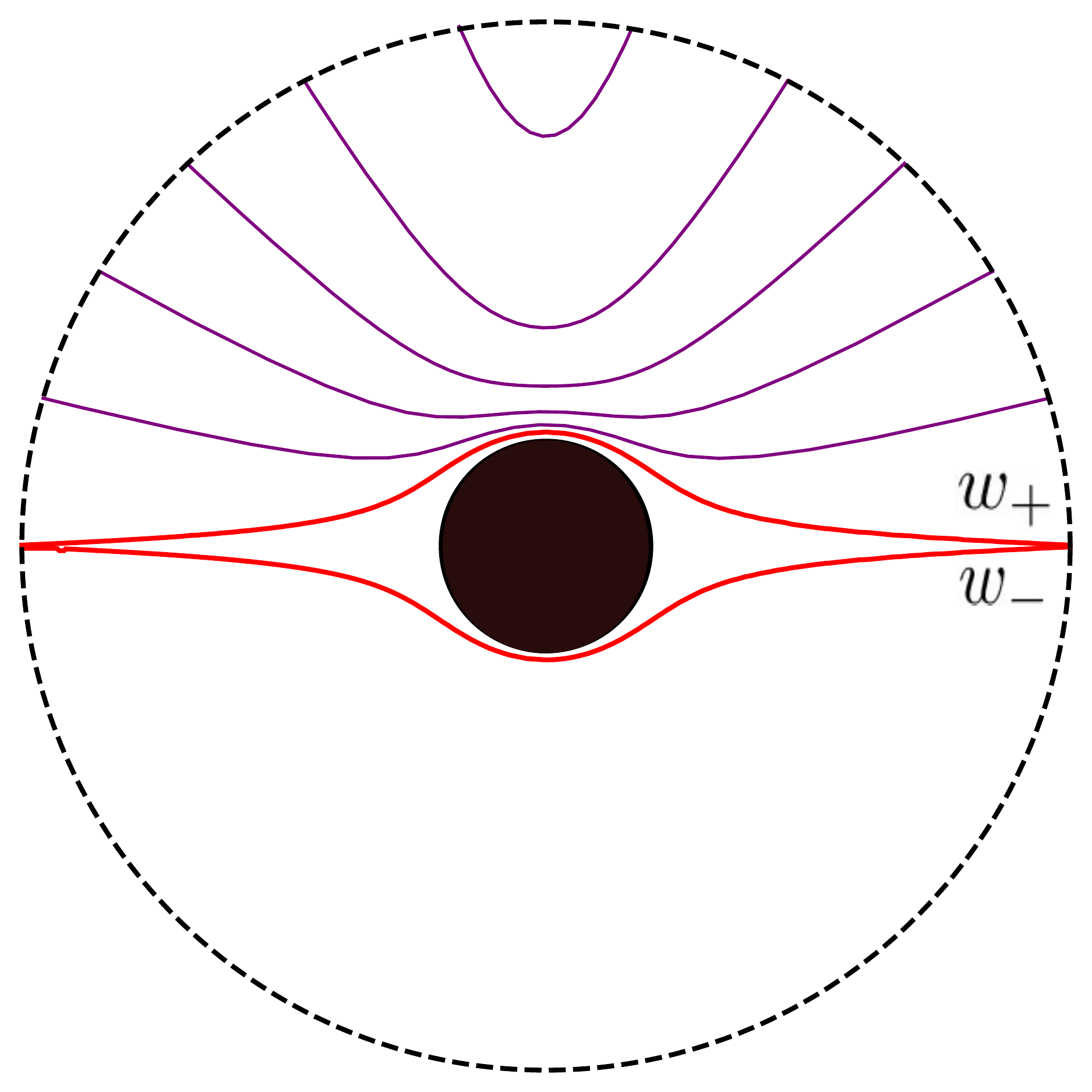}
\caption{Direct equal-time geodesics in the upper half of the initial time slice}
\label{direct-upper}
\end{figure}

\textbf{Proof}. To prove this proposition, we observe that the geodesics $w_\pm$ in BTZ coordinates shown on the Fig.\ref{schw-cartoon} are themselves parts of equal-time boundary-to-boundary geodesics. A direct geodesic between two given boundary endpoints can cease to exist if it somehow reaches $w_\pm$. The depth of the geodesic, i. e. minimum radial distance from the origin to the geodesic in the bulk, is given by the formula (\ref{Gamma+}):
\be
\Gamma_+^2 = R^2 \frac{1+\tanh^2 \frac{R \Delta t}{2}}{\tanh^2 \frac{R \Delta t}{2} + \tanh^2 \frac{R (\Delta \varphi)}{2}}\,; \label{Gamma+1}
\ee
where we set $n=0$ since we are only interested in direct geodesics. We can directly compare the $\Gamma_+$ to the distance $r_\pm$ from origin to $w_\pm$, which we can determine from the equations (\ref{W+-Schw}). Solving in terms of the radial variable, one gets 
\be
r^2 = R^2 \left(1-\left(\frac{\cosh R \varphi \mp \tanh \frac{\pi R}{2} \sinh R \varphi}{\cosh R t}\right)^2\right)^{-1}\,.
\ee
The r.h.s. is minimal at $\varphi = \pm \frac{\pi}{2}$, which is indeed clear from the symmetry of the wedge, see Figs.\ref{schw-collision-3D}-\ref{schw-cartoon}. It gives the distance
\be
r_\pm^2 = \frac{R^2}{\tanh^2 R t\ (1-\tanh^2 \frac{\pi R}{2}) + \tanh^2 \frac{\pi R}{2}}\,.\label{rpm}
\ee
First, let us restrict ourselves to the case of equal-time geodesics, $\Delta t =0$. In this case $r_\pm$ equals $\Gamma_+$ for $\Delta \varphi = \pi$. Since by assumptions of the proposition we consider only points in upper or lower parts of the boundary, $\Delta \varphi < \pi$, we have $r_\pm < \Gamma_+$ for all equal-time geodesics, as shown on Fig.\ref{direct-upper} in case of $t=0$, when the wedge takes up the most space in the bulk.  
Now we only have to prove that $r_\pm < \Gamma_+$ for non-equal-time geodesics. From equations of geodesics (\ref{r(l)}-\ref{t(l)}) it is clear that a geodesic reaches deepest into the bulk in the moment $t_0 = \frac12(t_a + t_b)$. Therefore it is this moment which makes sense as the edge case in (\ref{rpm}) when a geodesic could possibly try to reach $r_\pm$. Further, $\Gamma_+$ depends on $\tanh \frac{R \Delta t}{2}$, whereas $r_\pm$ depends on $\tanh R t$. It is true that $t \geq \frac{\Delta t}{2}$, and the inequality is saturated when either $t_a$ or $t_b$ is zero.  Both $r_\pm$ and $\Gamma_+$ are decreasing functions of their respective temporal arguments, and their initial values coincide if $\Delta t = 2 t_0$. The question that remains is how fast these functions decrease with time compared to each other. We have to compare two functions, which are proportional to the inverse of (\ref{Gamma+1}) and (\ref{rpm}), respectively: 
\be
f(x):=x ( 1-y) + y\,, \qquad g(x):= \frac{x+y}{1+x}\,;
\ee
where $x = \tanh^2 R t \in [0,\ 1]$ is a variable and $y = \tanh^2 R \frac{\pi}{2} < 1$ is a constant. Obviously $f(0) = g(0)$, but for $x > 0$ we have $f(x) > g(x)$, since one can expand $g(x)$ around $x = 0$ as follows: 
\be
g(x) = x(1 - x + O(x^2)) + y (1 - x +x^2 + O(x^3)) = f(x) - (1-y) x^2 + O(x^3)\,;
\ee 
(both functions are monotonic).
Thus we conclude that $r_\pm < \Gamma_+$ for any direct geodesic, and the proposition is proved \qed.
\\
$\,\,\,$
\\
Now let us consider the situation when the direct geodesics not always exist, namely when the boundary segment between the endpoints crosses the collision line of particles. Suppose that $a$ and $b$ are spacelike-separated points on the boundary such that $\varphi_a,\in (-\pi,\ 0)$ and  $\varphi_b \in (0,\ \pi)$, and $\Delta \varphi < \pi$ ($\Delta \varphi > \pi$). Then by definition the direct geodesic between points $a$ and $b$ exists if the geodesic does not intersect the identification wedge. When varying $t_a$ and/or $t_b$, we observe that the edge case is when it intersects the worldline of the particle $1$ (particle $2$). 

\begin{figure}[t]
\centering
\includegraphics[width=6cm]{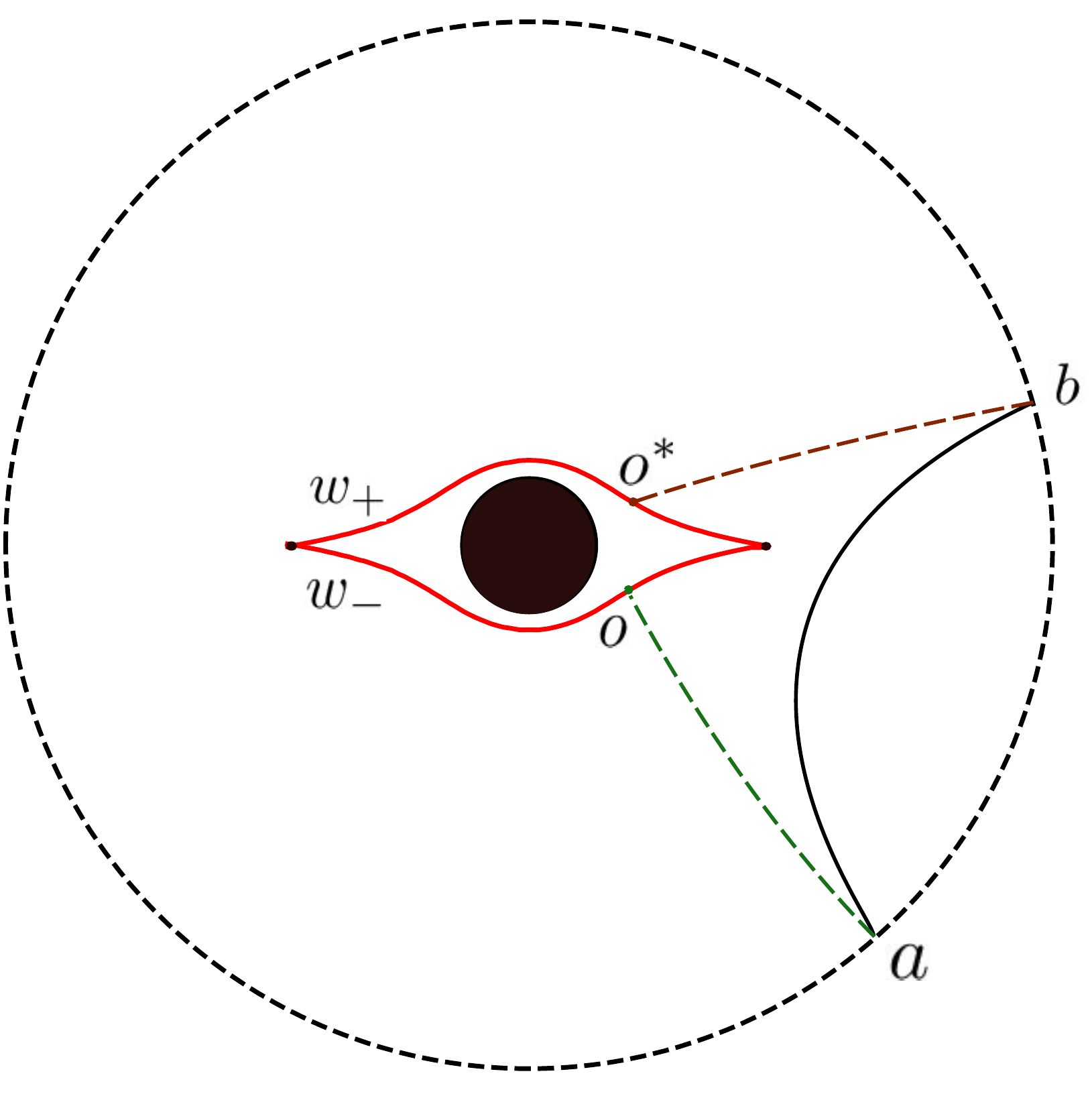}A.
\includegraphics[width=5cm]{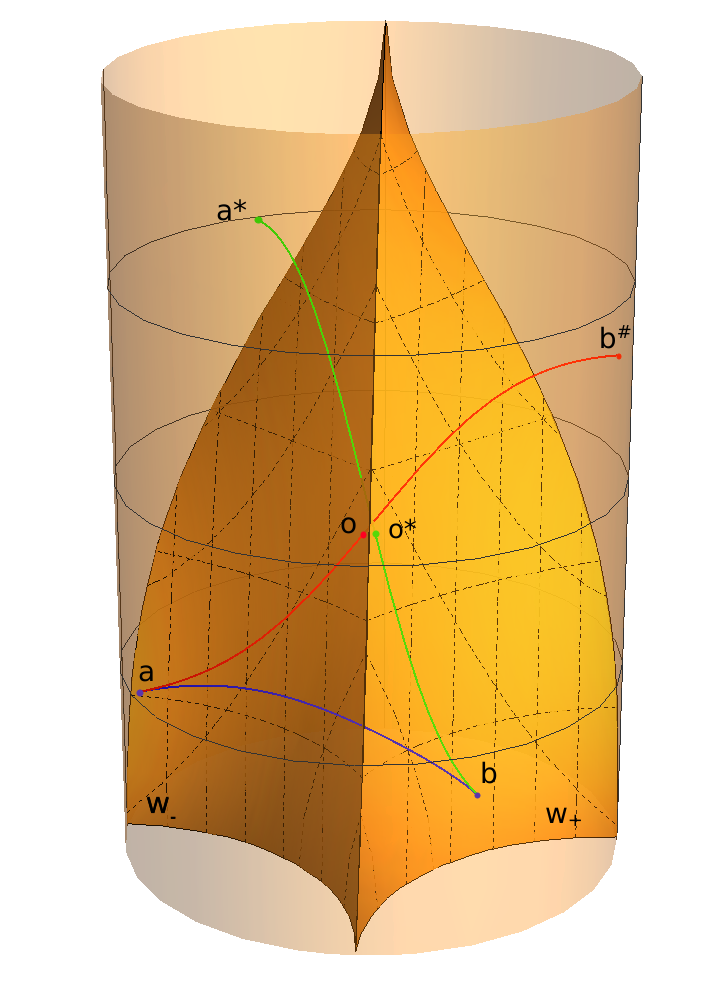}B.
\caption{\textbf{A}. Illustration of projection of direct and crossing geodesics on a time slice at $t > 0$. The solid black curve is a direct geodesic between $a$ and $b$. The dashed curves are pieces of the crossing geodesic between $a$ and $b$. \textbf{B}. A plot of direct geodesic and image geodesics which constitute the crossing geodesic in the $3$D global black hole rest frame picture. The BTZ identification surfaces as well as apparent horizon are not drawn to avoid cluttering.}
\label{crossing-basic}
\end{figure}
We have established the conditions of when the direct geodesics exist and when they do not. Further we will elaborate more on this in case of equal-time geodesics. For now we conclude this subsection by giving the expression of the regularized length of direct geodesic from (\ref{LdirTr}). Taking the trace, one obtains the expression (\ref{geodesicLength}) with $n = 0$:
\be
\mathcal{L}_{\text{dir}}(a,\ b) = \log \left[2(\cosh [R (\varphi_b - \varphi_a)] - \cosh [R(t_b - t_a)])\right]+2 \log \left(\frac{r_0}{R}\right)\,. \label{Ldirect}
\ee
This expression is valid only for $\Delta \varphi \leq \pi$. For $\Delta \varphi > \pi$, the direct geodesic and the $n = -1$ winding geodesic change places, so in that case the length of the direct (minimal) geodesic is given by 
\be
\mathcal{L}_{\text{dir}}(a,\ b) = \log \left[2(\cosh[R(\varphi_b - \varphi_a-2\pi)] - \cosh[R(t_b - t_a)])\right]+2 \log \left(\frac{r_0}{R}\right)\,. \label{Ldirect2}
\ee

\subsection{Crossing geodesics}
\label{CrossGeod}

The crossing geodesic consists of two pieces of geodesics going from the endpoints to the identification surfaces. More specifically, suppose we have the endpoints located as follows: $\varphi_a \in [-\pi, 0)$, $\varphi_b \in [0, \pi)$, and $t_a$, $t_b$ chosen in such a way that points are spacelike-separated. In this case the geodesic will consist of two pieces: a geodesic from $a$ to point $o \in W_-$ and a geodesic from $o^* \in W_+$ to point $b$, see Fig.\ref{crossing-basic}B\footnote{The image geodesics on this and other similar figures were plotted using parametric representation of geodesics and explicit formulas for the action of identification isometries in global coordinates. They are presented for reference in Appendix \ref{AppIsometries}.}. The surfaces $W_\pm$ are topologically identified, which is represented by the isometry, which acts according to the rule
\be
*:\quad X \mapsto X^* := {\bf u}_1^{-1} X {\bf u}_1\,; \label{Star}
\ee
We will also need the inverse identification isometry, which is defined as follows: 
\be
\# = *^{-1} :\quad X \mapsto X^\# := {\bf u}_1 X {\bf u}_1^{-1}\,; \label{Diez}
\ee
These isometries act on the identification surfaces $W_\pm$ as follows: 
\bea
&& *:\quad W_- \to W_+ = {\bf u}_1^{-1} W_- {\bf u}_1\,:  o \mapsto o^*\,, \\
&& \#:\quad W_+ \to W_- = {\bf u}_1 W_+ {\bf u}_1^{-1}\,:  o^* \mapsto o\,.
\eea
This enables us to use the geodesic image method \cite{AAT,AA,AKT} to find the length of the crossing geodesic. Since $*$ is an isometry, we have
\be
\mathcal{L}(a,\ o) = \mathcal{L}(a^*,\ o^*)\,;
\ee
On the other hand, we can define the inverse isometry $\#:\quad W_+ \to W_- = {\bf u}_1 W_+ {\bf u}_1^{-1}$, so that
\be
\mathcal{L}(b,\ o^*) = \mathcal{L}(b^\#,\ o)\,;
\ee
Therefore, the length of the crossing geodesic can be found as 
\be
\mathcal{L}_{\text{cross}}(a,\ b) := \mathcal{L}(a,\ o) + \mathcal{L}(b,\ o^*) =  \mathcal{L}(a,\ o) + \mathcal{L}(b^\#,\ o) = \mathcal{L}(a,\ b^\#)\,; \label{Labdiez}
\ee
or, equivalently,
\be
\mathcal{L}_{\text{cross}}(a,\ b) =  \mathcal{L}(a^*,\ o^*) + \mathcal{L}(b,\ o^*) = \mathcal{L}(a^*,\ b)\,. \label{La*b}
\ee
Thus, the length of the crossing geodesic between points $a$ and $b$ is equal to the length of the image geodesic from $a$ to $b^\#$ or from $a^*$ to $b$, and the crossing geodesic can be completely recovered from image geodesics and the identification surfaces. These image geodesics themselves are just regular geodesics in \textit{global} AdS$_3$ spacetime. In our case of colliding massless particles in AdS$_3$, we illustrate the behavior of image geodesics\footnote{In this subsection we discuss only images that are obtained by action of the massless particle holonomy ${\bf u}_1$.} in the black hole rest frame picture on the Fig.\ref{crossing-basic}B (BTZ identification is not shown on the picture). 

All information about the matter which produces the topological identification is encoded in the position of image points $a^*$ and $b^\#$. These points themselves can be generally located anywhere in AdS$_3$. Because of this, one has to exercise caution when working in any coordinate patch which does not cover the AdS$_3$ spacetime globally, such as the BTZ coordinate patch. While the endpoints $a$ and $b$ belong to the BTZ coordinate patch, the image points, generally speaking, do not. Since the image points generically do not belong to the BTZ coordinate patch, we are in a tricky situation: while the pieces $a o$ and $o^* b$ of a crossing geodesic do lie within the BTZ patch, image geodesics as a whole do not. However, image geodesics are the most convenient way to describe crossing geodesics, and we can make use of this machinery in global coordinates to prove some facts about crossing geodesics in BTZ coordinates. 

We begin again with establishing the conditions of existence of a crossing geodesic is that image geodesics \textit{must intersect (different) identification surfaces}. This ensures that the actual crossing geodesic will indeed run around the defect. Keeping this in mind, one can formulate some useful statements. Let us note that the above discussion of image geodesics representation for crossing geodesics is valid only for points located on different sides of the boundary. The following statement says that it is, in fact, the only set of situations when we encounter crossing geodesics.

\begin{Proposition} \label{nogoCrossing}
There are no crossing geodesics between the endpoints located to the same side of the identification wedge.
\end{Proposition} 
\textbf{Proof}. This statement is not completely obvious in BTZ coordinates, but it is almost trivial in global coordinates. All the geodesics in this case which we deal with are boundary-to-boundary spacelike geodesics in global AdS$_3$. A geodesic which starts from one side of the identification wedge (e.g. from the left on Fig.\ref{crossing-basic}B) and goes to $W_-$ comes out of the $W_+$ and has to go to the other side of the boundary, to the right in Fig.\ref{crossing-basic}B. Also a spacelike geodesic starting from the left cannot reach the $W_+$ surface first, before reaching the $W_-$. An important point to note is that once a geodesic leaves the identification wedge $W_\pm$, it cannot enter it once more. The boundary-to-boundary geodesics also cannot go through both $V_\pm$ and $W_\pm$ identifications, which is again evident from the black hole rest frame in the global coordinates. That means that winding geodesics cannot be also crossing, and vice versa. In the BTZ coordinates this is also clear from the fact that the part of winding geodesics which wraps around the horizon would have been lying inside of the cut out region close to the horizon, since the behavior given by formulas from the Appendix \ref{BTZgeodesics} dictates that winding geodesics always reach closer to the horizon than direct ones (more on that in the subsection \ref{windings}), and the surfaces $W_\pm$ can be considered as foliations of segments of direct geodesics in analogy to the considerations from the proof of the Proposition \ref{DirectSide}.
These arguments imply the uniqueness of the crossing geodesic constructed from the image method as above, as well as the statement of the proposition. An important corollary is that the minimal spacelike geodesic connecting two points on the same side of the boundary is always the direct one. \qed
\\
$\,\,\,$
\\
Now suppose that $a$ and $b$ are spacelike-separated points located on different sides of the boundary relative to the collision line. Then the crossing geodesic exists as long as image geodesics intersect the identification wedge\footnote{Since the particle itself moves along the lightlike geodesic and image geodesics have to intersect $W_\pm$, image geodesics which constitute an existing crossing spacelike geodesic are always spacelike as well.}. The edge case in this situation is when $o = o^*$ belongs to the worldline of the particle. We have similar picture for direct geodesics in this case, which do \textit{not} exist until they intersect the worldline. In the following subsection we begin to address this question in more detail in case of equal-time boundary endpoints.  

We conclude this subsection by calculating the length of crossing geodesic in terms of BTZ coordinates of endpoints. As discussed above, it equals the length of a winding geodesic, either $a b^\#$ or $a^* b$. Suppose that endpoints $a$ and $b$ are parametrized as matrices according to (\ref{AandB}). We write
\be
\mathcal{L}_{\text{cross}}(a,\ b) = \mathcal{L}(a,\ b^\#) = \log\ \tr A^{-1} B^\#\,, \quad \text{where} \quad B^\# := {\bf u}_1 B {\bf u}_1^{-1}\,;
\ee
We again note that $b^\#$ does not necessarily belong to the BTZ patch, however it does not matter since we do not have to calculate the actual coordinates of $b^\#$. To calculate the length, we take the trace: 
\be
\mathcal{L}_{\text{cross}}(a,\ b)= \log\ \tr A^{-1}{\bf u}_1 B {\bf u}_1^{-1}\,; \label{LcrossingTrace}
\ee
where ${\bf u}_1$ is the holonomy of the particle given by (\ref{U2}). We introduce auxiliary notations 
\bea
&& \Delta t = t_b - t_a\,; \qquad t_0 = \frac12(t_b + t_a)\,; \\
&& \Delta \varphi = \varphi_b - \varphi_a\,; \qquad \varphi_0 = \frac12(\varphi_b + \varphi_a)\,;
\eea
Using the formula (\ref{U2}) for the holonomy and the matrix parametrization of the endpoints (\ref{AandB}), we come to the resulting expression:
\bea
&& \mathcal{L}_{\text{cross}}(a,\ b) = \log\left[2 \left((-1+\mathcal{E}^2) \cosh R \Delta t+(1 + \mathcal{E}^2) \cosh R \Delta\varphi + \mathcal{E}^2 \cosh 2 R \varphi_0 +\right.\right.\nn\\&& \left.\left. \mathcal{E}^2 \cosh 2 R t_0 + 4\mathcal{E} \cosh R t_0 \cosh R \frac{\Delta t}{2} \cosh R \varphi_0 \left(\sinh R \frac{\Delta\varphi}{2} - \mathcal{E} \cosh R \frac{\Delta\varphi}{2}\right) +\right.\right.\\\nn && \left.\left. 4\mathcal{E} \sinh R t_0 \sinh R \frac{\Delta t}{2} \sinh R \varphi_0 \left(\mathcal{E}\sinh R \frac{\Delta\varphi}{2} - \cosh R \frac{\Delta\varphi}{2}\right)  - 2 \mathcal{E} \sinh R \Delta \varphi\right)\right]\,. \label{Lcrossing}
\eea
This expression has some symmetries: 
\begin{enumerate}
\item The $\ZZ_2$-symmetry of the bulk background under reflection with respect to the collision line: $\varphi_a \to -\varphi_b$, $\varphi_b \to -\varphi_a$;
\item Replacing particle $1$ with particle $2$: $\varphi_a \to \varphi_b-\pi$, $\varphi_b \to \varphi_a + \pi$;
\item The symmetry between temporal and spatial coordinates of the center of the boundary segment on which the geodesic is anchored: $t_0 \leftrightarrow \varphi_0$.
\end{enumerate}
The first two symmetries are not surprising, but the last one is a somewhat unexpected unique feature of our bulk spacetime geometry. It is also worth noting that $\mathcal{L}_{\text{cross}}$ is is a monotonically increasing function of $t_0$ and $\varphi_0$. 

\subsection{ETEBA geodesics}
\label{HRTGeod}

The holographic entanglement entropy of a subsystem in the dual of a non-stationary bulk spacetime is calculated using the HRT prescription \cite{HRT} as minimal surface anchored on the boundary region where the subsystem lives. In our case, this surface is a geodesic in BTZ coordinates anchored onto a segment $[\varphi_a,\ \varphi_b]$ on the boundary and with $t_a = t_b = t_0$. In this subsection, we focus on such geodesics with equal-time boundary endpoints, which were labeled by Hubeny and Maxfield \cite{Hubeny2013} as "equal-time-endpoint boundary anchored", or ETEBA geodesics. We will follow \cite{Hubeny2013,Ziogas,Ageev17} and use this terminology. We are particularly interested in their behavior during time evolution. The HRT geodesic which computes the entanglement entropy is the minimal ETEBA geodesic which can connect a given pair of endpoints. For this reason, we also address the issue of existence of ETEBA geodesics of different types to know when they can and cannot participate in the HRT prescription. 

\subsubsection{Direct equal-time geodesics}
These geodesics can be parametrized using equations (\ref{r(l)},\ref{phi(l)},\ref{t(l)}). In the equal-time case, they have the form (we set $\lambda_0$ to zero):
\bea
r(\lambda)^2 &=& R^2 + (\Gamma_+^2 - R^2) \cosh^2 \lambda \label{r(l)eqt}\,,\\
\varphi(\lambda) &=& \varphi_0 + \frac{1}{R} \arctanh \left( \frac{R}{\Gamma_+} \tanh \lambda \right)\,,\label{phi(l)eqt}\\
t(\lambda) &=& t_0\,;\label{t(l)eqt}
\eea
where \be
\Gamma_+ = \frac{R}{\tanh \frac{R (\Delta \varphi)}{2}}\,. \label{G2}
\ee
From these equations it follows that these geodesics lie completely in the time slice $t_0$, and their shape or length given by (\ref{Ldirect}) does not depend on time. The question which we are interested in is at what times a direct geodesic exists between given $\varphi_a \in (-\pi, 0)$ and $\varphi_b \in (0, \pi)$. 
The direct geodesic does not exist when it crosses surfaces $W_\pm$. The worldlines of particles go into the bulk towards the horizon, and they are the closest points of the identification wedge to the boundary in any given time slice, see Fig.\ref{schw-cartoon}. From the initial moment of time, the direct geodesic would have to intersect $w_\pm$ and pass through the dead zone inside the wedge, therefore they do not exist for some time. However, as we evolve the system, the wedge gets smaller, and eventually it will get small enough to let the direct geodesic run clear of the identification fully inside the fundamental domain. From the shape of direct geodesics (which is constant under simultaneous evolution of both boundary points), shown e.g. in Fig.\ref{initial-entangle}B, it is clear that the worldlines of particles are the last points of the shrinking wedge which geodesics can touch before going completely out of the dead zone. The moment of emergence of a direct geodesic is illustrated in Fig.\ref{crossing-basic}A. The geodesic $a_2 b_2$ there touches the worldline. After this moment, the particles will move further into the bulk, the wedge will shrink more and the geodesic will lie completely in the fundamental domain of the identification, similarly to the smaller geodesic $a_1 b_1$. Note that this argument holds both in case when we vary $t_0$ keeping $\Delta t=0$, and when we vary e.g. $t_a$ with fixed $t_b = 0$. The first scenario is what relevant to the evolution of HEE, and the second scenario is relevant for time dependence of correlation functions. 

Now let us discuss the appearance of the direct geodesic in a bit more detail for the case of equal-time direct geodesics with $\Delta t = 0$, since they are important for the calculation of holographic entanglement entropy. We can find explicitly the moment of time $t_{cr} = t_a = t_b=t_0$ when the direct geodesic between points with given $\varphi_a$ and $\varphi_b$, located according to the assumptions of the proposition, crosses the worldline of e.g. particle $1$ with $\varphi = 0$. Suppose this happens at the time $\tilde{t}$ in the point with radial coordinate $\tilde{r}$. We use the parametrization of geodesics given by (\ref{r(l)eqt},\ref{phi(l)eqt},\ref{t(l)eqt}). 
The worldline equation is given by (\ref{BTZworldline}). Plugging it into (\ref{r(l)eqt}) and solving in terms of $\lambda$, we find the value of the affine parameter at the intersection point as a function of $t_{cr}$ and $\Delta \varphi$: 
\be
\cosh^2 \tilde{\lambda} = \frac{\coth^2 R t_{cr} -1}{\coth^2 \frac{R \Delta \varphi}{2}-1}\,. 
\ee
Now we plug this into (\ref{phi(l)eqt}), requiring that $\varphi(\tilde{\lambda})=0$. This gives the equation, which we solve in terms of $t_0$:
\be
\cosh R t_{cr} = \frac{\cosh \frac{R \Delta \varphi}{2}}{\cosh R \varphi_0}\,. \label{t0}
\ee
This is the moment of emergence of equal-time direct geodesic. Note that for symmetric segments, i. e. $\varphi=0$, this expression reduces to 
\be
t_{cr} = \frac{\Delta \varphi}{2}\,.  \label{t0sym}
\ee
The length of the direct equal-time geodesic is given by the expression (\ref{Ldirect2}) with $t_a = t_b$: 
\be
\mathcal{L}_{\text{dir}}(\varphi_a, \varphi_b) = 2 \log \left(2 \sinh \left(R \frac{(\varphi_b - \varphi_a)}{2}\right)\right)+2 \log \left(\frac{r_0}{R}\right)\,. \label{LRT}
\ee
This expression is time-independent. 

\subsubsection{Crossing geodesics with equal-time endpoints}
The evolution of crossing ETEBA geodesics is the key to non-equilibrium dynamics of entanglement in our holographic bilocal quench setup. As explained in the previous discussion, the direct equal-time geodesics do not always exist. The following proposition establishes when crossing ETEBA geodesics exist and the chronology between vanishing of the crossing geodesic and emergence of the direct geodesic. 

\begin{Proposition} \label{CrossingEvolution}
For endpoints located as follows: $\varphi_a \in [-\pi, 0)$, $\varphi_b \in [0, \pi)$, and $t_a$, $t_b$ chosen in such a way that points are spacelike-separated, it is true that:
\begin{enumerate}
\item For $t_a = t_b=0$ the crossing geodesic always exists;
\item When increasing $t_a = t_b = t$, the crossing geodesic disappears only after the corresponding direct geodesic appears;
\item In the moment when the crossing geodesic disappears, $\mathcal{L}_{\text{crossing}}(a,\ b) \geq \mathcal{L}_{\text{direct}}(a,\ b)$.
\end{enumerate}
\end{Proposition}

\textbf{Proof}. As mentioned earlier, we have to work in global coordinates to prove most of this proposition. 
To prove the point 1), we have to show that the image points are located in such a way that the image geodesics have to intersect the wedge at $\tau = -\frac{\pi}{2}$ (which corresponds to $t = 0$ in BTZ coordinates. The global time slice $\tau = -\frac{\pi}{2}$ is special because it is closed under the action of the isometries, that is $(\tau = -\frac{\pi}{2})^{*,\#} = -\frac{\pi}{2}$ (see eqs. (\ref{tau*},\ref{tauDiez})). Therefore, we only have to prove that the angular coordinates of image points take desirable values. 
Specifically, if we have $\phi_a \in (2\pi-\theta, 2\pi)$ and $\varphi_b \in (0, \theta)$, then we want to prove that $\varphi_b^\# \notin (2\pi-\theta, 2\pi) \mod 2\pi$ and $\varphi_a^* \notin (0, \theta) \mod 2\pi$, where the angle $\theta$ is defined by
\be
\sin \theta = \tanh\pi R\,.
\ee
More intuitively, we have to prove that the image of a point $b$ ($a$ from the upper (lower) half of the living space on Fig.\ref{falling_particles}A does not end up in the lower (upper) half under a single action of the isometry $\#$ ($*$) defined according to (\ref{Diez}) ((\ref{Star})). Let us focus on the point $b$ and look at where does it go under the action of $\#$.

\begin{figure}[t]
\centering
\includegraphics[width=6cm]{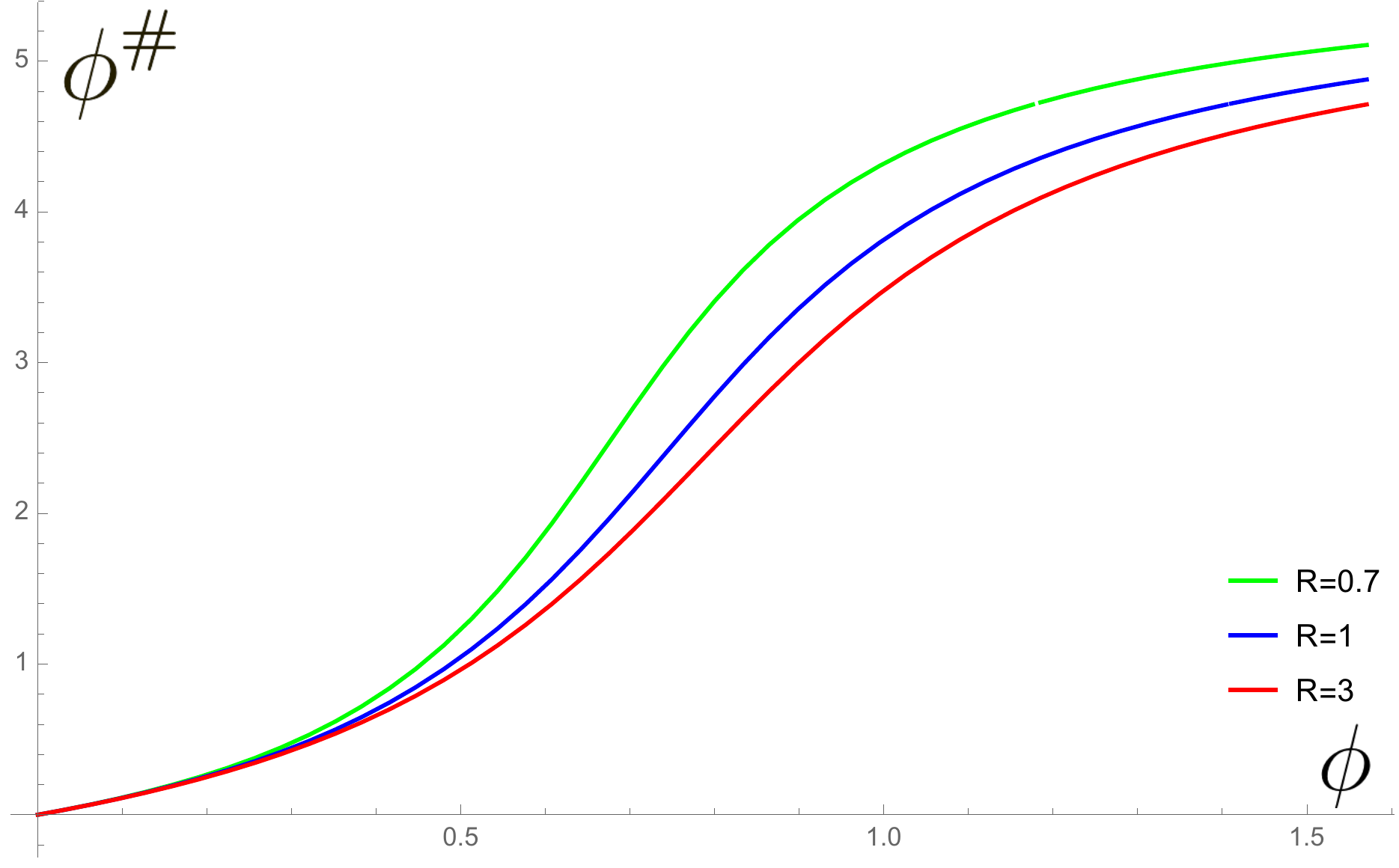}
\caption{Location of the $\#$-image point on the initial time slice for different values of the horizon radius.}
\label{phiDiezPic}
\end{figure}
We use the formula (\ref{phiDiez}) for the angle of the image point $\phi_a^*$, setting in that formula $\chi \to \infty$ and $ \tau = -\frac{\pi}{2}$:
\be
\tan \phi_b^\# = - \frac{-2 \mathcal{E} + (2 \mathcal{E} \cos \phi_b+\sin \phi_b) }{-2\mathcal{E}^2 + ((2 \mathcal{E}^2-1)\cos \phi_b +2 \mathcal{E} \sin \phi_b)}\,; \label{phiDiezinitial}
\ee
The dependence of $\phi_b^\#(\phi_b)$ is illustrated on Fig.\ref{phiDiezPic}. The coordinate of the image point is monotonically increasing function. We know that at $\tau = -\frac{\pi}{2}$ there is a fixed point of the isometry located at $\phi = 0$, which is where the particle sits, so $\phi^\#(0) = 0$. On the other hand, by definition (\ref{Diez}) we have $\theta \in W_+$ $\Rightarrow$ $\phi^\#(\theta) = 2\pi-\theta \in W_-$. By continuity and monotonicity, all image points with $\phi_b \in [0, \theta)$ will thus have angular coordinates $\phi_b^\# \in [0,\ 2\pi-\theta)$. In other words, the actions of the $\#$-isometry results in a rotation counter-clockwise with some angle less than $2\pi - 2\theta$. As a result, the $\phi_b^\#$ will never end up in the interval $(2\pi-\theta, 2\pi)$. This is exactly what we needed to prove, and the argument for $\phi_a^*$ points is completely analogous. 

\begin{figure}[t]
\centering
\includegraphics[width=6cm]{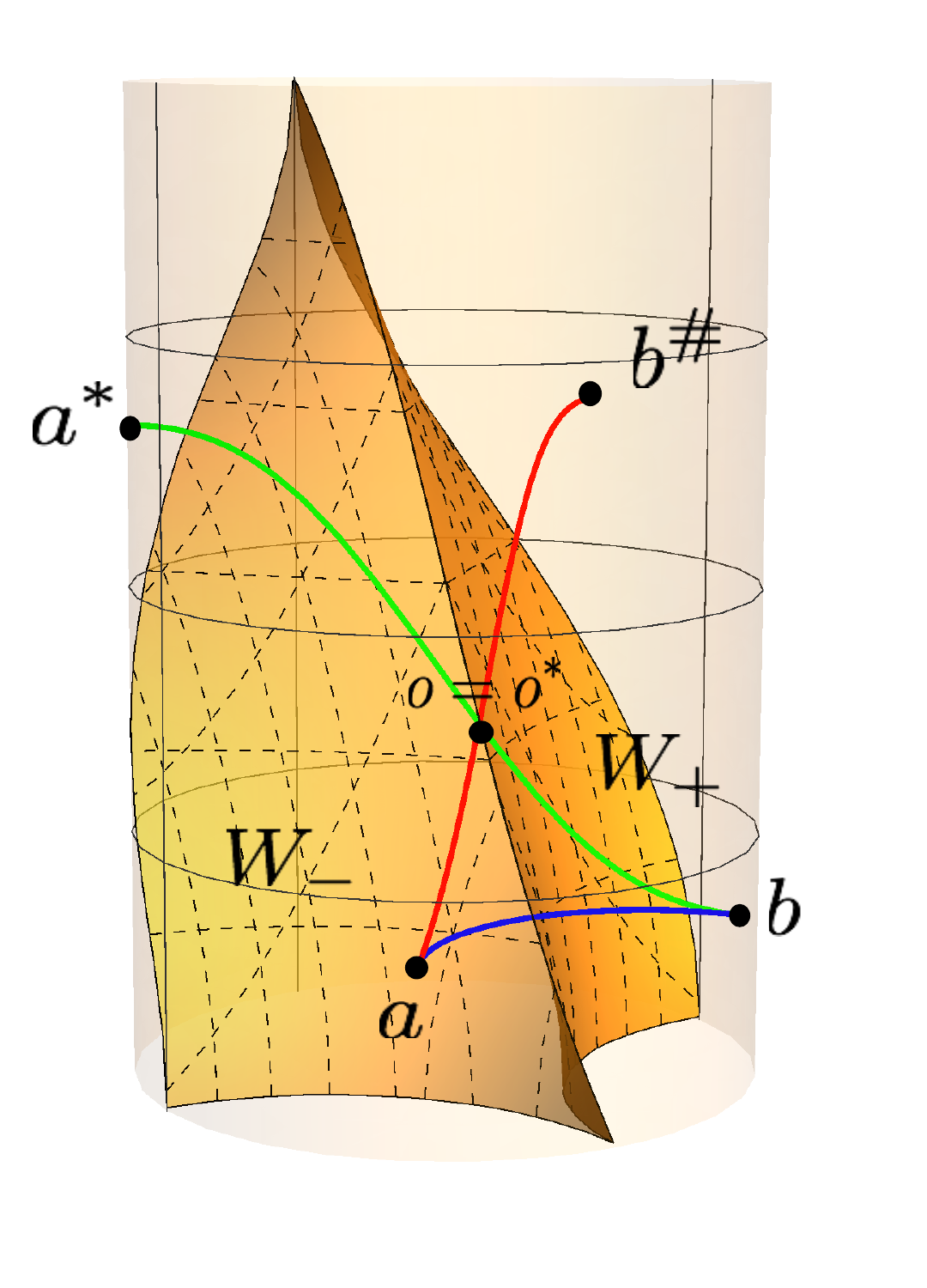}A. 
\includegraphics[width=6cm]{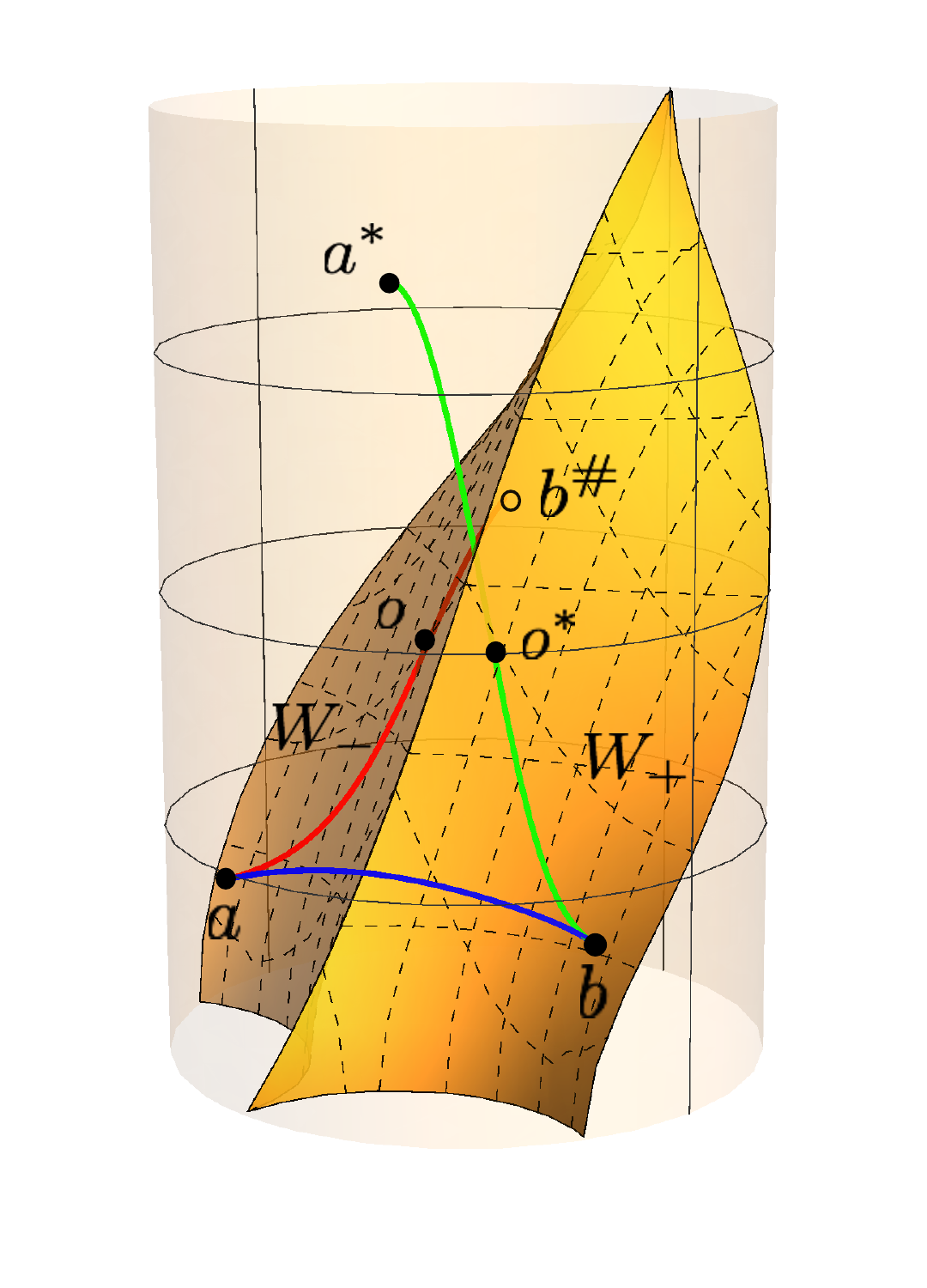}B.
\caption{Direct and crossing geodesics in global coordinates. \textbf{A.} The moment $\tau = -0.86$ when the symmetric direct geodesic with $\phi_b = -\phi_a = 0.614$ emerges coincides with the moment of vanishing of the corresponding crossing geodesic. Here $R = 5$. \textbf{B}. The case of non-symmetric geodesics: the direct geodesic with $\phi_a = -0.1$, $\phi_b = 1.182$, $\tau_a = -0.816$, $\tau_b = -0.928$ exists, whereas the corresponding crossing geodesic has not yet vanished as well; $R = 0.5$.}
\label{geodCrossingDirectFig}
\end{figure}

The point 2) concerns the time evolution of crossing ETEBA geodesics. From the transformation formulas from global to BTZ coordinates \ref{Sch-transf} we find that
\be
\tanh R t = \coth \chi \frac{\cos \tau}{\cos \phi}\,;
\ee
which means that on the boundary at $\chi \to \infty$ time evolution in BTZ coordinates corresponds to time evolution in global coordinates with angle-dependent rate. That means that once we fix the angles of endpoints in the initial moment, we can consider the evolution in global time to describe the evolution in BTZ time. Note, although, that in general case under the BTZ time evolution an equal-time geodesic on the initial time slice will be mapped to a geodesic with non-equal-time endpoints in global coordinates. The time coordinates of the endpoints will depend on their angular coordinates. However, in a special case of symmetric intervals $\varphi_a = -\varphi_b$ the crossing ETEBA geodesic in BTZ coordinates will always remains an ETEBA geodesic in global coordinates. The statement 2) itself can be verified by plotting the geodesics in global coordinates and using the isometry formulas from Appendix \ref{AppIsometries}. The sample plots of geodesics are presented on the Fig.\ref{geodCrossingDirectFig}. For symmetric endpoints, the direct geodesic appears precisely in the same moment when the crossing geodesic vanishes, as shown on Fig.\ref{geodCrossingDirectFig}A. For general endpoints evolving in time, the direct geodesic appears before the crossing geodesic vanishes, as shown on Fig.\ref{geodCrossingDirectFig}B.
This argument can be strengthened by looking at isometry formulas and certain light cones. We know that the crossing geodesic vanishes after the image geodesics intersect the particle worldline in the same point $o = o^*$. First, consider formulas (\ref{tau*},\ref{tauDiez}) with $\chi \to \infty$: 
\be
\tan \tau^{*,\#} = \tan \tau (1+2 \mathcal{E}^2) + \frac{ 2 \mathcal{E}}{\cos \tau} (\mathcal{E} \cos \phi \mp \sin \phi)\,; \label{tauStarDiez}
\ee
For fixed $\phi$, it is clear that these are growing functions of $t$. Moreover, since the coefficient in front of the first term $1+2 \mathcal{E}^2 >1$, we observe that $\tau^{*,\#} \geq \tau$ for $\tau \in [-\frac{\pi}{2}, 0)$, where the inequality is saturated only in the initial moment. That means that generally for some $\tau \in (-\frac{\pi}{2}, 0)$
\be
\tau^{*}_a > \tau_a\,, \qquad \tau^{\#}_b > \tau_b\,.
\ee
Because of the continuity and monotonicity of geodesics in global time (see eq. (\ref{tau(l)})) that also means that 
\be
\tau_o > \tau_a\,, \qquad \tau_{o^*} > \tau_b\,.
\ee
This is confirmed by Figs.\ref{crossing-basic}B,\ref{geodCrossingDirectFig}. The crossing geodesic vanishes when $o = o^*$, and this happens at some moment in global time $\tau_o > \tau_a,\tau_b$. 
Next, for a given direct boundary-to-boundary spacelike geodesic in AdS$_3$ we can always imagine a certain future light cone, to which the said geodesic belongs. The origin point of such light cone would be located on the boundary somewhere to the past of the geodesic. Since the $*$ and $\#$ mappings are generated by parabolic Lorentz isometries, that means that both image geodesics $a^* b$ and $a b^\#$ also belong to the same light cone. For a moment let us focus on the case of symmetric endpoints. In this case the particle lightlike worldline also belongs to this light cone in the moment of time when the given direct geodesic intersects it, which means that the image geodesic will also intersect it in the same moment, so we get precisely the picture shown on Fig.\ref{geodCrossingDirectFig}A. For a case of general endpoints, the particle worldline intersects our imaginary light cone in the point of intersection of the direct geodesic and the worldline, when the direct geodesic appears, and then goes inside the light cone to the future, missing image geodesics. The time evolution with fixed angular coordinates of endpoints from that moment effectively means that we move the imaginary light cone upwards in time, while the worldline remains fixed. The intersection point between the imaginary light cone and the worldline will always move to the future, and inevitably will coincide with the point of intersection of two image geodesics between themselves, eventually. This will be precisely the moment when the crossing geodesic vanishes. Thus the point 2) is proved. 
\begin{figure}[t]
\centering
\includegraphics[width=4cm]{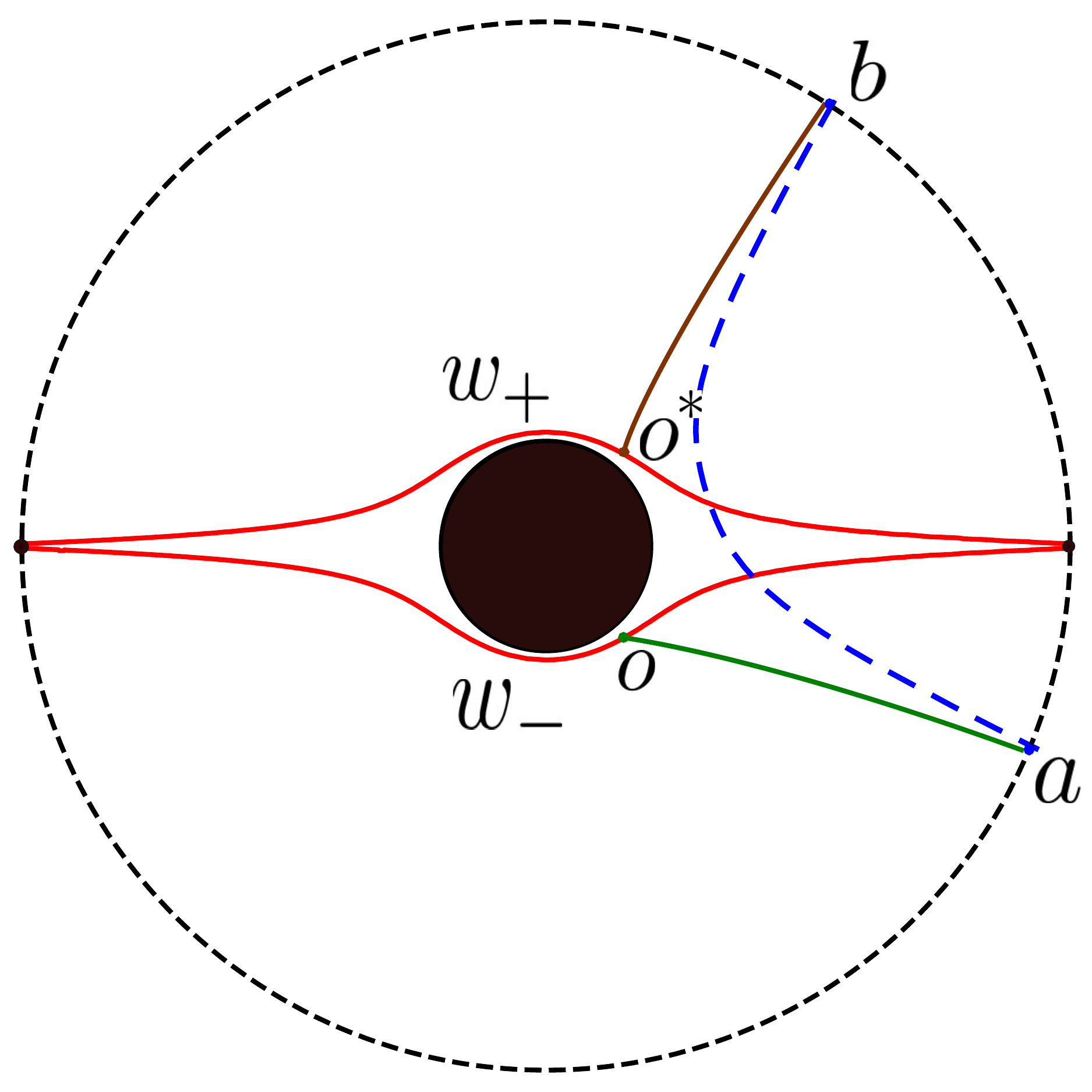}A.
\includegraphics[width=4cm]{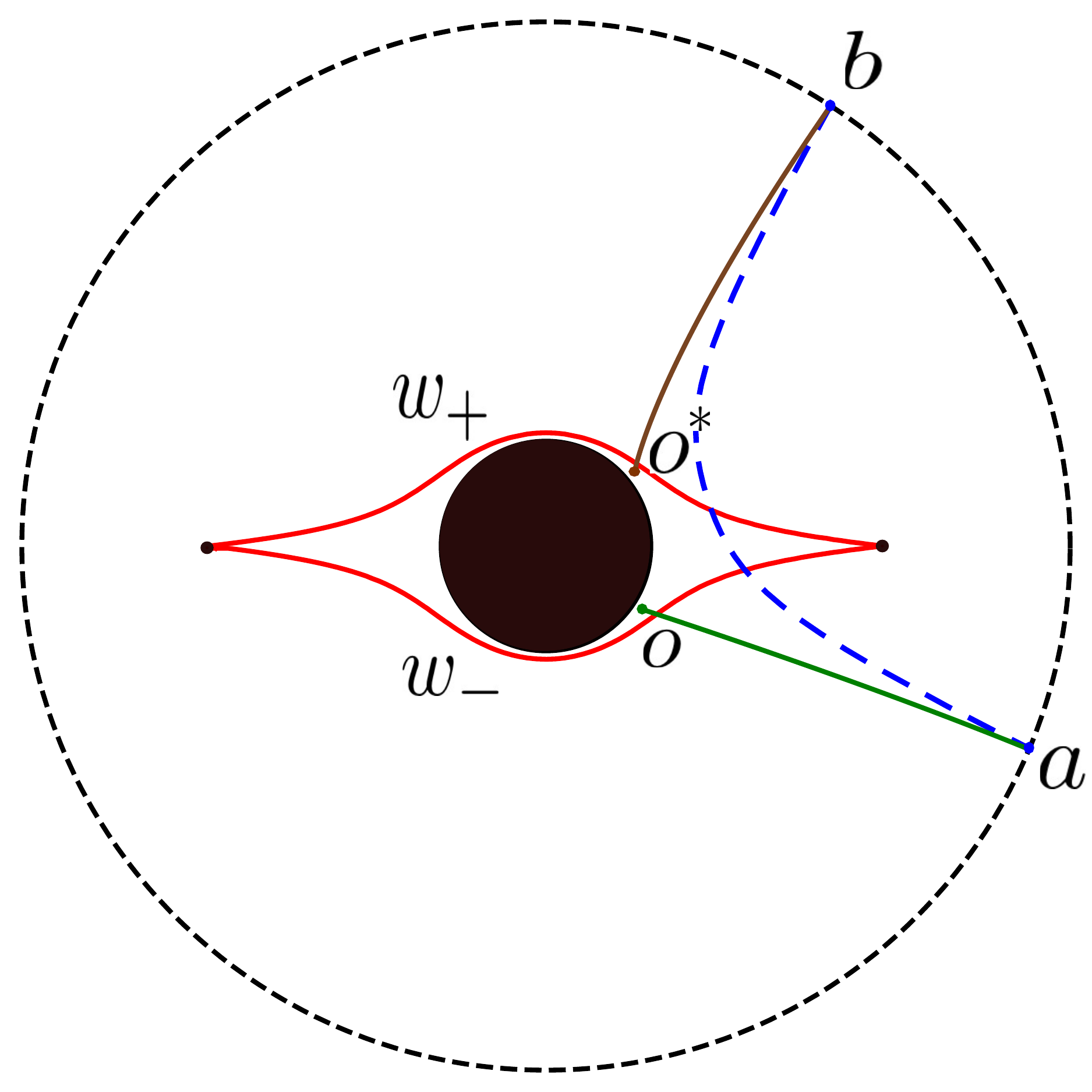}B.
\includegraphics[width=4cm]{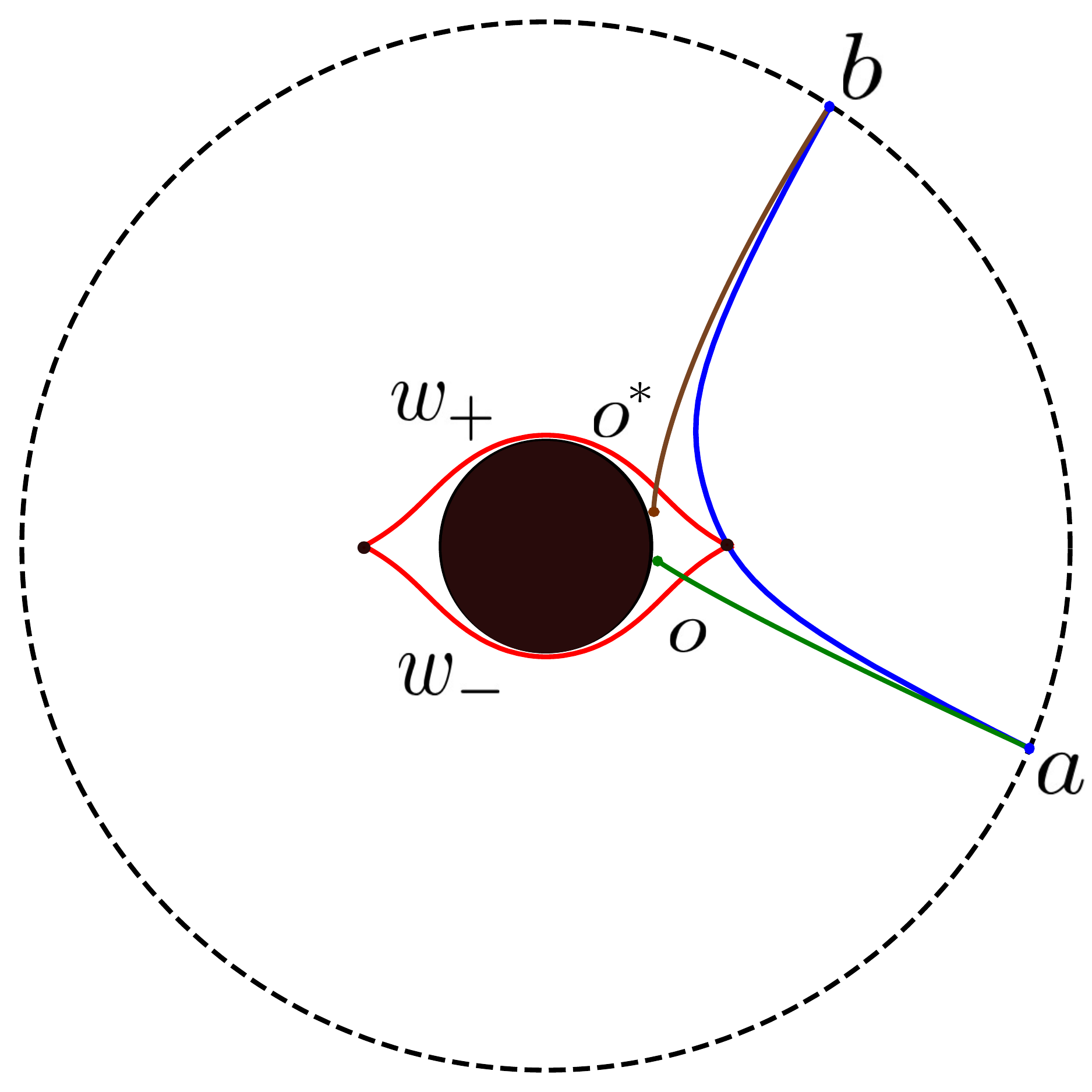}C.
\includegraphics[width=4cm]{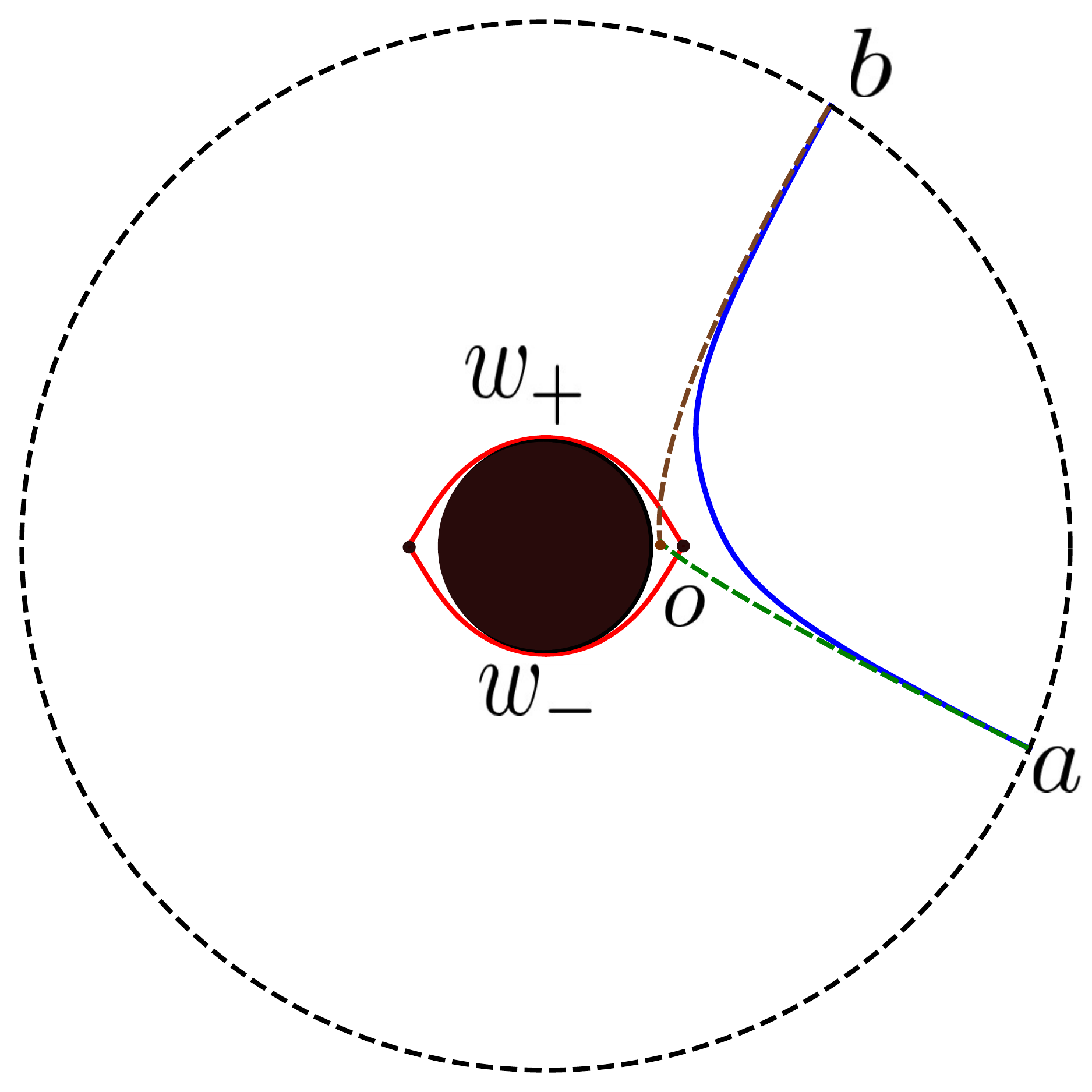}D.
\includegraphics[width=4cm]{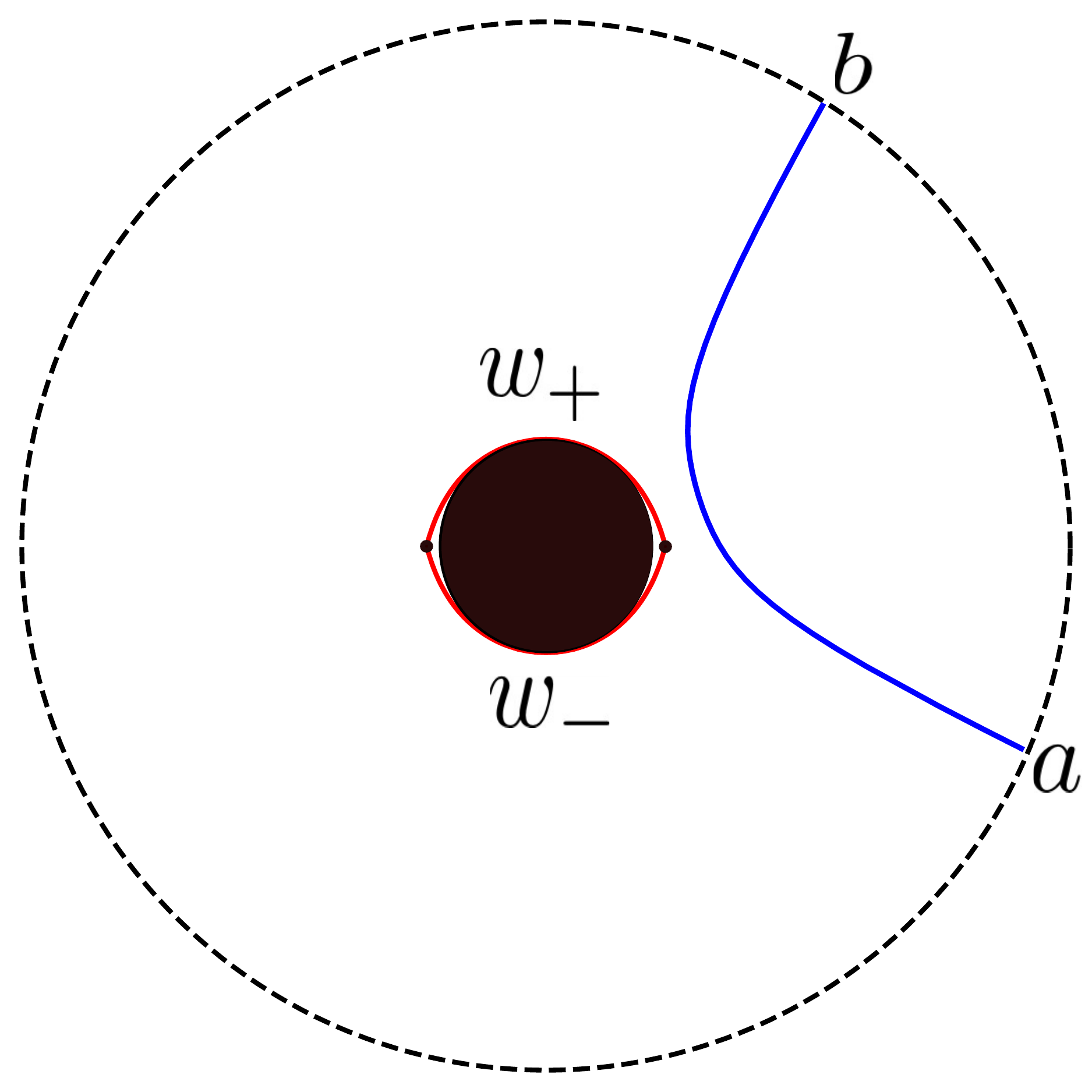}E.
\caption{Evolution of ETEBA geodesics which compete in the HRT prescription during the particle collision process.}
\label{crossing-basic-cartoon}
\end{figure}
The point 3) can be seen from the geometric picture of geodesics, as illustrated in Fig.\ref{crossing-basic-cartoon}D. The crossing geodesic at this point crosses the wedge in the position of the particle, and it consists of two AdS$_3$ geodesics joined together in that point in the bulk. Thus we have a curved triangle, all sides of which are spacelike geodesics in AdS$_3$. From the Lorentzian AdS version of the triangle inequality for spacelike-separated points, it is therefore correct that.
\be
\mathcal{L}_{\text{dir}}(a, b) \leq \mathcal{L}(a, o) + \mathcal{L}(b, o^*)|_{o^* = o}\,;
\ee
which is precisely what we needed to prove. Note even though we work with geodesics which have diverging lengths we still can use the triangle inequality, since two corners of the triangle rest on the boundary, and we have identical divergences from both sides of the above inequality, understanding it in regularized sense. \qed
\\
$\,\,\,$
\\
The length of the ETEBA crossing geodesic is given by (\ref{Lcrossing}) with $t_a = t_b = t$: 
\bea
&& \mathcal{L}_{\text{cross}}(\varphi_a,\ \varphi_b) = \log\left[2 \left((-1+\mathcal{E}^2)+(1 + \mathcal{E}^2) \cosh R \Delta\varphi + \mathcal{E}^2 \cosh 2 R \varphi_0 +\right.\right.\\\nn&& \left.\left. \mathcal{E}^2 \cosh 2 R t + 4\mathcal{E} \cosh R t \cosh R \varphi_0 \left(\sinh R \frac{\Delta\varphi}{2} - \mathcal{E} \cosh R \frac{\Delta\varphi}{2}\right) - 2 \mathcal{E} \sinh R \Delta \varphi\right)\right]\,. \label{LcrossingHRT}
\eea
This is the primary formula we will use for study of non-equilibrium behavior of entanglement in our model.

\subsection{Winding geodesics} \label{windings}

The BTZ black hole solution admits infinitely many geodesics between two given spacelike-separated endpoints on the boundary. One of these geodesics is the direct geodesic and all other geodesics wind around the horizon. In BTZ coordinates all geodesics are parametrized by equations (\ref{r(l)},\ref{phi(l)},\ref{t(l)}), direct and winding geodesics alike. For the direct geodesic $|\Delta \varphi| = |\varphi_b-\varphi_a| < \pi$, and for winding geodesics $\Delta \varphi > \pi$.

Our holographic dual to the bilocal quench is the AdS$_3$ spacetime with colliding particles in BTZ coordinates. Locally, the geometry of this spacetime is identical to that of the BTZ black hole, so in principle the geodesic equations also admit winding geodesic solution. However, the topological identification can actually interrupt the existence of winding geodesics, just like in case of direct geodesics. So, in order to answer the question whether the winding geodesic exists, we have to check if it crosses surfaces $W_\pm$.
\begin{figure}[t]
\centering
\includegraphics[width=4cm]{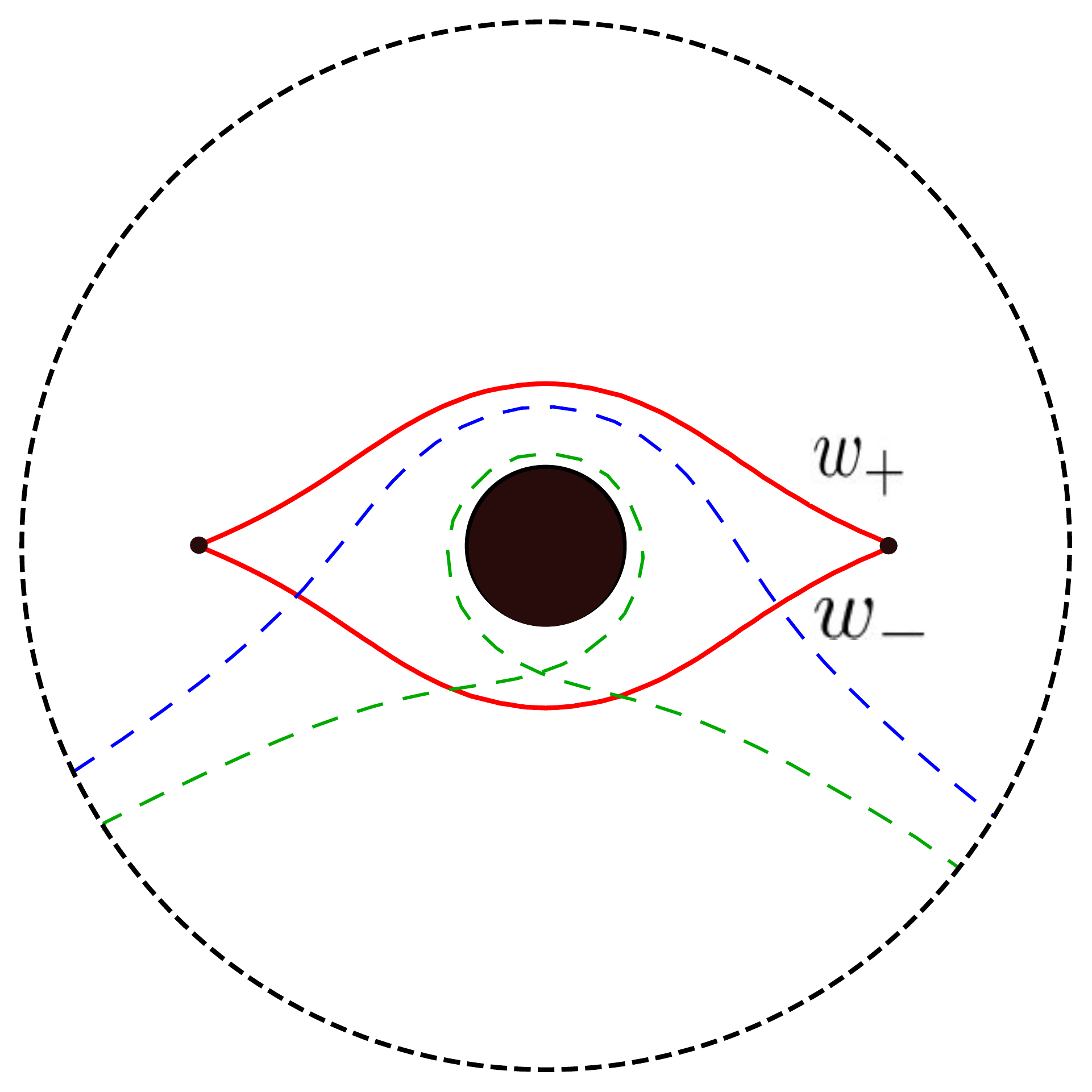}A.
\includegraphics[width=4cm]{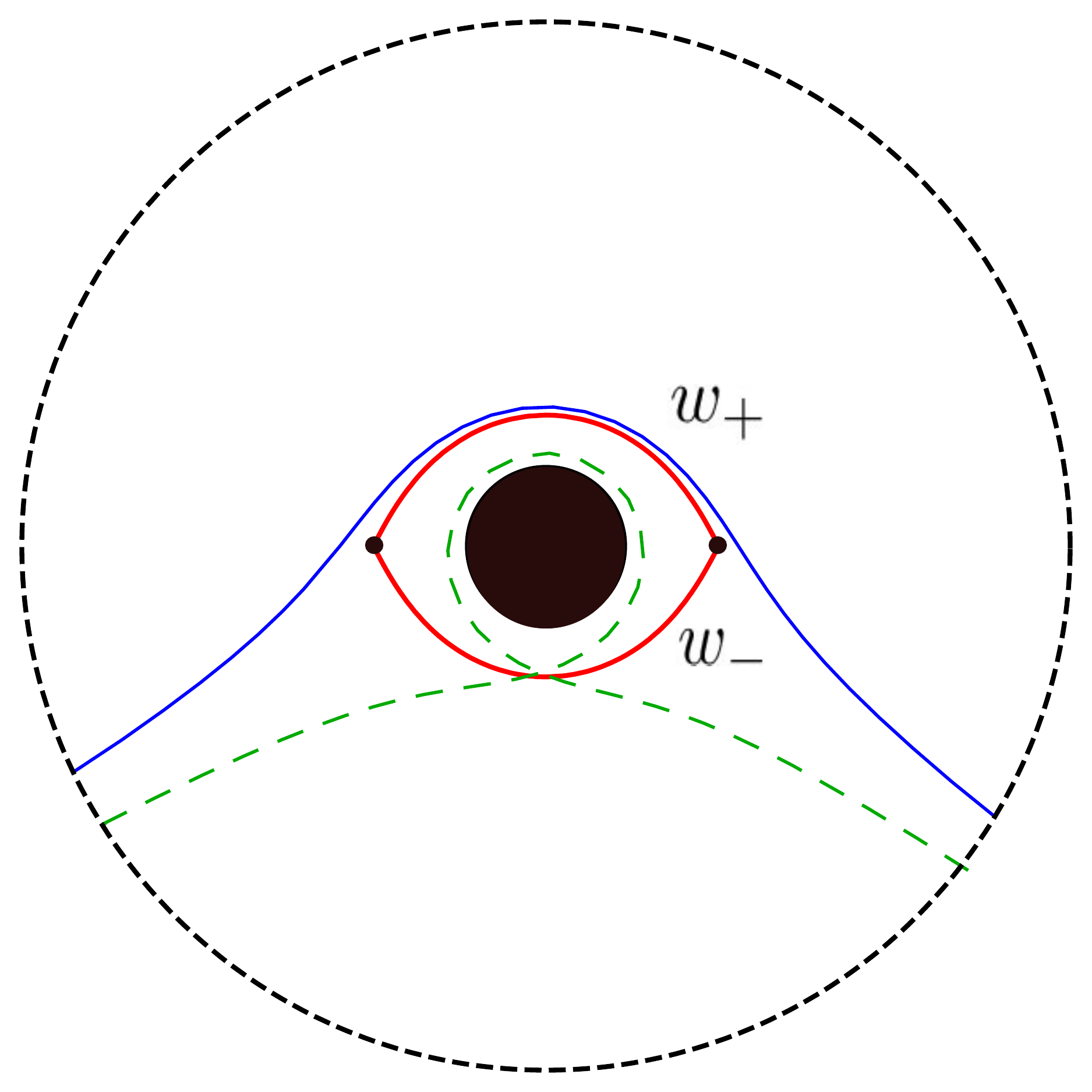}B.
\includegraphics[width=4cm]{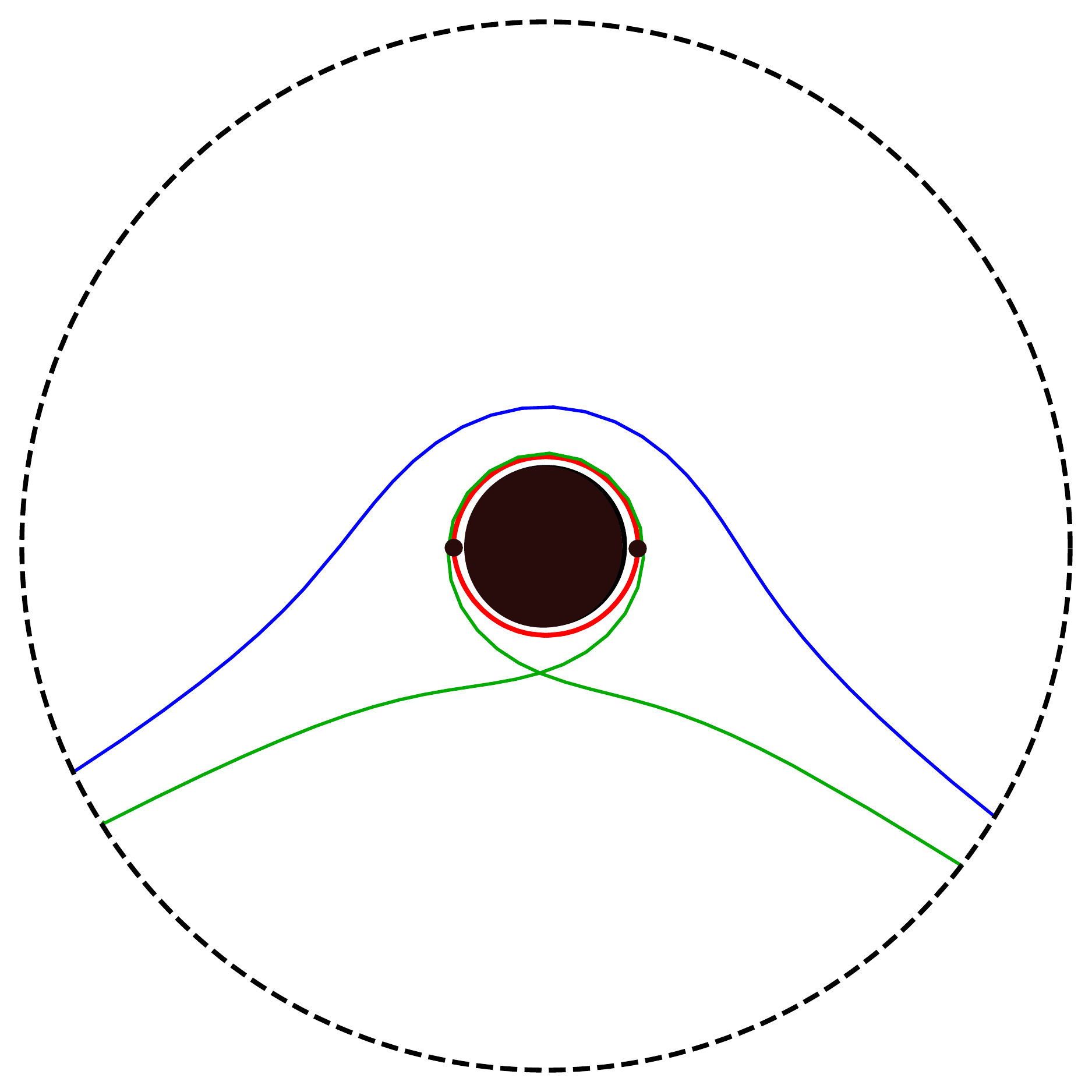}C.
\caption{Emergence of winding geodesics during the black hole creation. The blue geodesic has $n = 1$, and the green geodesic has $n = -1$. Here the disc depicts the bulk region $r < 1$, with the horizon at $r = R = 0.3$.}
\label{windings-emerge}
\end{figure}
A winding geodesic wraps around the horizon. It approaches to the horizon into radial distance given by eq. (\ref{Gamma+}):
\be
\Gamma_+ = R \sqrt{\frac{1+\tanh^2 \frac{R \Delta t}{2}}{\tanh^2 \frac{R \Delta t}{2} + \tanh^2 \frac{R (\Delta \varphi + 2\pi n)}{2}}}\,, \label{Gamma+Wind}
\ee
where $n \neq 0 \in \ZZ$ is the winding number. From this formula it is clear that between two given endpoints all winding geodesics approach to the horizon closer than direct geodesics. Moreover, there is a strict hierarchy between the depths of windings: a winding with higher $n$ always approaches closer to the horizon than any winding with lower $n$. Also, for smaller $\Delta \varphi$ winding geodesics approach the horizon closer. 

The identification wedge shrinks with time according to equation (\ref{W+-Schw}) (see Fig.\ref{schw-collision-3D}). Thus we arrive to the following statement: 

\begin{Proposition} \label{WindingExistence}
For given angular coordinates of endpoints $\varphi_a$, $\varphi_b$ and the winding number $n$ the corresponding winding geodesic exists only for $t_0 = \frac12(t_a + t_b) > \tilde{t}_n$, where $\tilde{t}_0$ is the moment of time when the winding geodesic crosses synchronous slices $w_\pm$ of the wedge faces at particle worldlines. 
\end{Proposition}

In other words, winding geodesics with given time separation between the endpoints will appear one by one as particle go deeper into the bulk towards the horizon. Geodesics with higher winding numbers will appear later than those with lower winding numbers. We illustrate the emergence of equal-time $n=\pm 1$ winding geodesics on Fig.\ref{windings-emerge}. On Fig.\ref{windings-emerge}A the wedge is still too large to accommodate both winding geodesics. Note that the direct geodesic at that time between two points on the lower half already exists from the beginning, according to proposition \ref{DirectSide}. On Fig.\ref{windings-emerge}B the identification wedge decreased in size enough to allow the blue winding geodesic to appear, but the green geodesic still intersects it. On Fig.\ref{windings-emerge}C even later moment of time is shown, when the green geodesic appears. In the initial moment $t=0$ there are no winding geodesics, and in the limit $t \to \infty$ the infinite amount of winding geodesic appears, which is the same situation as in the BTZ black hole spacetime. Note that these time scales of appearance of winding geodesics are generally much larger than the time scales of appearance of direct geodesics and vanishing of crossing geodesics. Holographically this means that winding geodesics are irrelevant for equilibration of entanglement, however they do contribute to late time behavior of correlation functions, as we will discuss later. 

To conclude this subsection, we remind that length of a winding geodesic is given by the formula (\ref{geodesicLength}):
\be
\mathcal{L}_{\text{wind}} (a,\ b) = \log\left(2(\cosh[R(\varphi_b - \varphi_a + 2\pi n)] - \cosh[R(t_b - t_a)])\right) +2 \log \left(\frac{r_0}{R}\right)\,. \label{geodesicLengthW}
\ee
The key observation is that lengths of winding geodesics between the given endpoints on the boundary are always larger than the lengths of corresponding direct and crossing geodesics. 

\section{Equilibration of entanglement} \label{sectionHEE}

We now have at our disposal all tools needed to perform the holographic computation of the entanglement entropy in the boundary CFT after the bilocal quench and analyze its time dependence during the thermalization process. We start our analysis with calculation of the holographic entanglement entropy (HEE). We then use it to investigate the equilibration and spreading of entanglement in subsystems in the boundary theory which live on segments of the circle during the particle collision process. We also compute holographic mutual information and discuss different possibilities of its non-equilibrium behavior, depending on the location of subsystems. 

\subsection{Holographic entanglement entropy}

Consider a subsystem in the boundary theory which lives on a segment $L$ of the circle, which is bounded by points $a$ and $b$. Then the entropy is calculated, according to the Ryu-Takayanagi proposal \cite{RT} generalized to the non-stationary case \cite{HRT}, as the minimal area of the codimension two surface in the bulk anchored on equal-time points $a$ and $b$. In an asymptotically AdS$_3$ spacetime, this surface is a geodesic, so we need to find the minimal geodesic between two given equal-time points on the boundary and calculate its length: 
\be
S(a,\ b) = \frac{\mathcal{L}_{\text{min}}(a,\ b)}{4G}\,; \label{RT}
\ee
Here $G = \frac{3}{2c}$ is the gravitational constant, and $c$ is the central charge of the boundary CFT. 
In our case, the bulk background is set up explicitly as a quotient of the AdS$_3$ spacetime by a non-trivial identification. That means that the metric itself is stationary in our case, but the identification is non-stationary, which is what makes the entire spacetime non-stationary and requires the use of the HRT proposal, which generalizes usual Ryu-Takayanagi prescription to the non-stationary case. Our bulk spacetime is arranged in such a way, that depending on the position of the endpoints, the crossing geodesic either participates in the HRT prescription, or does not. These two situations describe qualitatively different behavior of entanglement. 

In the BTZ black hole spacetime, which corresponds to the CFT at thermal equilibrium, the minimal geodesic is a direct geodesic. The length of such geodesic gives a result for HEE, and is obtained from (\ref{Ldirect}) for small subsystems  with size less than half of the circle, $\Delta \varphi \leq \pi$, by setting $t_a = t_b$. Introducing the UV cut-off in the boundary theory 
\be \epsilon := (R/r_0)^2\,, \label{UVcutoff}\ee
we have:
\be
S_{\text{eq}} (a,\ b)=\frac{c}{3} \log \left(\frac{2}{\epsilon} \sinh \left(R \frac{(\varphi_b - \varphi_a)}{2}\right)\right)\,; \label{Seq}
\ee
This is the entanglement entropy of the thermal equilibrium state with temperature given by $T = \frac{R}{2\pi}$ of a subsystem of size $\Delta \varphi = \varphi_b - \varphi_a < \pi$. For large subsystems with $\Delta \varphi > \pi$, one obtains the HEE from the expression (\ref{Ldirect2}) instead, which results in 
\be
S_{\text{eq}} (a,\ b)=\frac{c}{3} \log \left(\frac{2}{\epsilon} \sinh \left(R \frac{(\varphi_b - \varphi_a-2\pi)}{2}\right)\right)\,; \label{Seq2}
\ee
Since the geodesics which do not intersect $W_\pm$ are identical to those in the BTZ black hole spacetime, direct equal-time geodesics always govern the equilibrium regime in our model. Meanwhile, the length of a crossing ETEBA geodesic (\ref{LcrossingHRT}) is time dependent because of the shape of the identification wedge, hence crossing geodesics must govern the non-equilibrium regime. In the further discussion we will focus on the small subsystems with $\Delta \varphi \leq \pi$. The results for small subsystems can be related to large subsystems if one keeps in mind the subtraction of $2\pi$ in the equilibrium HEE formula (\ref{Seq2}) and symmetries of the formula (\ref{LcrossingHRT}). Since the horizon never appears in any finite time, we do not have to worry about its contribution in the Ryu-Takayanagi prescription.
Note also that the HEE of the complement of a subsystem is always given by the same geodesic as the HEE of the subsystem itself, because of the same reason. This is what we expect when considering the evolution of a pure state in the boundary CFT.

\subsubsection{Equilibrium in the initial state}

\begin{figure}[t]
\centering
\includegraphics[width=6cm]{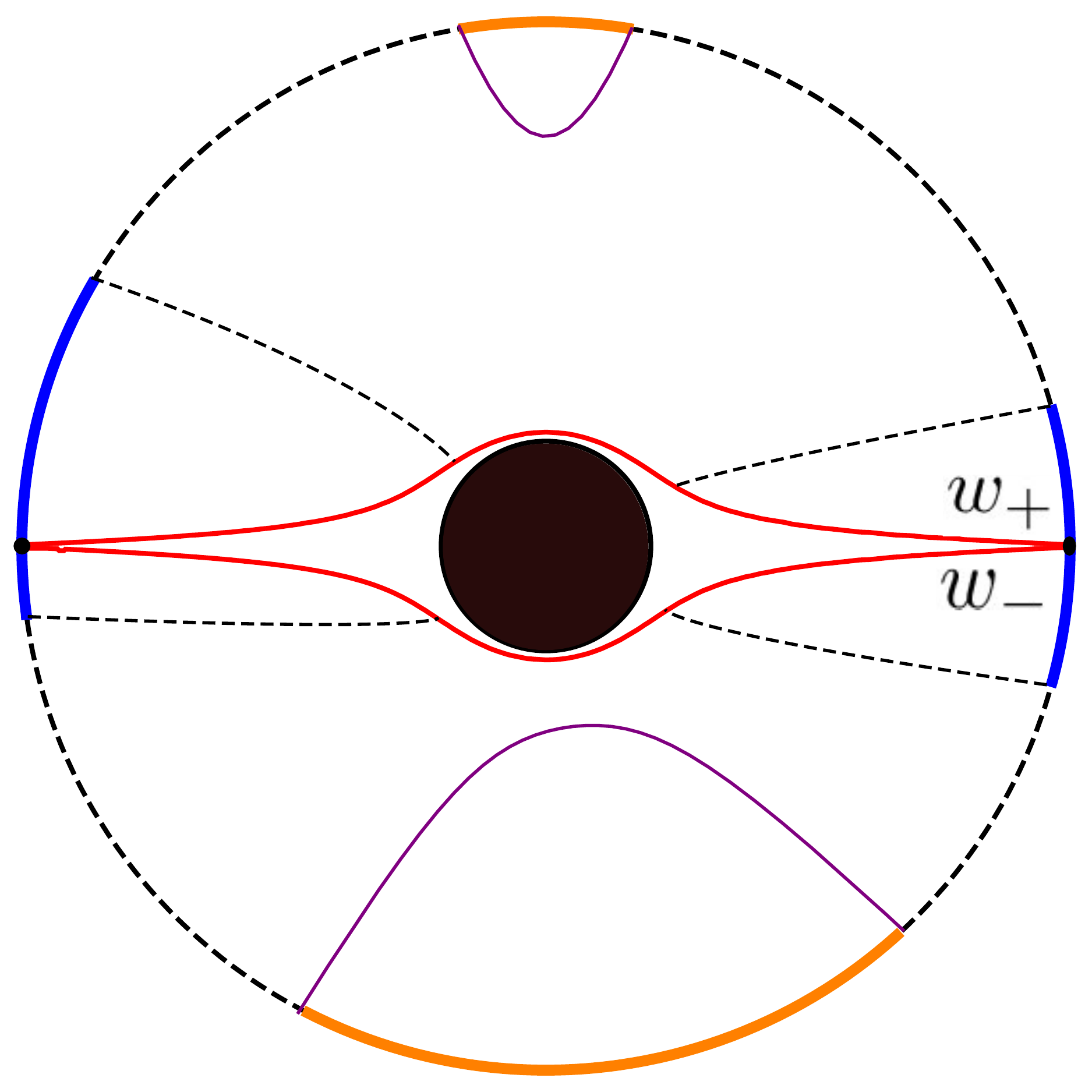}
\caption{Illustration of positions of subsystems which are in and out of equilibrium in the initial state. }
\label{initial-entangle}
\end{figure}

From the proposition \ref{DirectSide} it follows that one can anchor a direct geodesic on segments of the boundary spatial circle which lie to the side of the collision line at any moment of time. Therefore the entanglement entropy of a subsystem located in either upper or lower (with respect to the collision line) semi-circle is maximal and is given by the expression (\ref{Seq}), and is constant in time. Therefore we can come a conclusion that for subsystems which lie in between the initial excitations (orange subsystems on Fig.\ref{initial-entangle}) the single-interval HEE is exhibits constant equilibrium behavior. To study non-equilibrium dynamics in these subsytems, one might want to use more "fine-grained" observables, which would require information from the bulk beyond the minimal geodesic length. For example, one can consider  Renyi entropies \cite{Asplund13,Asplund14,Caputa141}, entwinement \cite{Bal14}, or contributions to holographic correlation functions from the non-minimal geodesics \cite{AAT,AKT,AA}. We will discuss the latter in the bilocal quench scenario in section \ref{sectionCorr}. 

\subsubsection{Crossing geodesics and non-equilibrium regime} \label{sectionNon-Eq}

\begin{figure}[t]
\centering
\includegraphics[width=14cm]{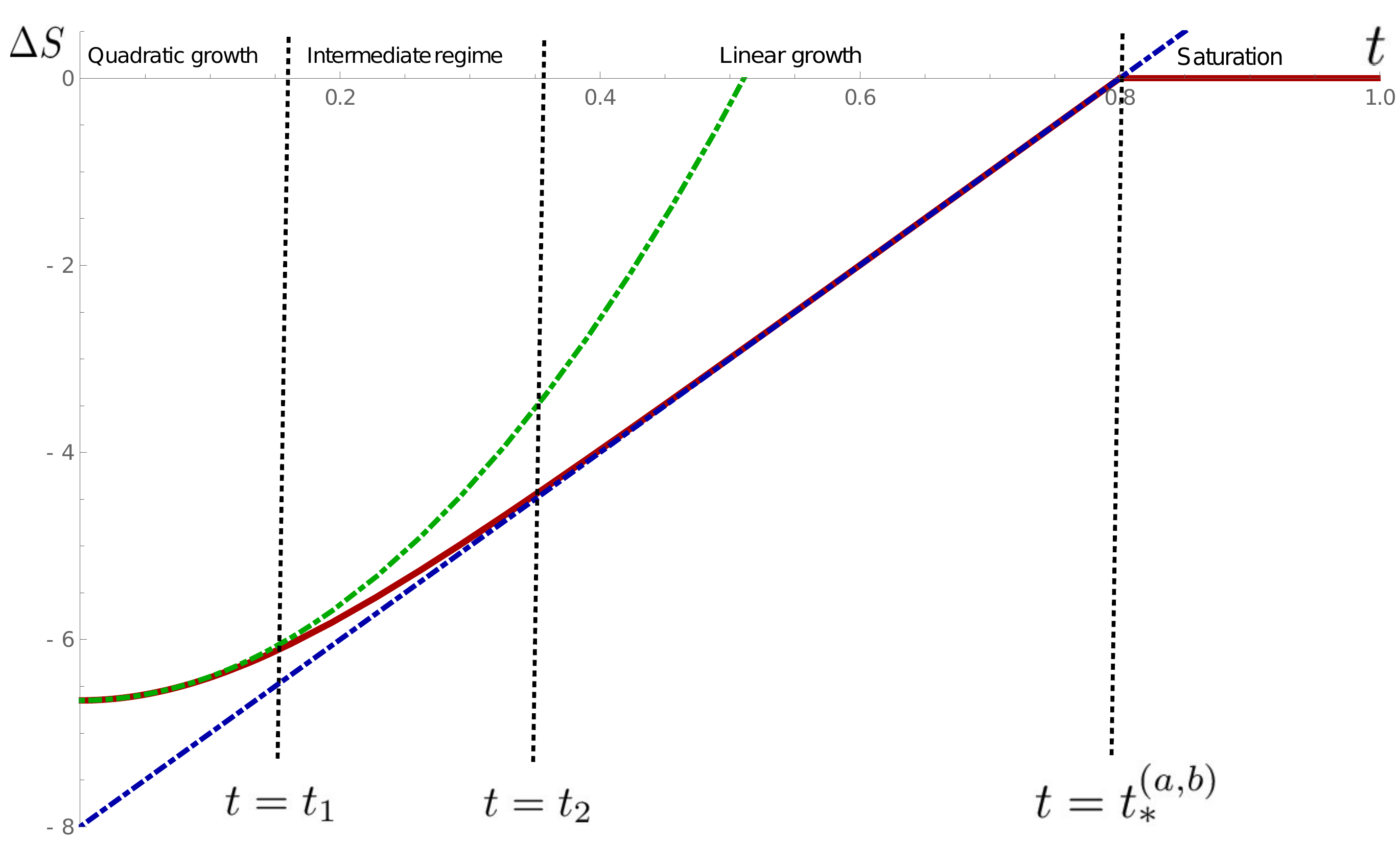}A.
\caption{Typical time dependence of HEE of a non-equilibrium subsystem. Here the dark red curve is the function $\Delta S(t)$, green dashed curve is the quadratic approximation given by the leading term in (\ref{HEEquadratic}), and blue dashed line is the linear asymptotic given by the leading term in (\ref{HEElinear}). The values of parameters are $R = 5$ and $\varphi_b = - \varphi_a = 0.8$.}
\label{HEE-evolution-detailed}
\end{figure}

On the other hand, the subsystems with endpoints on different sides of the boundary with respect to the collision line are initially out of equilibrium. This is because in the initial moment  the corresponding HRT geodesic is a crossing geodesic, and the evolution of crossing and direct HRT geodesics for such subsystems are described by the proposition \ref{CrossingEvolution}. These subsystems contain one of the excitations on the boundary in the initial moment, like illustrated by blue subsystems in Fig.\ref{initial-entangle}. Let us restrict our attention without loss of generality to subsystems which contain the particle $1$ moving along the $\varphi = 0$ worldline. Then we are interested in the case when $\varphi_a \in (-\pi, 0)$, and $\varphi_b \in (0, \pi)$. The evolution of the entanglement in such subsystems goes as follows, according to the proposition \ref{CrossingEvolution}:
\begin{itemize}
\item From the point 1) of the proposition, at $t=0$, the HRT geodesic has to be a crossing ETEBA geodesic, as shown on Fig.\ref{crossing-basic-cartoon}A. The length of the crossing ETEBA geodesic, given by (\ref{LcrossingHRT}), is initially smaller than the length of the direct geodesic and starts growing with time. The crossing geodesic length evolves with time with the identification, so the crossing geodesic corresponds to the non-equilibrium regime of HEE, as shown on Fig.\ref{crossing-basic-cartoon}B-D. 
\item For early times, the behavior of HEE is governed by the crossing geodesic. At the moment of time $t_{\text{cr}}$ given by (\ref{t0}) the direct geodesic emerges, see Fig.\ref{crossing-basic-cartoon}C. It competes with the crossing geodesic for being responsible for the behavior of HEE, when their lengths are equal. The transition to the direct geodesic in the leading HEE channel happens at the moment $t_*^{(a,b)}$, which we call the thermalization time of the subsystem. The points 2) and 3) of the proposition \ref{CrossingEvolution} ensure that the crossing geodesic still exists in that moment and the transition is continuous, as we will see below.
\item At late times, for $t > t_{*}^{(a,b)}$, the behavior of HEE is governed by the direct equal-time geodesic, which corresponds to the equilibrium regime and is expressed by the formula (\ref{Seq}). The point 3) of the proposition \ref{CrossingEvolution} ensures that the vanishing of the crossing geodesic does not influence the behavior of HEE (see Fig.\ref{crossing-basic-cartoon}D-E). 
\end{itemize}
Thus, the general formula for HEE of a crossing subsystem can be expressed as 
\be
S(a,\ b|t) = \min\left\{\begin{array}{ccc} 
S_{\text{non-eq}} (a,\ b | t)\\
S_{\text{eq}} (a,\ b)\,;
\end{array}\right. \label{S(t)}
\ee
where $S_{\text{non-eq}}$ is the contribution to HEE from the crossing ETEBA geodesic. It is obtained using the formula (\ref{LcrossingHRT}) with the boundary UV-cutoff introduced as in (\ref{UVcutoff}) for the length of the crossing ETEBA geodesic:
\bea
&& S_{\text{non-eq}} (a,\ b | t) = \frac{c}{6}\log\left[\frac{2}{\epsilon} \left(-1+\mathcal{E}^2+(1 + \mathcal{E}^2) \cosh R \Delta\varphi + \mathcal{E}^2 \cosh 2 R \varphi_0 + \mathcal{E}^2 \cosh 2 R t +\right.\right.\nn\\&& \left.\left. 4\mathcal{E} \cosh R t \cosh R \varphi_0 \left(\sinh R \frac{\Delta\varphi}{2} - \mathcal{E} \cosh R \frac{\Delta\varphi}{2}\right) - 2 \mathcal{E} \sinh R \Delta \varphi  \right)\right]\,. \label{Snon-eq}
\eea
where we remind that $\mathcal{E} = \coth \frac{\pi R}{2}$, $\Delta \varphi = \varphi_b - \varphi_a$ and $\varphi_0 = \frac12(\varphi_a + \varphi_b)$. At $\mathcal{E}=0$ one can recover the equilibrium result (\ref{Seq}). The initial value at $t = 0$ is given by 
\be
S_{\text{non-eq}} (a,\ b | 0) =\frac{c}{3}\log\left[\frac{2}{\epsilon} \left(\mathcal{E} (\cosh R \frac{\Delta \varphi}{2} - \cosh R \varphi_0) - \sinh R \frac{\Delta\varphi}{2}\right)\right]\,. \label{Snon-eq0}
\ee
As mentioned above, this quantity is smaller than the equilibrium value (\ref{Seq}) for the same segment. 
From the formula (\ref{Snon-eq}) it is clear that the quantity $S_{\text{non-eq}} (a,\ b| t)$ grows monotonically with time, until the crossing geodesic vanishes. However, as we evolve the system with time, after the moment $t=t_{\text{cr}}$ the direct geodesic appears as well and starts competing in the HRT prescription, and takes over at $t=t_*^{(a,b)}$, realizing the saturation of HEE at equilibrium. Now let us discuss the time dependence of the HEE in more detail. We illustrate the typical time dependence of $\Delta S(t) =  S_{\text{non-eq}}(t) - S_{\text{eq}}$ on Fig.\ref{HEE-evolution-detailed}. 

In the further discussion we assume that we deal with the formation of large black holes\footnote{Nevertheless, some illustrations are still presented with $R<1$. This is done to make certain features more prominent on the availible scale.} with $R > 1$, or in other words when the temperature is higher than the AdS$_3$ Hawking-Page temperature, $T > \frac{1}{2\pi}$. The formulae (\ref{Snon-eq}) and (\ref{Seq}) are still valid for $R < 1$, but the bulk geometry with a small black hole gives a subdominant contribution to the CFT path integral and thus is not a proper holographic dual for a thermal state \cite{Hubeny13}. 

\textbf{Early-time evolution}. We can expand the expression (\ref{Snon-eq}) in time around $t=0$. The leading terms of the expansion read: 
\be
S_{\text{non-eq}} (a,\ b | t) = S_{\text{non-eq}} (a,\ b | 0) +\frac{c}{6} R^2 \mathcal{E} \frac{\mathcal{E} + \cosh R \Delta \varphi \left(\sinh \frac{R \varphi_0}{2} - \mathcal{E} \cosh \frac{R \varphi_0}{2}\right)}{\left(\mathcal{E} \cosh R \Delta \varphi + \sinh \frac{R \varphi_0}{2} - \mathcal{E} \cosh \frac{R \varphi_0}{2}\right)^2} t^2 + O(t^4)\,;\label{HEEquadratic}
\ee
where $S_{\text{non-eq}} (a,\ b | 0)$ is given by the formula (\ref{Snon-eq0}). This behavior is the same as the so-called pre-local equilibration growth  \cite{Liu13,Liu2013}, which appears in the Vaidya global quench scenario. On the Fig.\ref{HEE-evolution-detailed} it is shown that the HEE behavior is well approximated by the quadratic expansion until a time scale $t = t_1$. We observe that
\be
t_1 = \frac{\Delta \varphi}{2R}\,.
\ee
We conjecture that this time scale has similar meaning as the "local equilibration" time scale in the Vaidya quench. However, the important difference of our case is that we do not make any assumptions about the size of the subsystem compared to the horizon radius, and due to this the time scale depends on $\Delta \varphi$, unlike in AdS-Vaidya case in Poincare coordinates \cite{Liu2013}.

\begin{figure}[t]
\centering
\includegraphics[width=7cm]{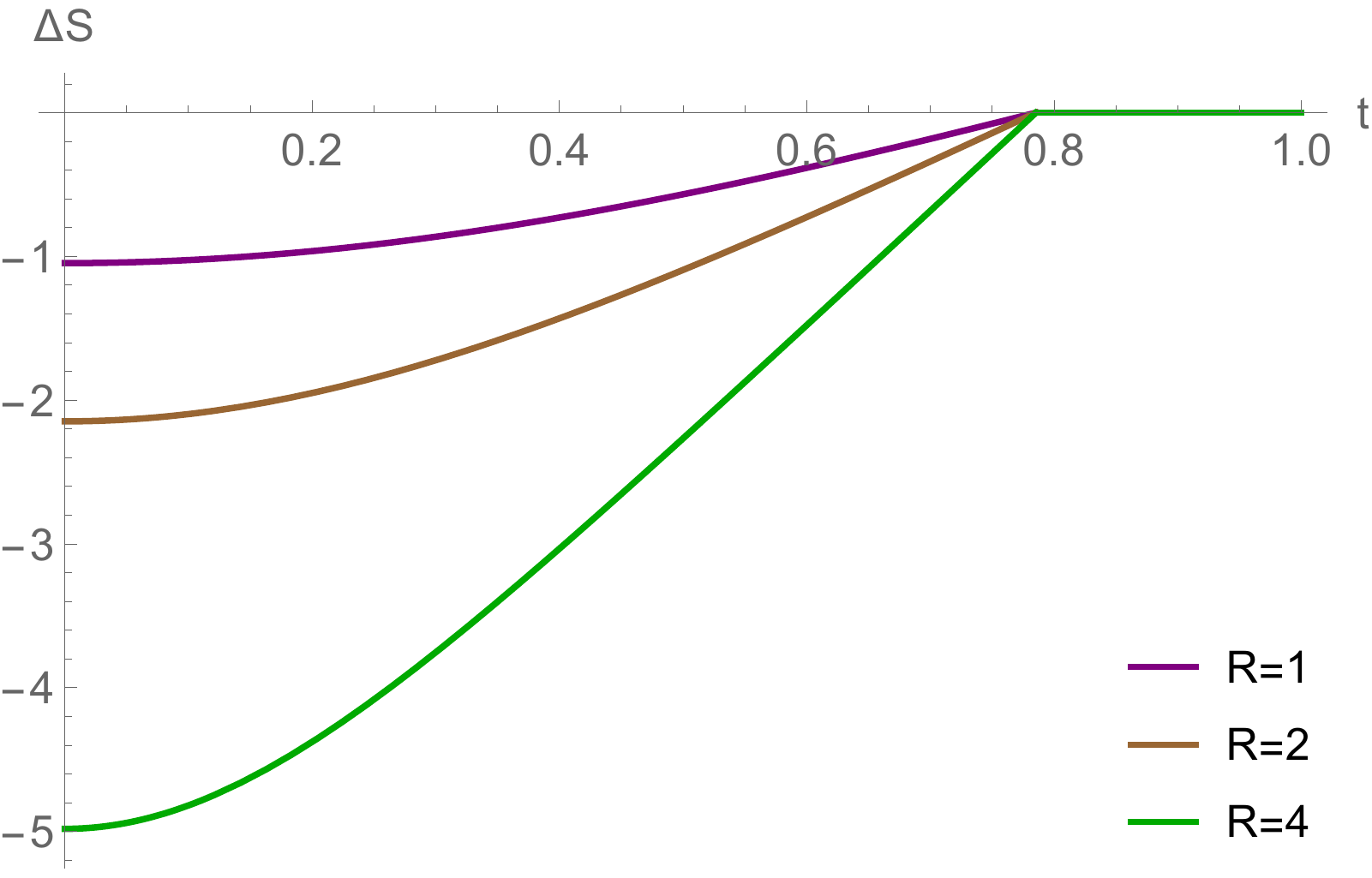}A.
\includegraphics[width=7.3cm]{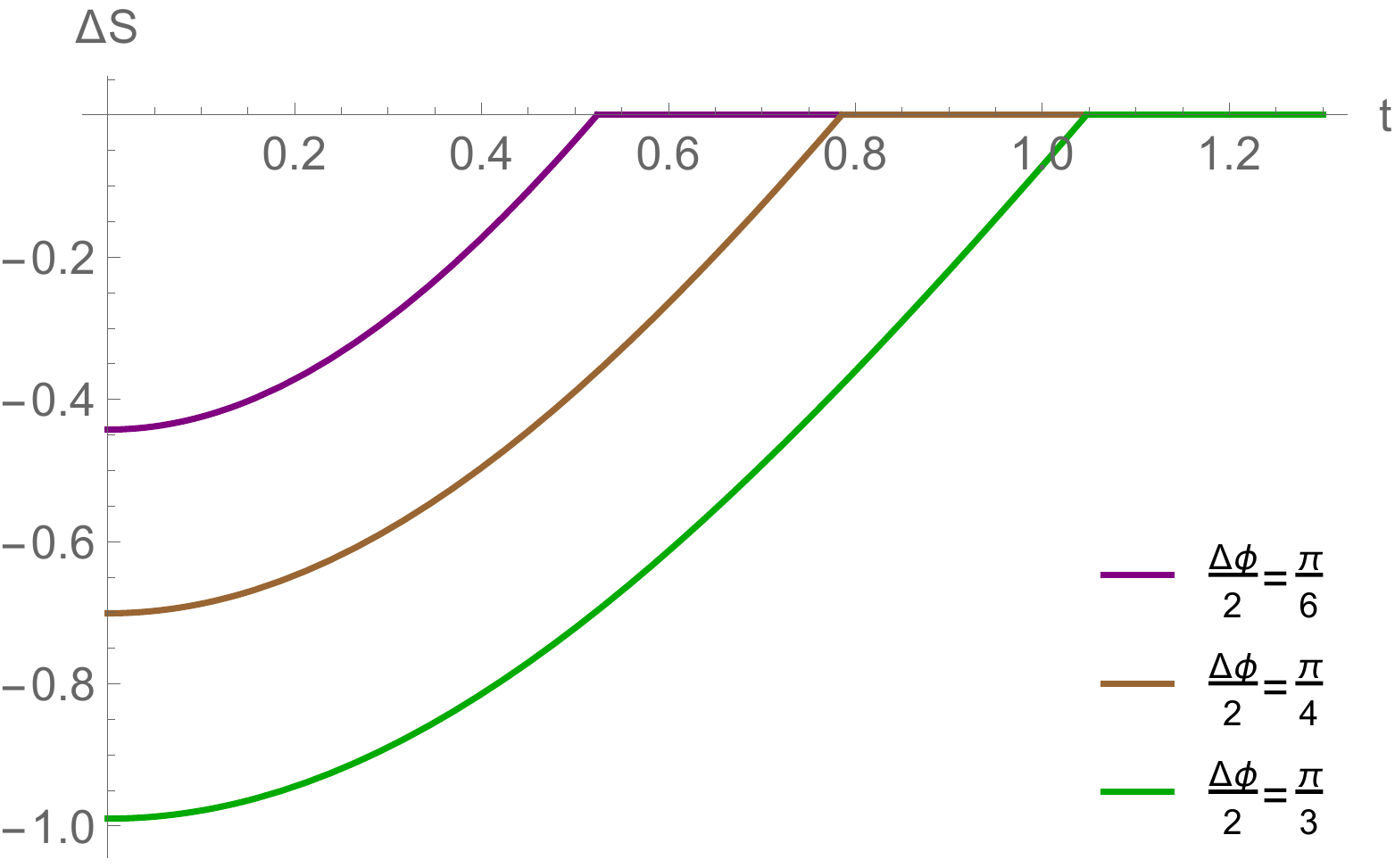}B.\\
\includegraphics[width=7cm]{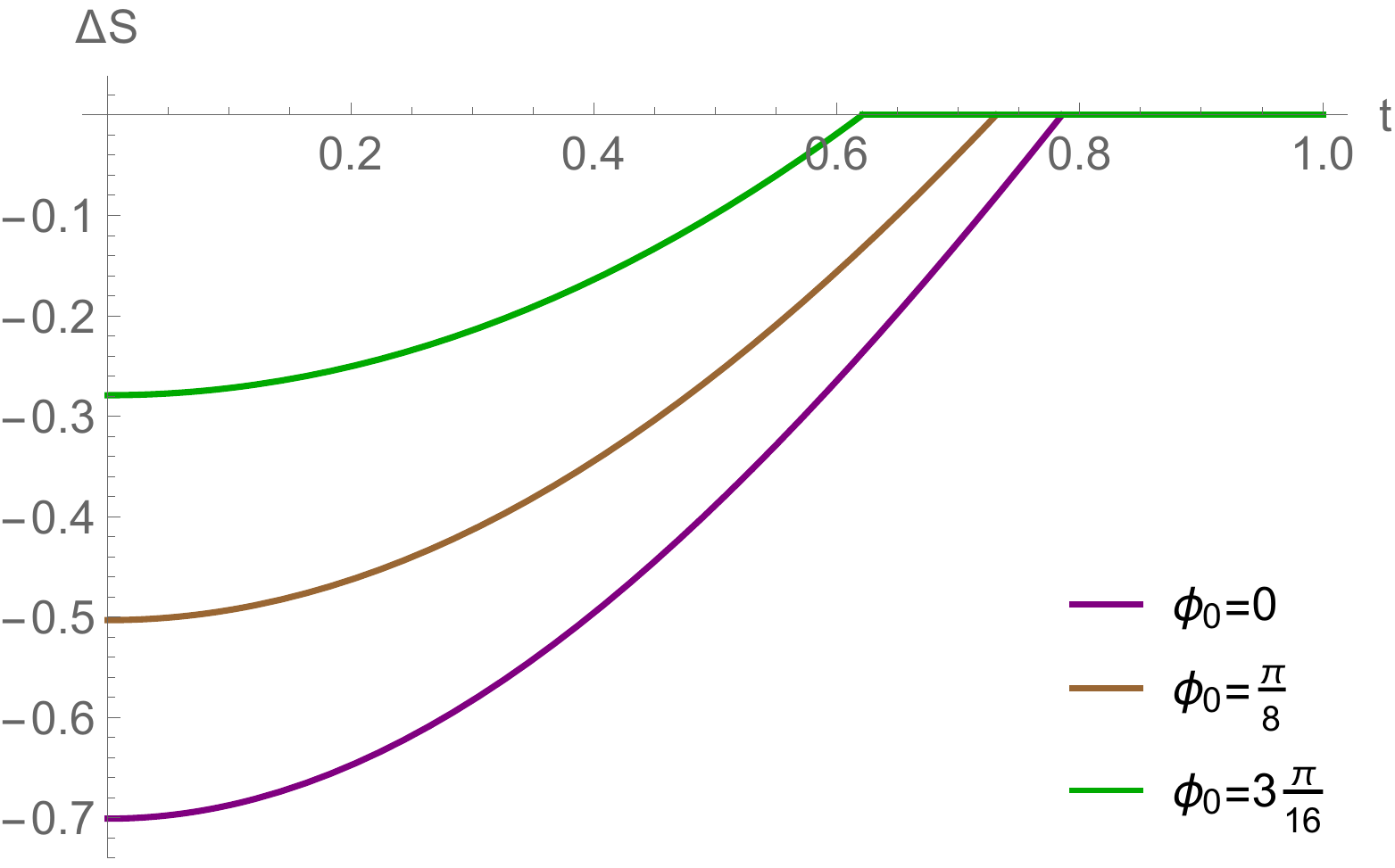}C.
\caption{\textbf{A}: Time dependence of HEE for different values of horizon radius. The parameters of the subsystem are $\varphi_0 = -\frac{\pi}{4}$, $\Delta\varphi = \frac{\pi}{2}$. \textbf{B}: Time dependence of HEE for subsystems of different size. Here $\varphi_0 =0$ and $R = 0.5$. \textbf{C}:
Time dependence of HEE for different positions of the subsystem. Here $\Delta \varphi = \frac{\pi}{2}$ and $R = 0.5$.}
\label{HEE-evolution}
\end{figure}

\textbf{Intermediate regime and linear growth}. For $t > t_1$ the behavior of HEE deviates from the quadratic growth. It starts approaching the linear regime, and reaches the linear asymptotic growth at a time scale $t = t_2$. The time scale $t_2$ depends inversely on the horizon radius $R$, which is shown on the plot \ref{HEE-evolution}A, and thus the higher the temperature, the more prominent is the linear growth regime. The leading linear asymptotic behavior can be established by expanding the difference $S_{\text{non-eq}}-S_{\text{eq}}$ in $e^{-R t}$. The result is the following\footnote{Note that this expansion is convergent only for high temperatures or long times, such that $e^{-R t}$ is small.}:
\be
\Delta S(t) = S_{\text{non-eq}}-S_{\text{eq}} =\frac{c}{3} R\ t+ \frac{c}{3} \log \left(\frac{\coth \frac{\pi R}{2}}{8 \sinh^2 R \frac{\Delta\varphi}{2}}\right) + O(\e^{-R t}) \,. \label{HEElinear}
\ee

The global quench models also exhibit the linear growth regime, which is evident from both CFT calculations \cite{Calabrese05,Calabrese16} and holographic calculations \cite{Abajo10,Bal10,Liu13,Liu2013,Li13,Hartman13,Hubeny2013,Mezei2016,Mezei16,Ziogas,Ageev17}. In the context of holographic thermalization global quench scenarios the linear growth regime is often referred to as \textit{entanglement tsunami} \cite{Liu13,Liu2013,Li13}. It was found \cite{Liu13,Liu2013,Hartman13,Li13} that the linear behavior in global quench models is universal and can be expressed as 
\be
\Delta S(t) = v_E s_{\text{eq}} A\ t + \dots\,; \label{LinearUniv}
\ee
Here $s_{\text{eq}}$ is the equilibrium density of HEE, $A$ is the surface area of the boundary of the subsystem, and $v_E$ is the entanglement velocity, which depends only on the dimension of the spacetime. We can make contact with our case in a similar way to the discussion in \cite{Ziogas}, if we consider the case $R\ \Delta \varphi / 2 \gg 1$. In this case the equilibrium entanglement entropy (\ref{Seq}) is given by pure area law (we omit the UV cutoff):
\be
S_{\text{eq}} \simeq \frac{c}{6} R \Delta \varphi\,;
\ee
from where we find that $s_{\text{eq}} = \frac{c}{6}R$. Taking into account that $A=2$, since our subsystem is bounded only by two points, we find that the asymptotic linear behavior (\ref{LinearUniv}) should look like 
\be
\Delta S(t) = v_E \frac{c}{3} R t + \dots\,; \label{LinearUniv2}
\ee
The universal result for global quenches in $2$d CFT is $v_E = 1$, and from comparing of (\ref{HEElinear}) to (\ref{LinearUniv2}) we see that this value for the entanglement velocity holds true for our bilocal quench scenario as well. Thus we obtain another argument for the fact that the notion of the entanglement velocity and its bounds is relevant not only for global quenches, but also for local quenches as well \cite{Rangamani15,Rozali17,Erdmenger17}. 
As it turns out, the linear growth regime in our case is directly related to the memory loss regime \cite{Liu13,Liu2013,Ageev17}, since the function $\Delta S$ in the leading order only depends on the difference $U = t = \frac{\Delta \varphi}{2}$. Another interesting fact that the time scale $t_2$ is related to the crossing HRT geodesic going inside the horizon, which is an evidence for the fact that the linear growth of HEE is related to the HRT geodesics probing the interiors of the black hole. We discuss these observations in more detail later in subsection \ref{Tsunami}. 

\textbf{Thermalization}. The transition to the equilibrium regime happens when the crossing geodesic and the direct geodesic have the same length. Hence thermalization time can be found from the condition 
\be
\mathcal{L}_{\text{crossing}}(t_*^{(a,b)}, \varphi_a; t_*^{(a,b)}, \varphi_b) = \mathcal{L}_{\text{direct}}(\varphi_a; \varphi_b)\,. \label{ThermalizationCondition}
\ee
where we have emphasized the time dependence in the length of the crossing geodesic. Using the formula (\ref{LcrossingHRT}) for l. h. s. and the formula (\ref{LRT}) for r. h. s., we find the expression for thermalization time: 
\bea \label{Ttherm}
&&\cosh R\ t_*^{(a,b)} = \cosh R \varphi_0 \left(\cosh R \frac{\Delta \varphi}{2} - \frac{1}{\mathcal{E}}\sinh R \frac{\Delta \varphi}{2}\right) \\\nn &&+ \sqrt{\cosh^2 R \varphi_0 \left(\cosh R \frac{\Delta \varphi}{2} - \frac{1}{\mathcal{E}}\sinh R \frac{\Delta \varphi}{2}\right)^2 - \cosh^2 R \varphi_0 - \sinh^2 R \frac{\Delta \varphi}{2} + \frac{1}{\mathcal{E}} \sinh R \Delta \varphi}\,;
\eea
It is important that this moment in time is later than the time of emergence of the direct geodesic $t_{\text{cr}}$ given by (\ref{t0}), but it is also before the time when the crossing geodesic vanishes, since the point 3) of the proposition (\ref{CrossingEvolution}) directly states that the crossing geodesic vanishes at some time when its length has grown higher than the length of the direct geodesic. Thus we have the continuous transition from the non-equilibrium growth to saturation of HEE happening at $t = t_*^{(a, b)}$. The formula \eqref{Ttherm} dictates that larger subsystems thermalize slower, see plot \ref{HEE-evolution}B, which is expected. Also let us note that for symmetric intervals $\varphi_0 = 0$ one can obtain from \eqref{Ttherm} and (\ref{t0}) the result 
\bea
t_* = t_{\text{cr}} = \frac{\Delta \varphi}{2}\,. \label{TthermSym}
\eea
This is the same thermalization time as in the AdS-Vaidya quench model \cite{Abajo10,Bal10,Liu13,Ziogas}. 
For a subsystem of the same given size $\Delta \varphi$ but non-zero $\varphi_0$ the thermalization time defined from (\ref{Ttherm}) will be smaller than $\frac{\Delta \varphi}{2}$, as shown on the plot \ref{HEE-evolution-detailed}C. 

Now let us discuss the character of the transition to saturation. While the HEE itself is continuous, which is what expected from the thermalization models, particularly those based on the global quench scenario \cite{Abajo10,Bal10,Bal2012,Liu13,Liu2013,Li13,Calabrese16,Anous16,Hartman13,Hubeny2013,Mezei2016,Mezei16,Ziogas, Ageev2017,Calabrese05,Ageev17}, the time derivative of the entanglement entropy is discontinuous at the transition point, which results in \textit{sharp transition} to saturation. This fact is the key difference of non-equilibrium dynamics in our model from dynamics in holographic Vaidya \cite{Abajo10,Bal10,Bal2012,Liu13,Liu2013,Li13,Anous16,Hartman13,Mezei2016,Mezei16,Ziogas,Ageev2017,Ageev17} and end-of-world brane \cite{Hartman13,Mezei16} quantum quench models in $2$d CFT. However, the sharp transition to saturation, when the linear growth regime lasts all the way to thermalization, is remarkably similar to that of the quasiparticle picture of entanglement spreading in $2$d CFT \cite{Calabrese16} and to equilibration of a strip subsystem after the global Vaidya quench in $d \geq 3$ \cite{Liu13,Mezei16}. 

\subsubsection{Emergent light cone}

On the Fig.\ref{HEE-3D}A we plot $\Delta S$ as a function of time and $\varphi = \Delta \varphi / 2$ in the special case of centered subsystems, $\varphi_0 = 0$. The expression (\ref{Snon-eq}) in this case simplifies to 
\be
S_{\text{non-eq}}^{\text{sym}} (\varphi | t) =\frac{c}{3}\log\left[\frac{2}{\epsilon} \left(\mathcal{E} (\cosh R \varphi - \cosh R t) - \sinh R \varphi\right)\right]\,. \label{Snon-eqSym}
\ee
From this picture and for the formula for the thermalization time $t = \varphi$ (\ref{TthermSym}) we can see that the entanglement spreads along an effective light cone. The sharp saturation, makes the light cone prominent, again hinting at similarity with the quasiparticle picture of entanglement spreading \cite{Calabrese16}. The effective light cone velocity is related to the \textit{butterfly velocity} $v_B$, which is the speed of propagation of quantum chaos in thermal state \cite{Shenker13,Maldacena15,Mezei2016,Qi17}. In the setting of global quench in two-dimensional holographic CFT it is true that $v_{LC} = v_B = v_E =1$ \cite{Mezei2016}. From the plot \ref{HEE-3D}A we observe that $v_{LC} = v_E =1$ also holds in our case of the bilocal quench.

One can also verify that $v_{LC} = v_B = 1$ in our quench scneario using considerations analogous to those in section 5 of \cite{Mezei2016}, which relate thermalization of perturbations with the concept of entanglement wedge reconstruction. The spreading of information on the boundary can be characterized by the rate of growth of a boundary region which contains the information about an initial excitation (for concreteness, we consider the particle $1$). In analogy with \cite{,Mezei2016}, we assume that the full information about the infalling excitation is contained in the centered boundary segment for which the direct geodesic crosses the wordline of the particle in the given moment of time. The bulk subregion bounded by this segment and the direct geodesic which crosses the worldline of the particle is the \textit{entanglement wedge} \cite{Czech12,Wall12,Headrick14,Dong16}. The corresponding boundary segment encodes all information about local physics in this bulk subregion \cite{Dong16}. In this case the rate of growth of this boundary segment is given by the formula for the thermalization time (\ref{TthermSym}). From this formula it follows that $v_{LC} = 1$. 

Another point to note is that the dependence of $S_{\text{non-eq}}(0)$ as a function of $\varphi_0$ in (\ref{Snon-eq0}) and the time dependence in (\ref{Snon-eqSym}) are identical. Moreover, if one considers the full entropy at $t = 0$ as a function of $\varphi_0$, then the dependence will be the same as the time dependence of the entropy for $\varphi_0 =0$. In the first case the saturation which happens for $|\varphi_0| > \frac{\Delta \varphi}{2}$ corresponds to the case when the segment completely lies in the upper or lower half of the boundary spatial circle and doesn't contain an initial excitation - as we discussed, all such subsystems are at equilibrium from the beginning. This symmetry between temporal and spatial coordinates of the segment appeared a symmetry 3) in the formula for the length of the crossing geodesic (\ref{Lcrossing}). In the case of HEE dynamics, this symmetry can be qualitatively explained using the quasiparticle model \cite{Calabrese16}. In $2$d CFT after the quench the spreading entanglement can be approximately modeled as propagation of EPR pairss of particles, which are created in the moment $t = 0$ by the quench and propagate with the speed of light \cite{Calabrese16,Hartman13,Casini15}. In our case these EPR pairs are created in two separate points: $\varphi = 0$ and $\varphi = \pi$. A subsystem with a given size $\Delta \varphi = 2\varphi$ will equilibrate once the quasiparticles will reach the its boundaries. If the subsystem is centered on the source of particles, i. e. $\varphi = 0$, then a given moment of time $t$ the quasiparticles will reach the coordinate $\pm \tilde{\varphi} = \tilde{t}$. On the other hand, this is the same situation as if particles were created at the point $\pm \tilde{\varphi}$ at $t = 0$. When $t = \varphi$, the particles reach the boundary of the segment, or, equivalently, the source of particles is at the boundary, thus equilibration happens. 

This quasiparticle explanation hints that this symmetry is caused by the production of entanglement by a point source from a bilocal quench. Consequently, there is no analogue of such symmetry in any global quench model because there is always full translational invariance to work with, and EPR pairs are produced uniformly in every point of space. It is also unlikely that this symmetry would hold in a higher-dimensional generalization, because in general velocities which characterize the spread of entanglement are slower in higher dimensions \cite{Liu13,Liu2013,Shenker13,Mezei2016}. However, we can expect that a CFT calculation of correlation functions after bilocal quench in $2$d CFT using the techniques along the lines of \cite{Asplund14,Anous16} can help reveal its nature in relation to some deep symmetries in CFT such as modular invariance. 

\subsubsection{Entanglement tsunami and memory loss} \label{Tsunami}

As we mentioned in subsection \ref{222}, it is known that a closed system loses information about the details of the initial state upon thermalization. In the previos work \cite{Liu13,Liu2013,Li13,Mezei16,Ageev17} it was found that the memory loss phenomenon in the context of holographic global quenches is prominent in the late-time evolution of HEE. Namely, at late times for large enough subsystems it was found out that HEE in the leading order only depends on the difference of time and the subsystem size, and separate dependence on those variables are exponentially suppressed at late times. This is referred to as \textit{memory loss regime} of HEE, and it happens shortly (in relative terms) before saturation in the global quench scenario. In this subsection we argue that the model of holographic bilocal quench also exhibits a property of memory loss for large subsystems, and this memory loss seems to be related with the black hole interior. 
 
\begin{figure}[t]
\centering
\includegraphics[width=7cm]{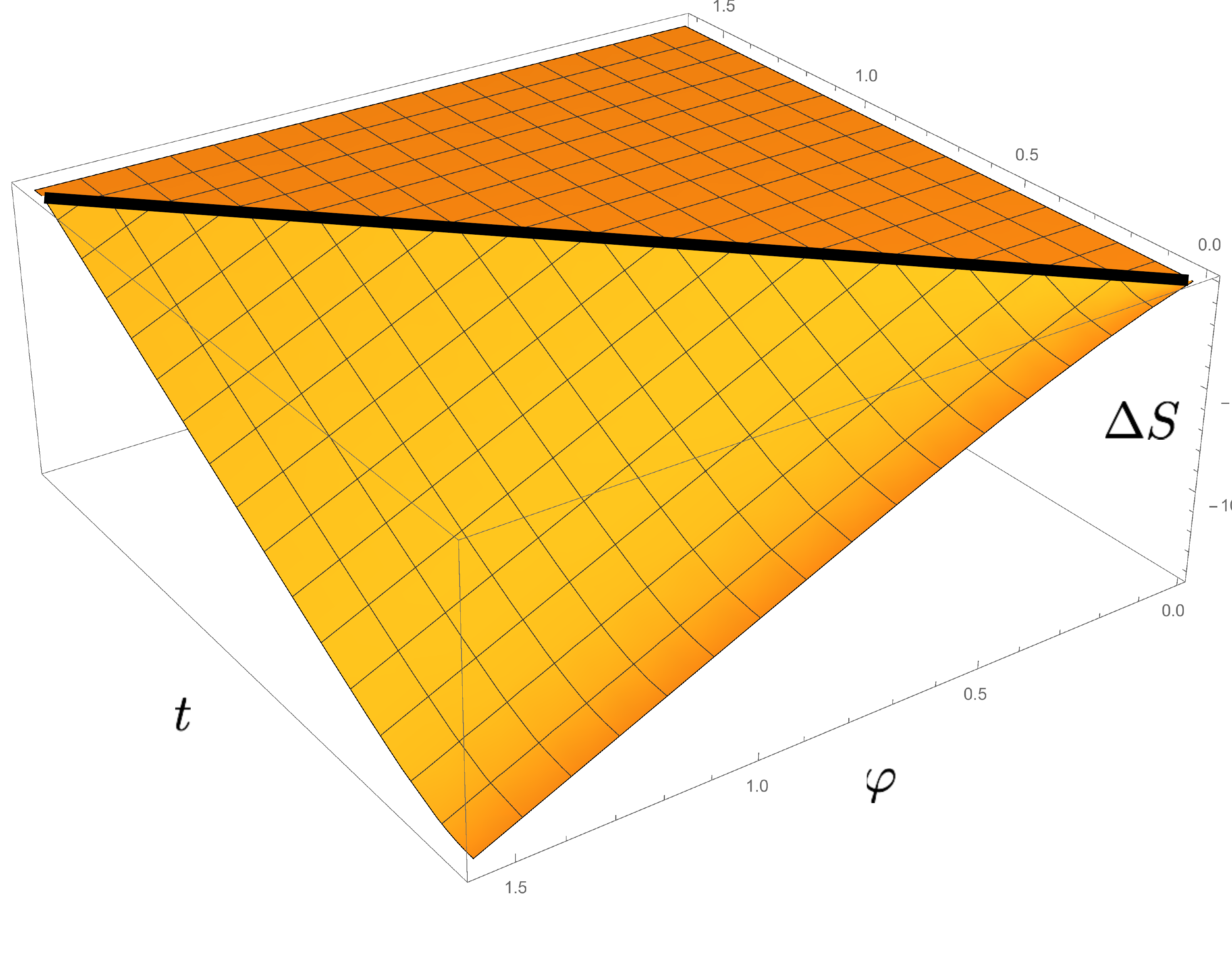}A.
\includegraphics[width=6cm]{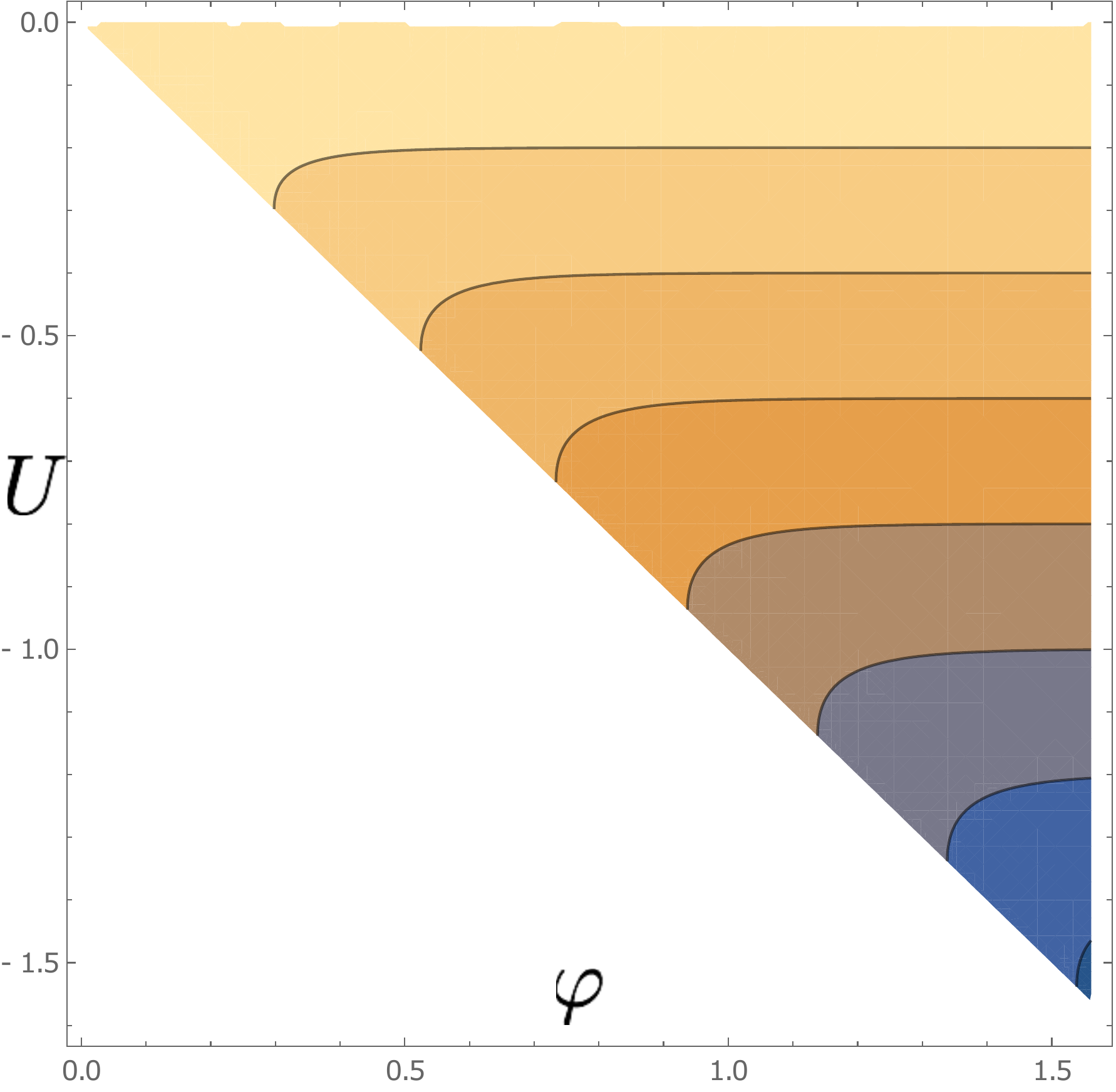}B.
\caption{\textbf{A}: Three-dimensional picture of entanglement spreading in case of symmetric intervals. The horizontal plateau represents the equilibrium regime. \textbf{B}: Density plot of non-equilibrium HEE as a function of $\varphi=\frac{\Delta \varphi}{2}$ and $U = t - \varphi$, with $R=5$. The horizontal lines signify the disappearance of dependence of $\varphi$, which is the memory loss regime.}
\label{HEE-3D}
\end{figure}

To observe it, we consider the expression for $\Delta S(t) = S_{\text{non-eq}}(t)-S_{\text{eq}}$, where $S_{\text{non-eq}}$ is given by (\ref{Snon-eq}) and $S_{\text{eq}}$ is given by (\ref{Seq}) in terms of coordinates of boundary points $\varphi_a$, $\varphi_b$ instead of $\Delta \varphi$ and $\varphi_0$. We find asymptotic expansion of $\Delta S$, taking $\e^{R \varphi_a} \sim \e^{-R \varphi_b} \ll 1$ and also considering the limit of late times, such that $\e^{-R t} \to 0$, but $\e^{R (t - \varphi_b)}$ and $\e^{R (t+ \varphi_a)}$ are kept fixed. The asymptotic has the form
\bea\nn
\Delta S (t) &\sim & \log \left[\left(\coth \left(\frac{\pi 
   R}{2}\right) \left(e^{R (\varphi_a+t)}-1\right)+1\right)
   \left(\coth \left(\frac{\pi  R}{2}\right) \left(e^{R
   (t-\varphi_b)}-1\right)+1\right)\right] \,; \label{MemoryLoss1}\eea
So we see that indeed the late-time dynamics of entanglement of large systems are characterized by the functional dependence of time and endpoints of the subsystem as light cone-like variable combinations $t-\varphi_b$, $t+\varphi_a$. This asymptotic is dominant for any values of the horizon radius. For large horizon radius $R > 1$, one can expand further to obtain the following leading behavior:
\be
\Delta S = R (t-\varphi_b) + R (t+\varphi_a) + 2 \log \coth \frac{\pi R}{2} + O(\e^{-R (t+\varphi_a)}) + O(\e^{-R (t-\varphi_b)}) \,; \label{MemoryLoss2}
\ee
This expression is the linear growth regime discussed above (\ref{HEElinear}), but with the additional requirement of large subsystem. Thus we conclude that for large subsystems during relaxation to equilibrium at high enough temperature the memory loss regime is basically the linear growth regime. On the Fig.\ref{HEE-3D}B we plot $\Delta S$ of symmetric intervals with $\varphi_b = -\varphi_a = \varphi$ as a function of $U = t-\varphi$ and $\varphi$. The region when the lines become horizontal is where the distinguishable dependence on $\varphi$ gets suppressed, which illustrates the memory loss regime. In other words, we observe that for large subsystems with boundaries far away from initial excitation at $\varphi=0$ the entanglement propagates as a wave front with retarded coordinate $t-\varphi$, and for high temperatures this wave-like behavior can be identified as the entanglement tsunami linear growth regime. 
This picture is largely similar to the global quench story \cite{Liu13,Liu2013} in $2$d. 

We also can take into account that initially we have two excitations, and segments of similar size and with positions that mirror each other on the boundary circle will equilibrate synchronously. That means that on upper or lower half of the boundary circle we will have two waves of entanglement propagating in opposite directions. One can therefore make a speculation that the instant equilibrium of HEE in segments which belong completely to the upper or lower half of the circle can be caused by something similar to a standing wave of entanglement, which emerges as a result of clash of two waves of entanglement from initial excitations.

\subsubsection{Linear growth and black hole interior}

The peculiar feature of non-local observables such as HEE in holographic non-equilibrium processes is that they can probe the region inside the apparent horizon. More specifically, in the holographic global quench thermalization scenario it was understood \cite{Hartman13,Li13} that HRT surfaces which calculate HEE in the boundary can probe the interior of the black hole, i. e. region of the spacetime inside the horizon. Moreover, it was found that the linear growth of HEE comes from growth of the piece of the HRT surface which lies inside the horizon. In the present subsection we discuss similar feature of the holographic bilocal quench model. 

We start by noting the fact that a crossing HRT geodesic which consists of two pieces of image geodesics, can reach inside the black hole. This happens because, as we discussed in section \ref{sectionGeodesics}, the image geodesics can span the entire global AdS$_3$ cylinder, and not just the fundamental domain of the topological identification. Because of this and the shape of the identification wedge relative to the image geodesics (see Fig.\ref{crossing-basic}), the image geodesics can probe the interior of the black hole, which is located behind the horizon (see Fig.\ref{falling_particles}). More important is the fact that the actual crossing geodesic, that is pieces $a o$ and $o^* b$ of the image geodesics, can reach inside the horizon as well. In other words, a crossing geodesic probes the interior of the black hole when points $o$ and $o^*$ lie inside the horizon. 
\begin{figure}[t]
\centering
\includegraphics[width=14cm]{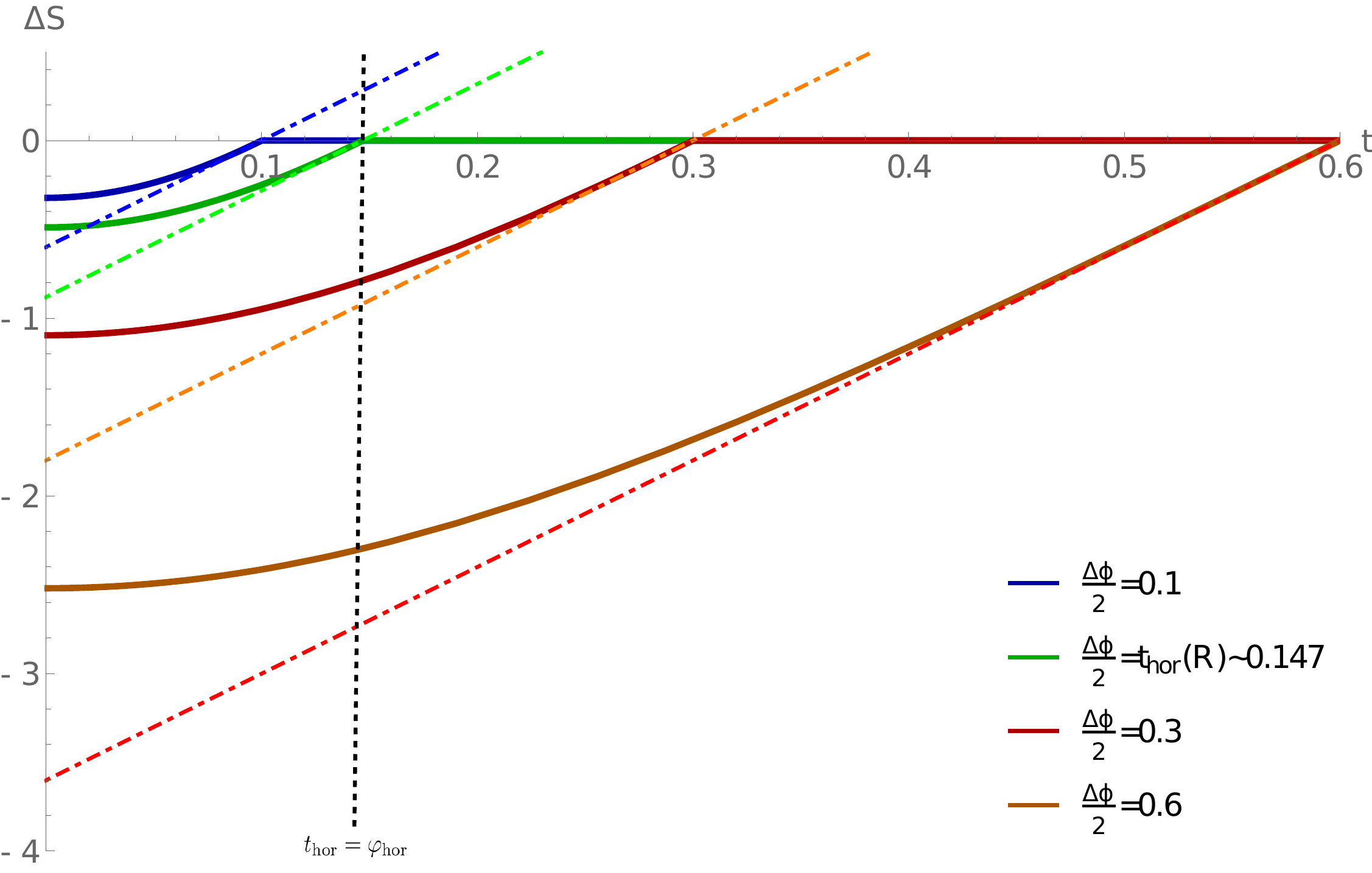}
\caption{Evolution of HEE of segments which probe the interior and do not probe the interior. }
\label{HEE-horizon}
\end{figure}
It appears that in our case the piece of the crossing geodesic lying inside the horizon is responsible for the linear growth of HEE. We were not able to prove this statement analytically like it was done in global quench scenario in \cite{Hartman13,Li13}, but we can establish this observation based on numerics. We also can obtain a bound for the subsystem size which separates subsystems which probe the black hole interior and thus exhibit the linear growth of HEE, from those which do not, in a special case. 

To accomplish this, let us consider symmetric segments only, with $\varphi_b = -\varphi_a = \varphi$. For these segments, as discussed in the proof of the point 3) of proposition \ref{CrossingEvolution}, the direct geodesic emerges in the same moment when the crossing geodesic vanishes, that is when $o = o^*$. That means that we can use the condition $o = o^*$ as a condition of thermalization in global coordinates for symmetric segments. Because intervals are symmetric, it is true that $\mathcal{L}(a, o)$ = $\mathcal{L}(o^*, b)$. But it is also true that $\mathcal{L}(o^*, b) = \mathcal{L}(o, b^\#)$. That means that when $o = o^*$, the point $o$ bisects the image geodesics in half. In particular, 
\be
\tau_{o=o^*} = \frac12 (\tau_a + \tau_{b^\#})\,.
\ee
As also discussed in the proof of the proposition \ref{CrossingEvolution}, the functions $\tau_{b^\#} (\tau_b)$ and, conssequently,  $\tau_{o} (\tau_b)$ are monotonically increasing faster than the value of $\tau_b = \tau_a$ itself. That means that we can detect the enetering of the crossing geodesic inside the horizon by looking first at when the point $o=o^*$ will enter the horizon in global coordinates. The point $o = o^*$ is also located on the worldline of the particle $1$, which enters the horizon at $\tau = -\frac{\pi}{4}$, which can be found from the worldline equation (\ref{Globalworldline}) and the horizon surface equation (\ref{hor-eq}) \cite{Matschull}. Thus we come to a condition
\be
\frac12 (\tau_a + \tau_{b^\#}) = -\frac{\pi}{4}\,; \label{HorCondition}
\ee
This is the moment of global time when a subsystem will thermalize exactly at the time when the deepest point of the HRT geodesic $o=o^*$ will just reach the horizon. The smaller subsystems which thermalize faster (because of the relation $t = \varphi$) will not probe the interior, and larger subsystems which thermalize longer will probe the interior. 

Keeping in mind that $\tau_a$, $\tau_{b^\#} \in [-\frac{\pi}{2}, \frac{\pi}{2}]$, we can rewrite this equation as 
\be
\tan \tau_a = \cot \tau_{b^\#} = \frac{1}{\tan \tau_b (1 + 2 \mathcal{E}^2) + \frac{2 \mathcal{E}}{\cos \tau_b} (\mathcal{E} \cos \phi_b + \sin \phi_b)}\,;\label{HorCondition1}
\ee
where we used the formula (\ref{tauDiez}) for $\tau^\#$ at the boundary $\chi_b \to \infty$. Next, we use the equalities $\phi_b = -\phi_a$, $\tau_b = \tau_a$ and transform the equation (\ref{HorCondition1}) into BTZ coordinates using the formulas (\ref{Sch-transf}: 
\be
(1 + 2\mathcal{E}^2) \cosh^2 R \varphi - 2 \mathcal{E}^2 \cosh R \varphi \cosh R t - \mathcal{E} \sinh 2R \varphi = \sinh^2 R t\,.
\ee
Then we substitute the expression (\ref{TthermSym}) for thermalization time for symmetric subsystems $t = \varphi$, the definition of $\mathcal{E} = \coth \frac{\pi R}{2}$, and then solve for $\varphi$. The result is 
\be
\varphi_{\text{hor}} = \frac{1}{2R} \arcsinh \tanh \frac{\pi R}{2}\,; \label{HorBound}
\ee
This is the lower bound for the size of the symmetric segment which will probe interior of a black hole of the given size at some point during its equilibration. We denote the corresponding thermalization time as $t_{\text{hor}}$

We plot time evolution of HEE for symmetric segments which do or do not probe the interior on the Fig.\ref{HEE-horizon}. The blue curve corresponds to a segment that doesn't probe the interior since its size is smaller than $\varphi_{\text{hor}}$. The green curve corresponds to the critical segment with $\varphi = \varphi_{\text{hor}}$, which just touches the horizon in the moment of saturation. Red and brown curves all correspond to large segments that probe the interior and, as seen on the plot, exhibit the linear growth regime. Larger segments have longer linear growth. The time scale $t_2$ which signifies the start of the linear growth (see Fig.\ref{HEE-evolution-detailed}) is the time scale when the crossing HRT geodesics penetrates the horizon. The formula (\ref{HorBound}) gives the lower bound for $t_2$ in case of symmetric segments for given $R$. 

Thus we see that the probing of the black hole interior by crossing HRT geodesics is related to the linear growth of HEE. In the previous subsection we also discussed that for large temperatures the linear growth is related to the memory loss regime for a given subsystem. We see that in the bilocal quench model it is evident that loss of memory of the initial state at thermalization for a given subsystem is related to the problem of probing the black hole interior. 

\subsubsection{Scrambling time and entanglement shadow}
\begin{figure}[t]
\centering
\includegraphics[width=7cm]{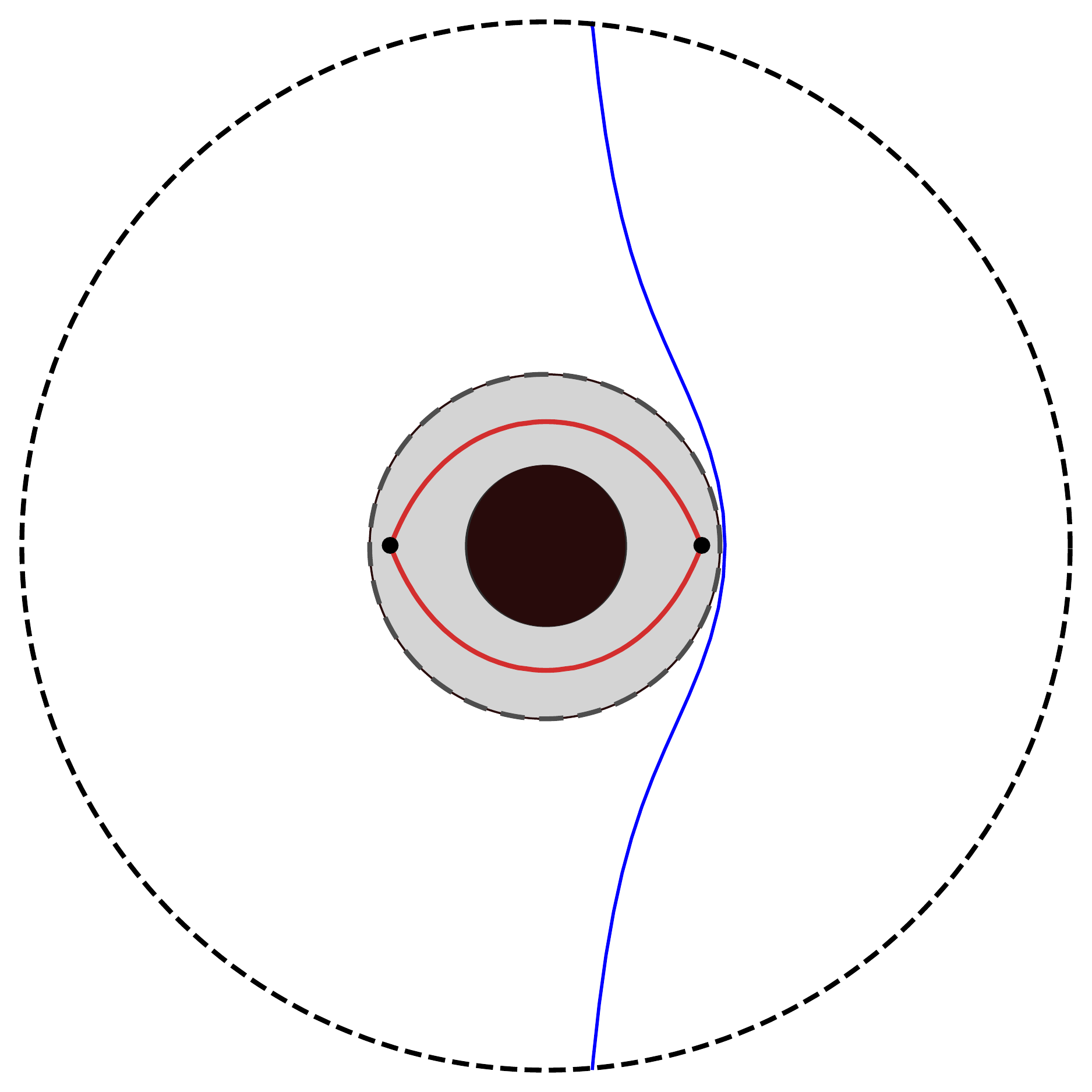}
\caption{The grey region is the entanglement shadow, which appears when $t > \frac{\pi}{2}$. The blue curve is the maximal RT geodesic.}
\label{shadow}
\end{figure}
Bigger subsystems thermalize longer, according to the formula (\ref{Ttherm}) (see Fig.\ref{HEE-evolution}B). On the other hand, among the segments of the same size $\Delta \varphi$ the segments which thermalize longest are the symmetric segments with $\varphi_0 = 0$.
From these considerations it follows that there are two distinct subsystems among all subsystems which thermalize last: it is the symmetric segment with $\varphi_b = -\varphi_a = \frac{\pi}{2}$ and its complement. The thermalization time of these subsystems can be calculated from the formula (\ref{TthermSym}) and equals to 
\be
t_* = \frac{\pi}{2}\,. \label{Tscrambling}
\ee
At this time, all subsystems, which include half of total degrees of freedom in the boundary theory or less, are thermalized. This timescale is often referred to as \textit{scrambling time} \cite{Sekino08,Lashkari13} in the context of relaxation of small perturbation of the thermal state. This is the time scale when the information about the initial state is scrambled so thoroughly that it cannot be restored from any small fraction of the total amount of degrees of freedom. 

Now let us ask the question: how the fact that after the scrambling time all small subsystems are thermalized is reflected in the bulk geometry? All small subsystems being at equilibrium means that their entnaglement entropy is governed by direct geodesics, which means that colliding particles are not probed by HRT geodesics anymore. In other words, beginning from $t = t_*$, particles are located in a region of the spacetime which is not probed by the entanglement. This signifies that the \textit{entanglement shadow} \cite{Bal14,Freivogel14} appeared in the bulk in the moment $t = t_*$, that is the region which is not probed by RT surfaces. This is a region between the horizon and the radius given by the depth of the direct geodesic with opening angle $\pi$ or, equivalently, from the position of the particle at $t = \frac{\pi}{2}$, see Fig.\ref{shadow}. Using the latter definition and the worldline equation (\ref{BTZworldline}), we get
\be
r_{\text{shadow}} = R \coth \frac{\pi R}{2} = R \mathcal{E}\,. \label{Rshadow}
\ee
Thus we get that the dimensionless radius of the entanglement shadow equals the energy of the particle $1$ in global coordinates. This entanglement shadow is precisely the same as the entanglement shadow of the BTZ black hole spacetime \cite{Hubeny13,Bal14}.

Thus we observe that during the evolution of a pure state after the quench the information about the initial state gets scrambled when the ingoing matter falls into the entanglement shadow, which is associated with the forming black hole. While the true mixed thermal state is unreachable during the unitary time evolution, we do get the state, where all the initial data is completely scrambled over the entire system. Holographically this means that the horizon does not form in finite amount of time, however the entanglement shadow appears after the scrambling time, which is finite for compact CFT. 

\subsection{Mutual information}

\begin{figure}[t]
\centering
\includegraphics[width=7cm]{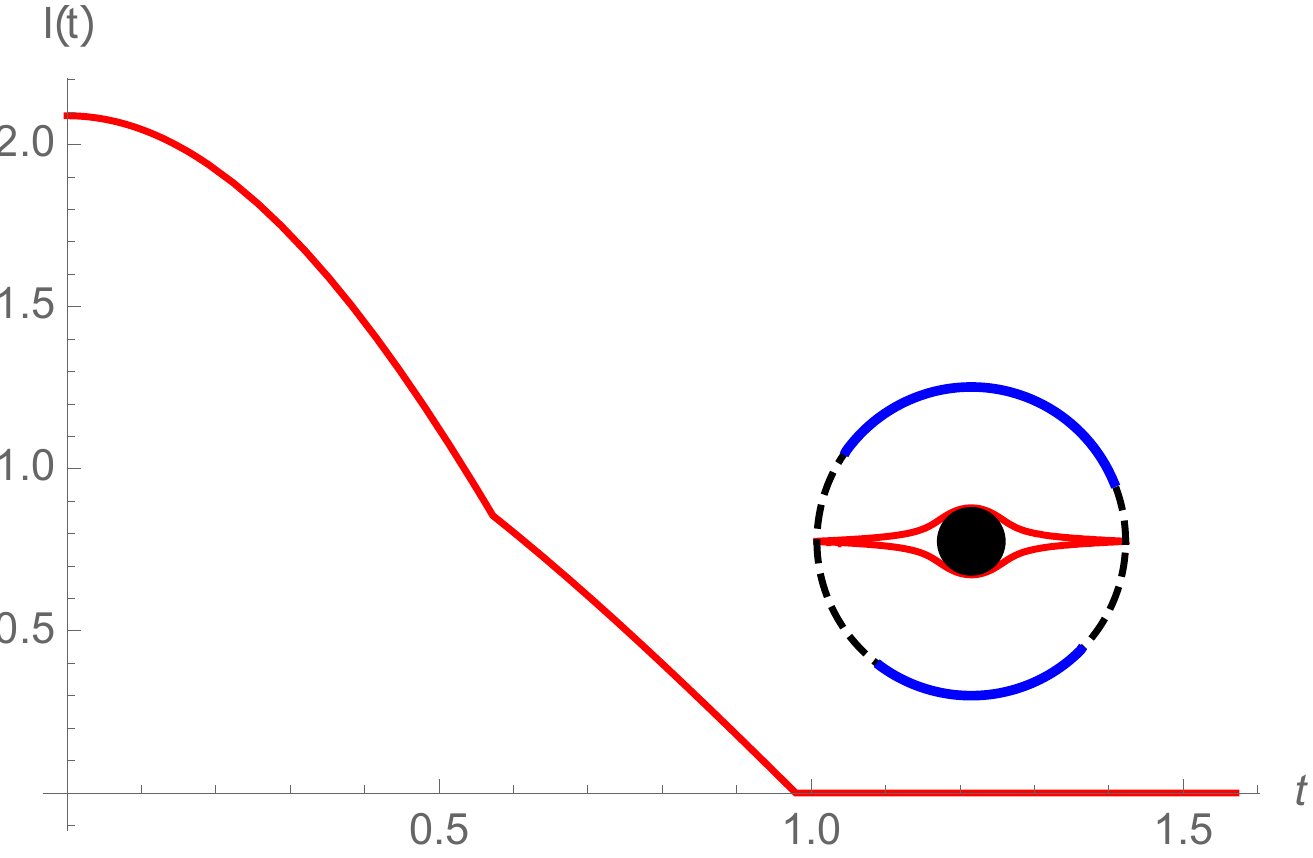}A. 
\includegraphics[width=7cm]{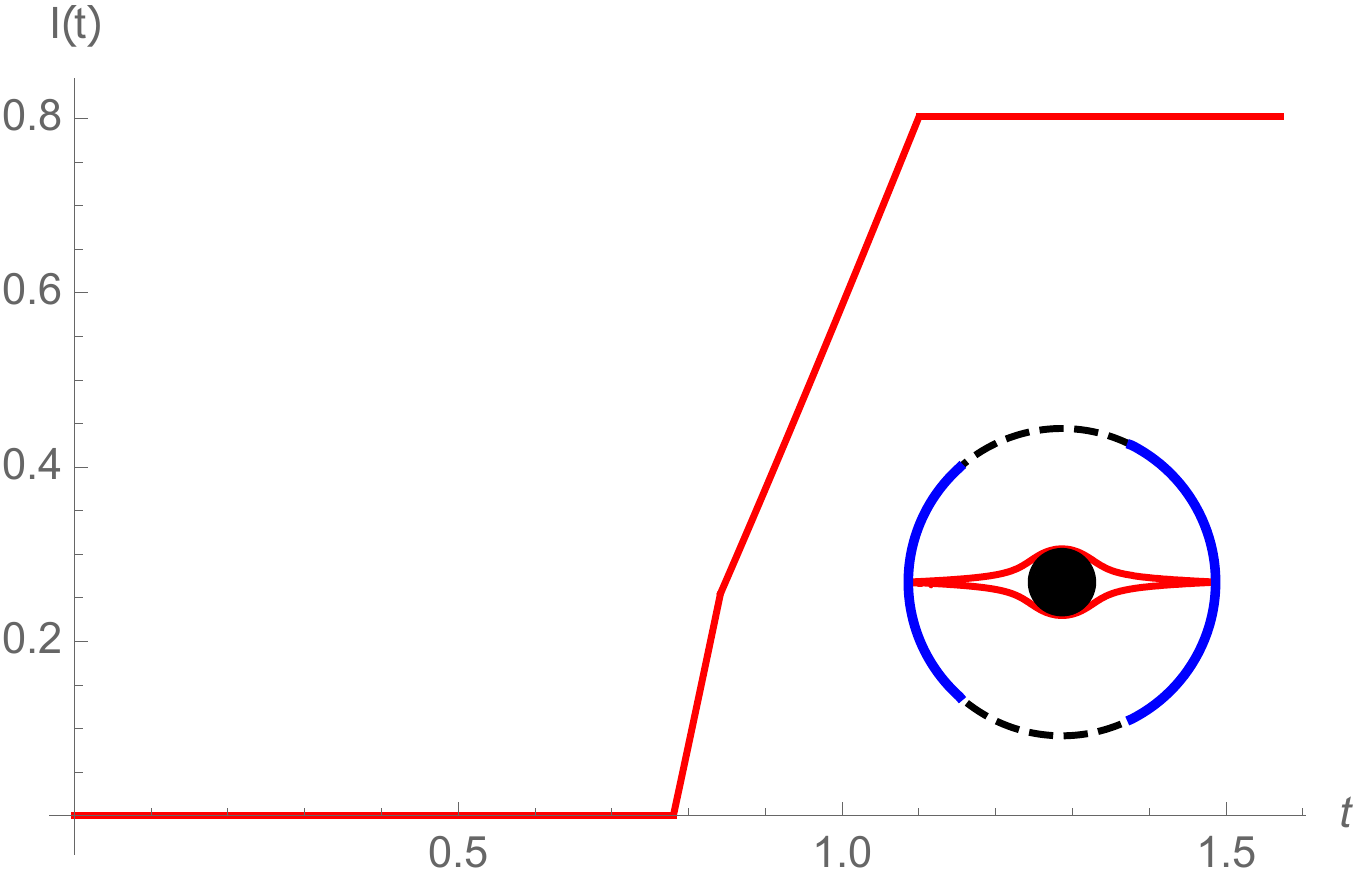}B.\\
\includegraphics[width=7cm]{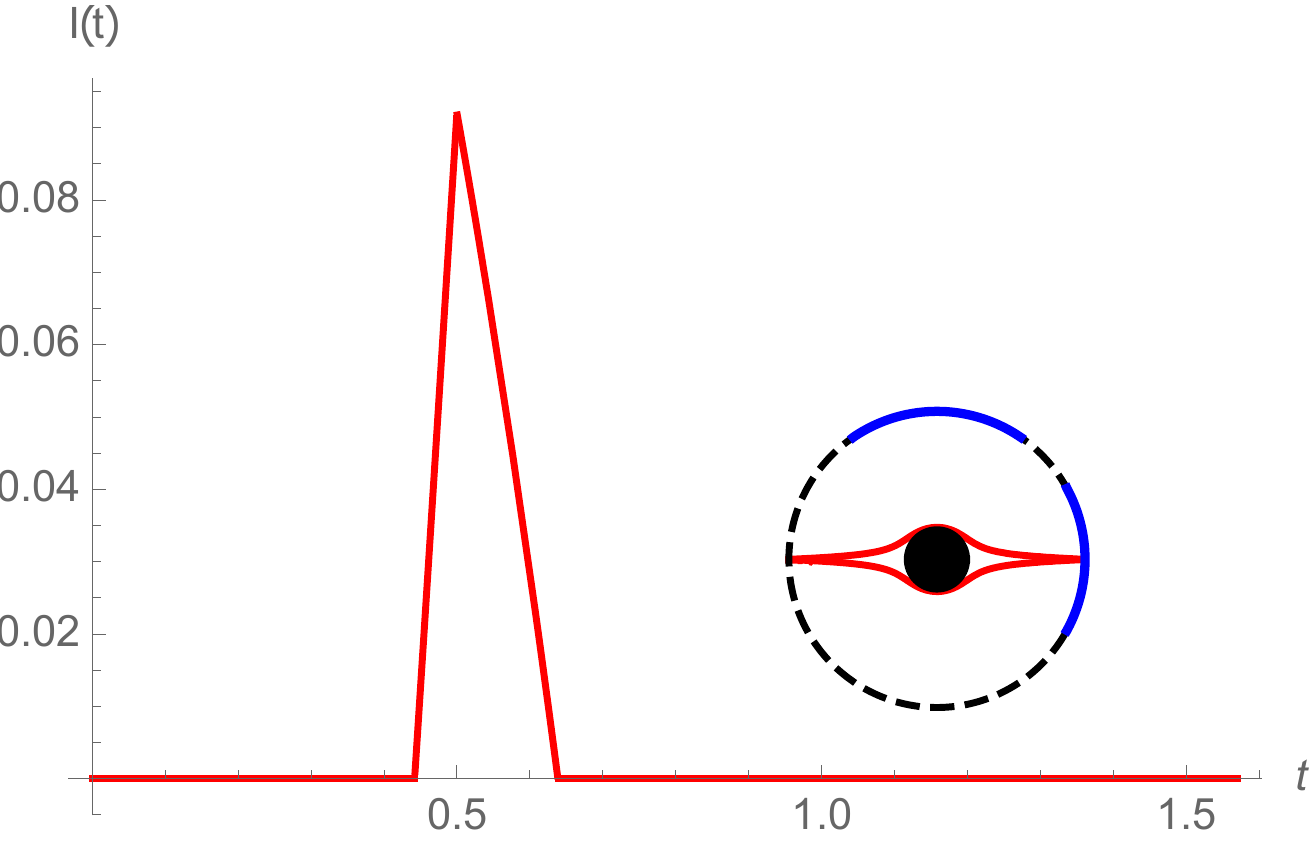}C.
\caption{Mutual information with dependence of time for different subsystems. Scales of subsystems: A. $\Delta\varphi_{up}=27\pi/40$, $\Delta\varphi_{down}=9\pi/20$; B. $\Delta\varphi_{left}=1.68$, $\Delta\varphi_{right}=2.2$; C. $\Delta\varphi_{up}=1.2$, $\Delta\varphi_{right}=1$ }
\label{MI-pics}
\end{figure}
In this section we consider the behavior of mutual information using HRT prescription. The mutual information for two disjoint regions is defined as:  
\bea\label{MI}
I(A, B) = S(A) + S(B) - S(A\cup B)
\eea
where $S(A\cup B)$ is a joint entropy, i.e. a minimal geodesic length between two possible types of geodesics. It means that the mutual information is zero if two regions are well separated. 

We compute the mutual information for different scales and locations of regions and analyze the time dependence. We are interested in equilibration of mutual information that can be described as reaching plateau behavior.

The mutual information in thermal equilibrium is either zero or a positive constant. We expect that the mutual information reaches equilibrium regime when all terms in the eq.(\ref{MI}) saturate. We present the most destinctive cases of mutual information behavior as function of time in the Fig.\ref{MI-pics}. We can see that the equilibration comes at the different moments of time that we called $t^{AB}_{*}$ that has been considered in several papers \cite{Asplund13,Asplund15,Caputa15, Ageev2017, Ziogas}. A value of this time scale depends on the location and size of the regions, but is bounded from above by the scrambling time of the entire system $t_* = \frac{\pi}{2}$ (see the discussion above). 

We have plotted the pictures of the mutual information in dependence of time up to scrambling time (\ref{Tscrambling}). The mutual information behavior in the system with two excitations is generally similar to the behavior in AdS-Vaidya geometry \cite{Ziogas}, however there are kinks that can be described differently which occur when the change of a regime happens in one of the terms in (\ref{MI}). In the Fig.\ref{MI-pics}A. two kinks correspond to the thermalization of the small and large joined geodesics. In the Fig.\ref{MI-pics}B. the first kink means that a small subsystem thermalizes as well as the second kink corresponds to the thermalization of the second subsystem. The mutual information in the last case, presented in the Fig.\ref{MI-pics}C., has a peak that corresponds to the thermalization of the subsystem which contains the initial excitation.

\section{Thermalizing correlators} \label{sectionCorr}

Another way to probe thermalization is to study two-point correlation functions in the state in the boundary CFT which is dual to the bulk spacetime with a forming black hole. In this section we discuss non-equilibrium two-point correlation functions of a scalar operator $\mathcal{O}_\Delta$. We calculate the Lorentzian two-point correlation functions in the framework of the geodesic approximation \cite{Bal99}, which holds for $\Delta \gg 1$. We apply the geodesic approximation in the BTZ coordinates of AdS$_3$ spacetime with colliding massless particles. According to this prescription, to calculate the two-point correlator between points $a$ and $b$ on the boundary of a given asymptotically AdS spacetime, one has to sum over all geodesics between these points. The original formulation \cite{Bal99} is valid in the Lorentzian signature only when the points on the boundary are spacelike-separated. In the case of timelike-separated points, the situation is more tricky, and a generalization is needed, e.g. see \cite{Bal2012,AKT}. Because of this issue, we consider two cases separately. The pole structure of Lorentzian correlators is recovered by introducing the phase factors which result in a desired $i\epsilon$-prescription. With that in mind, the schematic formula for the correlator between spacelike-separated points $a$ and $b$ in geodesic approximation reads:
\be
 G_\Delta^A (t_a, \varphi_a; t_b, \varphi_b) =\sum_n \e^{-\Delta \mathcal{L}_{\text{ren}}(a, b)}  \times \Phi_A^\Delta (a, b; n) \,; \label{CorrGeodesic}
\ee
where index $A$ stands for $ret$ (retarded), $F$ (Feynman) or $W$ (Wightman) correlators. The lengths of the geodesics are appropriately renormalized by subtraction of diverging part. The limits of summation are defined by the concrete geometry of the defect in the AdS$_3$ spacetime. It can include a finite number of winding geodesics, like in case of a point particle \cite{AB,AAT,AA,Arefeva2015,AKT,AK}, or an infintite amount of winding geodesics, like in case of the BTZ black hole \cite{Keski98}, see formula (\ref{GWeq}). In the general case, the set of values of the summation variable $n$ depends on the choice of points $a$ and $b$ as well. However, one can always distinguish the dominating contribution coming from a minimal geodesic. 

\subsection{Spacelike correlations}

\subsubsection{Leading contributions}

In the section \ref{sectionGeodesics} we have established that if points $a$ and $b$ are spacelike-separated, one can always distinguish a minimal geodesic among the set of all geodesics between them - this is either direct geodesic, or the crossing geodesic, depending on the location of the endpoints. 
So for the most of this section we will only consider the leading contributions to correlators: 
\be
 G_\Delta^A (t_a, \varphi_a; t_b, \varphi_b) \sim \e^{-\Delta \mathcal{L}_{\text{min}}(a, b)}  \times \Phi_A^\Delta (a, b; 0) \,. \label{CorrGeodesic0}
\ee
Here $\Phi_A$ is an appropriate factor for a given correlator dictated by the $i\epsilon$-prescription in the Lorentzian signature. The explicit formulas for different $\Phi_A$ are given by (\ref{PhaseW},\ref{PhaseRet},\ref{PhaseF}). Let us restrict ourselves to Wightman and Feynman correlators. Then for spacelike-separated points $\Phi_{F,W}=1$, and we omit these factors in this subsection. 

Analogously to HEE, we have two kinds of behavior of correlation functions depending on the location of the endpoints.

1. Suppose that endpoints are located to the same side of the collision line, that is either $\varphi_a$, $\varphi_b \in [0, \pi]$ or $\varphi_a$, $\varphi_b \in [-\pi, 0]$. Then the minimal geodesic is the direct geodesic, as stated by proposition \ref{DirectSide}. The length of the direct geodesic is given by (\ref{Ldirect}). We renormalize it by subtracting the $2 \log \frac{r_0}{R}$ piece, and thus the correlator is thus given by 
\be
 G_\Delta^A (t_a, \varphi_a; t_b, \varphi_b) \sim \left(\frac{1}{2 (\cosh R \Delta \varphi - \cosh  R \Delta t) }\right)^\Delta \,. \label{CorrDirect}
\ee
This is the same result as in case of thermal equilibrium (\ref{GWeq}), and it is expected since in section \ref{sectionHEE} we discussed that with these endpoints HEE, which is also defined by the minimal geodesic length, is at its equilibrium value.

2. Suppose that endpoints are located on different sides of the boundary with respect to the collision line. More specifically, suppose that $\varphi_a \in [-\pi, 0]$ and $\varphi_b \in [0, \pi]$. Then the minimal geodesic is either the direct geodesic, or the crossing geodesic. In the general case the lengths of these two geodesics are comparable, as evident from our discussion of HEE, so we take into account both these contributions on equal footing. However, as we discussed, generally crossing and direct geodesics for such choice of endpoints do not always exist, so we need to account for that as well. For this purpose we introduce auxiliary $\Theta$-functions: 
\bea
&& \Theta_{\text{cross}}(a, b) = \left\{\begin{array}{ccc}
1\,,& \quad \text{if $a^* b$ and $a b^\#$ intersect $W_\pm$}\\
0\,,& \quad \text{if $a^* b$ and $a b^\#$ do not intersect $W_\pm$}\,.
\end{array}\right.\,; \label{ThetaCross} \\
&& \Theta_{\text{dir}}(a, b) = \left\{\begin{array}{ccc}
1\,,& \quad \text{if the direct geodesic does not intersect $W_\pm$}\\
0\,,& \quad \text{if the direct geodesic intersects $W_\pm$}\,.
\end{array}\right.\,; \label{ThetaDir}
\eea
The idea is that $\Theta = 1$ if the corresponding geodesic exists, and $\Theta = 0$ if it does not. The crossing geodesic exists when the image geodesics cross the identification wedge, and the direct geodesic exists when it does not cross the identification wedge and fully belongs to the fundamental domain in the bulk. Hence the definition.

Now, using the formula for length of the crossing geodesic (\ref{Lcrossing}) and for length of the direct geodesic (\ref{Ldirect}) and again subtracting the divergence, we can write the expression for the correlator as: 
\bea
&& G_\Delta^A (t_a, \varphi_a; t_b, \varphi_b) \sim \left(2 \left((-1+\mathcal{E}^2) \cosh R \Delta t+(1 + \mathcal{E}^2) \cosh R \Delta\varphi + \mathcal{E}^2 \cosh 2 R \varphi_0 +\right.\right.\nn\\&& \left.\left. \mathcal{E}^2 \cosh 2 R t_0 + 4\mathcal{E} \cosh R t_0 \cosh R \frac{\Delta t}{2} \cosh R \varphi_0 \left(\sinh R \frac{\Delta\varphi}{2} - \mathcal{E} \cosh R \frac{\Delta\varphi}{2}\right) +\right.\right.\nn\\\nn && \left.\left. 4\mathcal{E} \sinh R t_0 \sinh R \frac{\Delta t}{2} \sinh R \varphi_0 \left(\mathcal{E}\sinh R \frac{\Delta\varphi}{2} - \cosh R \frac{\Delta\varphi}{2}\right)  - 2 \mathcal{E} \sinh R \Delta \varphi\right)\right)^{-\Delta}\times\nn\\&& \times\Theta_{cross}(a, b)  + 
 \left(\frac{1}{2 (\cosh R \Delta \varphi - \cosh R \Delta t) }\right)^\Delta \times \Theta_{dir}(a, b) \,. \label{CorrCrossing}
\eea
This is a discontinuous function because of the fact that different geodesics not always exist. We will discuss the issue of these discontinuities in geodesic approximation in more detail in the next subsection. For now, let us consider the correlation function (\ref{CorrCrossing}) in more detail in particular case of equal-time points $t_a = t_b = t$. In this case we can write $\Theta$-functions as Heaviside step functions of time: 
\bea
&& \Theta_{\text{dir}}(a, b) = \theta (t - t_{cr})\,;\\
&& \Theta_{\text{cross}}(a, b) = \theta (-t+ t_{o = o^*})\,;
\eea
The first equality reflects the fact that the direct geodesic exists after $t = t_{cr}$ given by formula (\ref{t0}). The second line means that the crossing geodesic exists until the time when the corresponding image geodesics cross $W_\pm$ in the same point $o = o^*$ located on the worldline of the particle $1$. The equal-time correlator then reads
\bea
&& G_\Delta^A (t, \varphi_a; t, \varphi_b) \sim \left(2 \left((-1+\mathcal{E}^2)+(1 + \mathcal{E}^2) \cosh R \Delta\varphi + \mathcal{E}^2 \cosh 2 R \varphi_0 +\right.\right.\nn\\&& \left.\left. \mathcal{E}^2 \cosh 2 R t + 4\mathcal{E} \cosh R t \cosh R \varphi_0 \left(\sinh R \frac{\Delta\varphi}{2} - \mathcal{E} \cosh R \frac{\Delta\varphi}{2}\right) +\right.\right.\nn\\\nn && \left.\left.  - 2 \mathcal{E} \sinh R \Delta \varphi\right)\right)^{-\Delta}\times\theta(-t+t_{o = o^*}) +  \left(\frac{1}{2 (\cosh R \Delta \varphi - 1) }\right)^\Delta \times \theta(t-t_{cr}) \,. \label{CorrCrossingEqT}
\eea
In the section \ref{sectionNon-Eq} we have calculated the thermalization time $t_*^{(a, b)}$ of a segment between $a$ and $b$ (\ref{Ttherm}). From the perspective of correlation functions, this is the time when two terms in (\ref{CorrCrossingEqT}) exchange dominance: for $t < t_*^{(a, b)}$ the first term is leading and we have non-equilibrium regime, and for $t>t_*^{(a, b)}$ the second, equilibrium term dominates. Note that from proposition \ref{CrossingEvolution} it follows that $t_{cr} < t_*^{(a, b)}< t_{o = o^*}$. That means that in general there are two small intervals in the end of non-equilibrium regime and in the beginning of the equilibrium regime when both terms contribute. This is a feature which arises in correlation functions, but the HEE is always defined by a single minimal geodesic and therefore is insensitive to such details. 

\subsubsection{Contribution of windings}

In the section \ref{windings} we have discussed that in the bulk between the two given boundary points winding geodesics emerge at late times. These winding geodesics give subleading contributions to two-point functions in geodesic approximation, according to the general formula (\ref{CorrGeodesic}). The correlators dual to BTZ black hole spacetime include the contribution of infinite amount of winding geodesics (\ref{GWeq}), whereas in case of our bulk spacetime with particles the identification wedge prohibits the existence of any winding geodesics at least at early times. Therefore, it is interesting to look at the emergence of winding geodesics as subleading contributions to the correlation functions, as they tell us how thermalizing correlation functions approach the equilibrium correlation functions (\ref{GWeq}) at late times. 

In the proposition (\ref{WindingExistence}) we have discussed that a winding geodesic between points $a$ and $b$ with the winding number $n$ exists after the time $\tilde{t}_n$, when it is able to wind around the horizon without intersecting $W_\pm$, see Fig.(\ref{windings-emerge}). We thus can define the $\Theta$-functions which regulate the presence of winding contributions in the correlation functions:
\bea
\Theta_{\text{winding}}(a, b; n) = \left\{\begin{array}{ccc}
1\,,& \quad \text{if the $n$-th winding geodesic does not intersect $W_\pm$}\\
0\,,& \quad \text{if the $n$-th winding geodesic intersects $W_\pm$}\,.
\end{array}\right.\,; \label{ThetaWind}
\eea
where $n \neq 0$ is an integer. Note that $\Theta_{\text{winding}}(a, b; |n|) \neq \Theta_{\text{winding}}(a, b; -|n|)$ (as can be seen from Fig.\ref{windings-emerge}). According to the proposition \ref{WindingExistence}, we can simply write
\be
\Theta_{\text{winding}}(a, b; n) = \theta (t_0 - \tilde{t}_n)\,;
\ee
and also in the limit $t_0 \to \infty$ it is true that $\Theta_{\text{winding}}(a, b; n) = 1$ for any $n$ and any angular coordinates of endpoints $a$ and $b$. Using these functions and the expression for the length of a winding geodesic (\ref{geodesicLengthW}), we conclude that the general form of the two-point correlation functions with spacelike-separated points in geodesic approximation is the following, depending on the location of the endpoints.
\begin{itemize}
\item For $\varphi_a$, $\varphi_b \in [0, \pi]$ or $\varphi_a$, $\varphi_b \in [-\pi, 0]$, we have 
\bea
&& G_\Delta^A (t_a, \varphi_a; t_b, \varphi_b) =\left(\frac{1}{2 (\cosh R \Delta \varphi - \cosh R \Delta t) }\right)^\Delta +\nn\\&&
 \sum_{n \in \ZZ,\ n \neq 0} \left(\frac{1}{2(\cosh [R (\Delta \varphi + 2 \pi n)] - \cosh [R \Delta t]}\right)^\Delta \theta (t_0 - \tilde{t}_n)\,; \label{CorrWinding1}
\eea
\item For $\varphi_a \in [-\pi, 0]$ and $\varphi_b \in [0, \pi]$, we have 
\bea
&& G_\Delta^A (t_a, \varphi_a; t_b, \varphi_b) \sim \left(2 \left((-1+\mathcal{E}^2) \cosh R \Delta t+(1 + \mathcal{E}^2) \cosh R \Delta\varphi + \mathcal{E}^2 \cosh 2 R \varphi_0 +\right.\right.\nn\\&& \left.\left. \mathcal{E}^2 \cosh 2 R t_0 + 4\mathcal{E} \cosh R t_0 \cosh R \frac{\Delta t}{2} \cosh R \varphi_0 \left(\sinh R \frac{\Delta\varphi}{2} - \mathcal{E} \cosh R \frac{\Delta\varphi}{2}\right) +\right.\right.\nn\\\nn && \left.\left. 4\mathcal{E} \sinh R t_0 \sinh R \frac{\Delta t}{2} \sinh R \varphi_0 \left(\mathcal{E}\sinh R \frac{\Delta\varphi}{2} - \cosh R \frac{\Delta\varphi}{2}\right)  - 2 \mathcal{E} \sinh R \Delta \varphi\right)\right)^{-\Delta}\times\nn\\&& \times\Theta_{cross}(a, b)  + 
 \left(\frac{1}{2 (\cosh R \Delta \varphi - \cosh R \Delta t) }\right)^\Delta \times \Theta_{dir}(a, b) + \nn\\&&
 \sum_{n \in \ZZ,\ n \neq 0} \left(\frac{1}{2(\cosh [R (\Delta \varphi + 2 \pi n)] - \cosh R \Delta t}\right)^\Delta \theta (t_0 - \tilde{t}_n) \,. \label{CorrWinding2}
\eea
\end{itemize}
From these formulas (\ref{CorrWinding1}-\ref{CorrWinding2}) it is clear that in the limit $t \to \infty$ one recovers the equilibrium two-point function (\ref{GWeq}). The geodesic approximation dictates that the subleading terms in the image sum appear one by one with time, as the particles move towards each other in the bulk. 

The presence of $\Theta$-functions in the correlators obtained from geodesic approximation when there are multiple geodesics possible between two given endpoints are a common occurence in cases of locally AdS$_3$ spacetimes with singularities which admit multiple geodesics. A prime example for such occurence is the AdS$_3$ with a point particle \cite{Bal99,AB,AAT,AKT,AK}. Our bulk spacetime is a special case of AdS$_3$ with two massless point particles. Correlators dual to AdS$_3$ with two particles were studied in global coordinates in \cite{Arefeva2015,AA}, and they also exhibit such discontinuities. On the example of massive static particle in AdS$_3$ it was shown in \cite{AK} that geodesic approximation gives a continuous result which coincides with the correlators obtained from GKPW dictionary, if the spacetime is an AdS$_3$ orbifold. In the general case, however, the geodesic correlators are discontinuous, and these discontinuities are smoothened by corrections to the geodesic approximation, which is evident if one calculates correlation functions using e.g. full GKPW dictionary \cite{AK}. In the orbifold case, these corrections to the geodesic approximation are exactly zero.

In our case, the bulk spacetime itself is not an orbifold. However, as the time goes on, the spacetime geometry approaches that of the BTZ black hole spacetime, which is an oribfold. So one could say that in our case the bulk spacetime looks more and more like an AdS$_3$ orbifold. So if we assume that considerations from \cite{AK} can be extended for our case of the AdS$_3$ spacetime with two massless particles, then it can be expected that the corrections to the geodesic approximation will be diminishing at late times. 

\subsection{Timelike correlations}

Now let us turn to the discussion of the time dependence of two-point correlation functions. We consider the case of endpoints $a = (0,\ \varphi_a)$ and $b = (t,\ \varphi_b)$ (with $t  >0$) and study the time dependence of correlation functions in the approximation of the leading term in geodesic approximation (\ref{CorrGeodesic0}). First, let us note that when $t$ is small enough such that $ t^2 - \Delta \varphi^2 < 0$, then the interval between the endpoints is spacelike and we can just use the geodesic approximation as discussed above. However, for timelike-separated points on the boundary $t^2 - \Delta \varphi^2 > 0$ there is no real smooth geodesic between them. The geodesic approximation requires some continuation to the case of timelike-separated points. 

This issue previously was tackled in the study of holographic global quench models \cite{Abajo10,Bal10,Bal2012} and other non-stationary backgrounds \cite{AB,AAT,AKT}. Several methods to continue the geodesic approximation to the timelike case were proposed in \cite{Bal2012}: a specific analytic continuation, complexified geodesics and quasigeodesics. 
Here we use the method based on reflection mapping \cite{AAT,AKT}, which can be considered as a streamlined version of method of quasigeodesics adapted for topological quotients of AdS$_3$ in global coordinates. 

The idea is the following \cite{AKT}. Let us define a mapping $R$ which acts on the boundary of AdS$_3$ in global coordinates: 
\be
R:\; (\tau, \phi) \mapsto (\tau+\pi, \phi+\pi)\,; \label{reflection}
\ee
This mapping has the key property that is useful for us. Consider a timelike interval $a b$ on the boundary. If one acts with $R$ on the point $a$, then the resulting interval $a^R b$ is spacelike. This is because in global coordinates $\phi \sim \phi + 2\pi$. 
\begin{figure}[t]
\centering
\includegraphics[width=6cm]{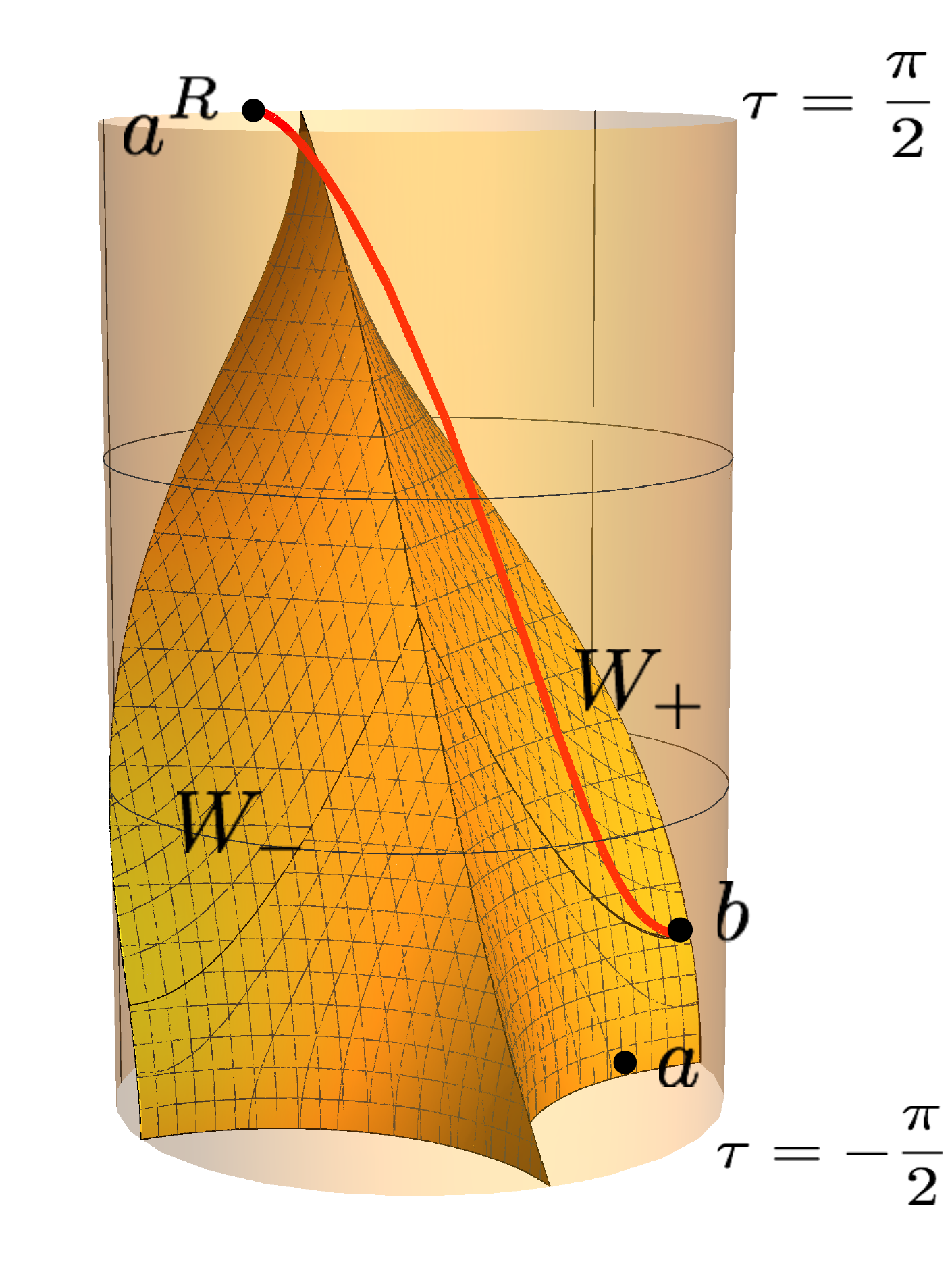}
\caption{A reflection geodesic in relation to the wedge of the infalling particle.}
\label{reflectionFig}
\end{figure}
Now, since $a^R$ and $b$ are spacelike-separated, we can consider the AdS$_3$ geodesic between points $a^R$ and $b$. We will call it the reflection geodesic. Its length can be found from the matrix formula (\ref{LdirTr}). We parametrize the points $a$ and $b$ by matrices $A$ and $B$, according to (\ref{SL2}), by global coordinates (\ref{barrel}), keeping in mind that they are located near the boundary at $\chi_0 \to \infty$:
\bea
&&A = \e^{\chi_0} \begin{pmatrix} \cos \tau_a + \sin \phi_a
  & \sin \tau_a + \cos \phi_a \\
-\sin \tau_a + \cos \phi_a & \cos \tau_a - \sin \phi_a \\
 \end{pmatrix}\,;\label{AandBglobal}\\&&
B = \e^{\chi_0} \begin{pmatrix} \cos \tau_b + \sin \phi_b
  & \sin \tau_b + \cos \phi_b \\
-\sin \tau_b + \cos \phi_b & \cos \tau_b - \sin \phi_b \\
 \end{pmatrix}\,;\nn
\eea
Using (\ref{reflection}), we write the matrix parametrization for coordinates of reflected point $a^R$: 
\be
A^R = \e^{\chi_0} \begin{pmatrix} -\cos \tau_a - \sin \phi_a
  & -\sin \tau_a - \cos \phi_a \\
\sin \tau_a- \cos \phi_a & -\cos \tau_a + \sin \phi_a \\
 \end{pmatrix} = -A \,; \label{Areflected}
\ee
This formula is explicitly invariant and we will use it later for points $a$ and $b$ parametrized by BTZ coordinates as in (\ref{AandB}).
Now we can use (\ref{LdirTr}) and (\ref{Areflected}) to write the expression for the reflection geodesic length in terms of matrices of initial endpoints $A$ and $B$:
\be
\mathcal{L}(a^R, b) = \log\ \tr (A^R)^{-1} B = \log \left(-\tr A^{-1} B\right)\,; \label{LreflectTr}
\ee
In our case we deal with timelike-separated endpoints on the boundary, which in global coordinates are parametrized as $a=(-\frac{\pi}{2}, \phi_a)$ and $b = (\tau, \phi_b)$. To find the two-point correlation function between these points, we use the reflection prescription \cite{AKT} and replace regular direct geodesic with the reflection geodesic for timelike-separated points:  
\be
G_{\Delta}^{A}(a, b) \sim \e^{-\Delta \mathcal{L}_{\text{ren}}(a^R, b)} \times \Phi^\Delta_A (a, b)\,; \label{GtimelikeGeneral}
\ee
Note that we also take into account the $\Phi$-factors of Lorentzian correlators, which are given in Appendix \ref{Lorentzian} by formulas (\ref{PhaseW},\ref{PhaseRet},\ref{PhaseF}). For our choice of endpoints, the reflection geodesic is a regular direct geodesic. This follows from the fact that by definition of reflection mapping $\tau_{a^R} = \pi$. Meanwhile, in global coordinate the particle worldline is a light-like geodesic which goes from the point $(0, 0)$ to the point $(\pi, \pi)$ on the boundary, and the identification surfaces $W_\pm$ are located beneath the worldline, see Fig.\ref{reflectionFig} (the BTZ identification surfaces $V_\pm$ are not shown for clarity). This means that all reflection geodesic will pass freely above the wedge, without intersecting it. Moreover, $\phi_{a^R} = \phi_a + \pi$ means that the point $a^R$ is located in the second external region with respect to the BTZ black hole identification, and thus the reflection geodesic is a direct geodesic that goes through the wormhole from the external region under consideration in our problem into the other exterior region which is located in the identification dead zone. Thus, the direct geodesic which goes through the wormhole is the only possible reflection geodesic. 

Now we can calculate its length using the formula (\ref{LreflectTr}) and parametrizing points $a$ and $b$ by BTZ coordinates according to (\ref{AandB}). The result is (we subtract the diverging part):
\be
\mathcal{L}_{\text{ren}}(a^R,\ b) = \log \left[2(-\cosh(\varphi_b - \varphi_a) + \cosh(t_b - t_a))\right]\,. \label{Lreflection}
\ee
Note how it is different from the expression for the length of direct spacelike geodesic (\ref{Ldirect}) by opposite sign of the expression under the logarithm, as expressed by the formula (\ref{LreflectTr}. 
Now we substitute this expression into (\ref{GtimelikeGeneral}) to obtain the two-point correlation functions between timelike-separated points $a = (0,\ \varphi_a)$ and $b = (t,\ \varphi_b)$, $t > 0$:
\be
 G_\Delta^A (0, \varphi_a; t, \varphi_b) \sim \left(\frac{1}{2 (-\cosh R \Delta \varphi + \cosh  R t) }\right)^\Delta  \Phi^\Delta_A (a, b)\,. \label{CorrDirectTimelike}
\ee
This expression establishes the fact that the leading temporal behavior of the two-point correlation functions after the quantum quench is the same as in thermal equilibrium (\ref{GWeq}) and does not carry any details about the infalling matter which forms the black hole. The same result was obtained in the case of Vaidya global quench \cite{Bal2012}. Note, however, that if one takes the temporal coordinate of one endpoint before the quench, e.g. $t_a < 0$, then the time dependence of the correlation function will be different \cite{Bal2012,Anous16}. In the context of our holographic bilocal quench model of colliding particles forming a black hole this calculation is out of the scope of this paper. However we can speculate that in such situation we would have to consider reflection geodesics which somehow pass through the identification $W_\pm$ like spacelike crossing geodesic. Such reflection geodesic will carry information about the holonomy of the infalling particle and will lead to a more distinct time dependence of the correlation function. 

\section{Conclusions and discussion} \label{sectionDiscussion}

In this paper we have investigated black hole formation from a collision of particles as a model of holographic thermalization in the boundary theory after the bilocal quantum quench in the framework of AdS$_3$/CFT$_2$-correspondence. This model holographically describes unitary time evolution of a non-homogeneous initial state with two excitations on the antipodal points of the CFT spacetime cylinder. We have probed the thermalization process by studying the non-equilibrium behavior of entanglement and two-point correlation functions. More specifically, we studied the dynamics of entanglement by using the time-dependent generalization of the RT prescription \cite{RT,HRT} to calculate the holographic entanglement entropy and mutual information, and investigating their time dependence, and we used the geodesic approximation and its extension to timelike case to study real-time two-point functions. The key results are the following.
\\
$\,\,\,$
\\
\textbf{1}. In the initial state of the boundary theory the subsystems which are located between the excitations exhibit constant thermal behavior of the entanglement entropy (\ref{Seq}), since $t = 0$. Because we study the evolution of a pure state, their complements, which contain both initial excitations, are also thermalized, with HEE given by (\ref{Seq2}). 
We  expect that this partial equilibrium in the initial state from the HEE perspective is one of the features of the bilocal quench in $2$d specifically. One can think that this partial instant equilibration of the entanglement entropy is caused by the long-range entanglement induced just after the quench. The emergence of long-range entanglement is a characteristic feature of holographic local quenches, as discussed in \cite{Nozaki13} and also observed in \cite{Asplund14}. It happens as a result of the fact that we use the time reversal-invariant classical bulk dynamics for holographic description of a non-reversible process, in which we instantly inject a large amount of energy.  Meanwhile, the subsystems which at $t = 0$ contained one of the initial excitations, display non-trivial non-equilibrium dynamics of HEE (\ref{Snon-eq}) and two-point correlation functions (\ref{CorrCrossing}). This picture suggests that the bilocal quench setup perhaps could work as a toy model for diffusion of two quantum fluids in a finite-volume vessel at finite temperature which are initially divided by two walls, which is analogous to a joining quench where two systems at the same finite temeprature are brought into contact in two points (for holographic consideration of joining quench of two thermal systems in a single point, see \cite{Erdmenger17}). As explained in \cite{Asplund14}, the holographic description of the local joining quench is similar to the description of an operator local quench. In our case we observe the same on qualitative level.  
\\
$\,\,\,$
\\
\textbf{2}. We have obtained the explicit formula (\ref{Snon-eq}) for the non-equilibrium behavior of the entanglement for subsystems which contained one of the initial excitations. The time dependence of HEE governed by the formula (\ref{Snon-eq}) is substantially similar to the global quench scenarios \cite{Abajo10,Bal10,Bal2012,Liu13,Liu2013,Calabrese16,Anous16,Hartman13,Hubeny2013,Mezei16,Ziogas,Ageev2017,Ageev17}.
In our case the time evolution of HEE also display the early-time quadratic growth and the linear growth regime, see Figs \ref{HEE-evolution-detailed}-\ref{HEE-evolution}. This confirms the assumptions about universality of non-equilibrium growth of entanglement entropy with respect to the choice of initial out-of-equilibrium state in strongly-coupled systems proposed in previous work \cite{Liu13,Liu2013,Mezei2016,Mezei16} for our special choice of initial state with two excitations. However, there is a strong difference in the character of  transition to saturation: in our case the HEE undergoes a sharp transition to saturation with discontinuous first derivative, as the direct geodesic becomes dominating in the HRT prescription over the crossing geodesic. This behavior is similar to a first order phase transition\footnote{This analogy with classical thermodynamics is not complete, because in the case of thermodynamical first order transition the entropy itself a discontinuous first derivative of a thermodynamic potential.}, which also happens in higher-dimensional global quench Vaidya models and in holographic models of formation of a charged black hole \cite{Liu13,Mezei16}. Contrary to this, in the global quench setup in two dimensions the saturation transition is smooth. The sharp transition to saturation also reveals that in our case we can observe a sharp emergent light cone (see Fig.\ref{HEE-3D}), which hints at similarity with quasi particle picture of entanglement spreading. 
\\
$\,\,\,$
\\
\textbf{3}. Because of the similarities in non-equilibrium dynamics of HEE between the bilocal quench and global quench setups, we have observed that the universal characteristics of entanglement grwoth in the global quench models, namely the entanglement velocity $v_E$ and the emergent lightcone velocity $v_{LC}$ are relevant characteristics of the entanglement growth for the equilbration after the bilocal quench as well. We have confirmed that in our $2$d model they equal to the speed of light, which is in agreement to their respective definitions in the case of $2$d global quench. The fact that such velocity characteristics of entanglement propagation are meaningful in the case of certain local quenches was established in the work \cite{Rangamani15,Rozali17,Erdmenger17}, where authors deal with local quenches which drive the system out of the initial thermal equilibrium. We have also discussed that the argument about the relation between $v_{LC}$, $v_B$ and entanglement wedge subregion duality by Mezei and Stanford \cite{Mezei2016} also can be extended to our case. While it is significantly simplified by the fact that we work solely in $(2+1)$d bulk spacetime, and the RT surfaces are just direct geodesics, the interesting point is that in our case we deal not with small perturbations of the equilibrium state that fall into the black hole in the bulk, but with strong perturbations of vacuum which are responsible for the creation of the black hole and thermalization of the pure state. However, the bulk geometry of the particle creation itself in global coordinates (see Figs.\ref{falling_particles},\ref{Matschull3D}) looks similar to a setup when the particle $1$ falls into a black hole. Such situtations have been considered in the work \cite{Caputa142,Caputa15,Rangamani15,Rozali17} as holographic duals to local quench in the thermal state. We can say that the bilocal quench holographic model is a model of thermalization of a pure state which shares similarities with both global quench models and local quench models of equilibration of perturbations of the thermal state. 
\\
$\,\,\,$
\\
\textbf{4}. We have observed the relation between the scrambling time of the system $t_* = \frac{\pi}{2}$ and the emergence of the entanglement shadow in the bulk. Namely, once the state evolves part the scrambling time, the identification wedge gets fully covered by the entanglement shadow, which cannot be probed by the entangling surfaces, which by that time are all direct geodesics and cannot reach into the bulk deeper than the radius of the entanglement shadow given by (\ref{Rshadow}). In fact, after the scrambling time no subsystem can probe the infalling matter in the entanglement shadow, regardless of its particular features. That means that the same is true for e.g. an infalling Vaidya shell in AdS-Schwarzschild coordinates. We expect that the relation between the emergence of the entanglement shadow and scrambling time should be easily extendable to quantum quenches in higher-dimensional case. 
\\
$\,\,\,$
\\
\textbf{5}. We have discussed that the linear growth of HEE at large distances from initial excitations is identified with the regime of memory loss and wave-like spreading of entanglement. We also have provided some evidence that, similarly to the holographic global quench case, the linear growth is governed by HRT geodesics that probe the interior of the forming black hole. Therefore, the behavior of HRT crossing geodesics hints that evidently there is a connection between the memory loss of details of the initial state during the thermalization and the interior of the black hole. We have also discussed that there are some hints that the thermalizing state loses detailed information about the initial configuration that can be seen purely from the geometry of the bulk spacetime. First, as we discussed in the end of section \ref{sectionBHcreation}, the shape of the identification wedge, defined by holonomies of the colliding particles, approaches cylindrical at the limit $t \to \infty$. The cusps at the particle worldlines, which break the rotational symmetry, are gradually smoothed as the state approaches thermal equilibrium. 
The other point is related to which geodesics govern equilibrium and non-equilibrium regimes in our model. The non-equilibrium physics in our model is governed by crossing geodesics. The information about crossing geodesics is encoded in image geodesics. These geodesics are constructed using the identification isometry of the matter which forms the black hole, and can go anywhere in the global AdS$_3$. Meanwhile, the equilibrium regime is described by direct geodesics which are completely restricted to the fundamental domain of the identification and do not carry any information about the infalling matter which describes the initial state. We can say that after the thermalization the system has lost all information about the infalling matter and about the orbit of its identification isometry which is located outside of the fundamental domain. All of these points lead us to a conclusion that we have a strong parallel between the memory loss during thermalization and information loss in the black hole formation, and that this parallel is prominent in the behavior of holographic entanglement entropy. 
\\
$\,\,\,$
\\
\textbf{6}. The holographic mutual information can exhibit a variety of different behaviors, depending on positions and sizes of the subsystems $A$ and $B$. The most notable feature is the kinks in the time dependence of the mutual information, which happen when segments in the formula (\ref{MI}) thermalize. These kinks have the same origin as the sharp transition to saturation in the time dependence of HEE. 
\\
$\,\,\,$
\\
\textbf{7}. Similarly to the HEE, the two-point correlation functions with spacelike-separated points in the minimal geodesic approximation exhibit non-trivial non-equilibrium behavior in the case when the collision line is located between the endpoints. However, the subleading contributions from winding geodesics appear with time for any choice of endpoints at late times, and they signify the approach to equilibrium of the system as a whole. The time dependence of the correlation functions after the quantum quench coincides with the thermal behavior, same as in case of the global quench. To verify this fact, we were able to use a particularly simple extension of the geodesic approximation to the timelike case based on the reflection mapping, which is suitable for dealing with AdS$_3$ topological quotients. It is remarkable that the reflection geodesics which provide the continuation for correlation functions have to go through the black hole interior into the second asymptotic region. This shows similarity between the reflection geodesics prescription and analytic continuation of Lorentzian amplitudes in the bulk background of the eternal BTZ black hole discussed in earlier work \cite{Kraus2002}. Contrary to the latter case, in our setup we deal with a pure state, and the second exterior region is completely unphysical, yet it still seems to play substantial role in description of late time behavior of holographic observables.
\\
$\,\,\,$
\\
Now let us discuss some possible future directions of the work.
\\
$\,\,\,$
\\
(\textit{i}) First of all, we have not actually given the precise proof that the bulk spacetime with two colliding massless particles creating a black hole can indeed holographically describe the bilocal quench, realized by two operator excitations on the CFT side. To do that, one would need to calculate the $6$-point correlation functions using CFT techniques and match the results to our holographic computation. We expect that one could use the method similar to that of \cite{Anous16}, where such computation was performed for the boundary dual of the Vaidya global quench, and also to other earlier work, e.g. \cite{Asplund14,Fitzpatrick14,Hijano15}. Specifically, one could consider a $6$-point function of two light operators in the background generated by two "heavy" operators, and calculate it in the approximation of the vacuum Virasoro conformal block. The main difference from the calculation based on the monodromy method in \cite{Anous16} would be that in our case we have a finite amount (namely two) operators which produce excitations represented by massless particles in the bulk. Because of this, the limiting procedure which would allow for perturbative solution of monodromy equations at large central charge, will be different. The vacuum Virasoro conformal block of the correlation function can holographically be obtained from the length of the minimal geodesic \cite{Fitzpatrick14,Asplund14,Alkalaev16}, so we expect that such CFT computation would reproduce our holographic expressions for correlation functions (\ref{CorrDirect}) and (\ref{CorrCrossing}) in the leading order of the semiclassical expansion. Also, throughout the paper we have been implicitly assuming that the energy of initial excitations must be high enough in order to form a BTZ black hole instead of just a static conical defect. These two cases are easily identified and separated, if one considers the collision of particles in the center of mass frame, as shown by the formula (\ref{EnergyBound}) (see \cite{Matschull}), but the black hole rest frame bulk geometry doesn't have a well-defined continuation to the case of the formation of a conical defect, so in our holographic computations this energy threshold is hidden. The CFT computation of correlators could clarify this issue as well.
\\
$\,\,\,$
\\
(\textit{ii})
An interesting related question is study of the bilocal quench away from the holographic limit $c \to \infty$. Specifically, it could be interesting to see how the entanglement scrambles in examples when the cental charge is finite. Such studies were conducted for local quench \cite{Asplund13,Asplund14}, as well as global quench \cite{Asplund15} scenarios. The key observation in the latter case is that in some cases at finite central charge there are memory effects, which interrupt the equilibrium saturation of HEE and mutual information at late times. We could expect that in the bilocal quench scenario one would also observe similar memory effects for finite (small) $c$ theories. Moreover, one could consider the $N$-local quench protocol generalization for more than $2$ excitations in order to find out how the late time memory effects in HEE depend on the fraction of initial excitations located inside the given subsystem, or simply speaking how much memory do these memory effects carry about the initial state. However, of course the study of quenches produced by multiple localized excitations in a CFT at finite $c$ seems to be a challenging problem. 
\\
$\,\,\,$
\\
(\textit{iii}) Another related question is the study of the $1/c$ corrections. We have observed that HEE in our model exhibits sharp transition to saturation. We could expect that the perturbative semiclassical  corrections in $1/c$ would smooth out the transition. Even more interesting are the non-perturbative, $\e^{-c}$ corrections. These are considered to be the corrections which should restore the information lost in black holes in holographic correlation functions \cite{Fitzpatrick16}. In case of our setup that would mean that the $\e^{-c}$ corrections to HEE would carry information about the topological identification even after the saturation. That fact could be used to investigate the question: just how much the entanglement entropy actually knows about the bulk geometry beyond the semiclassical approximation and out of the entanglement wedge?
\\
$\,\,\,$
\\
(\textit{iv}) A possibly promising direction of further study is the higher-dimensional generalizations of the bilocal quench to the cases of AdS$_4$/CFT$_3$ and AdS$_5$/CFT$_4$. The holographic dual for thermalization after bilocal quench in those cases should be the AdS spacetime with colliding shockwaves which create a black hole \cite{ArefevaQGP}. From the present work, as well as from previous work on local quenches in $3$d \cite{Rangamani15,Rozali17} and bounds on the entanglement propagation \cite{Mezei2016,Mezei16} we can expect that the non-equilibrium dynamics of entanglement will be more rich in higher dimensions. Also, the bilocal quench in AdS$_5$/CFT$_4$ could serve as a viable more realistic holographic model of thermalization in heavy ion collisions \cite{ArefevaQGP,Arefeva12}.
\\
$\,\,\,$
\\
(\textit{v}) Also an interesting direction is to consider the collision of multiple particles. The shape of the identification surfaces will be much more sophisticated, but we expect that we could see many similarities. In particular, it is plausible that if we consider the collision of $N$ particles, then we still will have the HEE of subsystems which are located in between the particles instantly thermalized, and HEE of subsystems which contain some (but not all) particles in the initial moment would demonstrate some sort of non-equilibrium nontrivial dynamcs. It was shown that the Vaidya shell can be obtained by taking the continuous limit $N \to \infty$ of $N$ colliding particles \cite{Lindgren15}, and the CFT dual of the Vaydia bulk spacetime in \cite{Anous16} was studied in a similar manner. It is plausible that we could use the bulk geometries with $N$ colliding particles to study the transition and similarities between $N$-local quenches and Vaidya global quench, and extend further the ideas from this work on the similarities between the local quenches and global quenches which lead to thermalization. 

\section*{Acknowledgments}

The authors are grateful to Dmitri Ageev, Douglas Stanford and Tadashi Takayanagi for useful discussions. This work is supported by the Russian Science Foundation (project 14-50-00005, Steklov Mathematical Institute).

\appendix 

\section{Geodesics and isometries in global coordinates}
\label{AppIsometries}

The metric of global AdS$_3$ reads (\ref{barrelmetric}):
\begin{equation}
 ds^2=-{\cosh}^2\chi\,d\tau^2+d{\chi}^2+{\sinh}^2\chi\,d\phi^2.
\end{equation}
Geodesics can be parametrized as follows (here we present slightly modified form compared to what was used in \cite{Arefeva2015,AAT}):
\bea
&& \tan (\tau -\tau_0)= -\frac{2E}{1-E^2 + J^2}  \coth \lambda\,; \label{tau(l)} \\
&&  \tan (\phi - \phi_0) = \frac{-1-E^2 + J^2}{2J} \tanh \lambda\,; \label{nonvarphi(l)}\\
&& \sinh^2 \chi = \frac{J^2}{E^2-J^2} \cosh^2 \lambda + \frac{(-1-E^2+J^2)^2}{4(E^2-J^2)} \sinh^2 \chi\,. \label{chi(l)}
\eea 
Here $\tau_0$, $\phi_0$, $E$, $J$ are constants. If we consider geodesics between the boundary points, then these constants are related to the coordinates of endpoints as follows: 
\bea
&& \tau_0 = \frac12(\tau_a + \tau_b)\,, \qquad \phi_0 = \frac{1}{2}(\phi_a + \phi_b)\,; \\
&& \frac{2E}{1-E^2 + J^2} = \tan \frac{\Delta \tau}{2}\,, \qquad \frac{-1-E^2 + J^2}{2J} = \tan \frac{\Delta \phi}{2}\,;\\
&& \frac{J^2}{E^2-J^2} = \frac{\tan^2 \frac{\Delta \phi}{2}}{\tan^2 \frac{\Delta \tau}{2} - \tan^2 \frac{\Delta \phi}{2}}\,, \qquad \frac{(-1-E^2+J^2)^2}{4(E^2-J^2)} = \frac{1}{\tan^2 \frac{\Delta \tau}{2} - \tan^2 \frac{\Delta \phi}{2}}\,; 
\eea
where we introduced $\Delta \tau = \tau_b - \tau_a$ and $\Delta \phi = \phi_b - \phi_a > 0$ for definiteness. 

In the main text we use the image geodesics in global AdS$_3$ with respect to the action of the isometry generated by the holonomy ${\bf u}_2$: 
\be
X \to X^* := {\bf u}_2^{-1} X {\bf u}_2\,, \qquad X \to X^\# := {\bf u}_2 X {\bf u}_2^{-1}\,;
\ee
where $X$ is parametrized on the group manifold language as (\ref{eq2.4}) and the holonomy is given by (\ref{U2}). Using the parametrization (\ref{eq2.4}) again for image points $X^*$ and $X^\#$, one can extract explicit formulas for coordinates of image points. For $X^*$, we have
\bea
&& \tan \tau^* = \tan \tau (1+2 \mathcal{E}^2) + 2 \mathcal{E} \frac{\tanh \chi}{\cos \tau} (\mathcal{E} \cos \phi - \sin \phi)\,; \label{tau*}\\
&& \tan \phi^* = \frac{2 \mathcal{E} \cosh \chi \sin \tau + \sinh \chi (2 \mathcal{E} \cos \phi-\sin \phi) }{2\mathcal{E}^2 \cosh \chi \sin \tau + \sinh \chi ((2 \mathcal{E}^2-1)\cos \phi -2 \mathcal{E} \sin \phi)}\,; \label{phi*}\\
&& \!\!\!\!\!\!\!\!\!\!\!\!\cosh^2 \chi^* = \cosh^2 \chi \cos^2 \tau + ((1+2 \mathcal{E})^2 \cosh \chi \sin \tau + 2 \mathcal{E} (\mathcal{E} \cos \phi - \sin \phi) \sinh \chi)^2\,; \label{chi*}
\eea
and for $X^\#$ we have 
\bea
&& \tan \tau^\# = \tan \tau (1+2 \mathcal{E}^2) + 2 \mathcal{E} \frac{\tanh \chi}{\cos \tau} (\mathcal{E} \cos \phi + \sin \phi)\,; \label{tauDiez}\\
&& \tan \phi^\# =- \frac{2 \mathcal{E} \cosh \chi \sin \tau + \sinh \chi (2 \mathcal{E} \cos \phi+\sin \phi) }{2\mathcal{E}^2 \cosh \chi \sin \tau + \sinh \chi ((2 \mathcal{E}^2-1)\cos \phi +2 \mathcal{E} \sin \phi)}\,; \label{phiDiez}\\
&& \!\!\!\!\!\!\!\!\!\!\!\!\cosh^2 \chi^* = \cosh^2 \chi \cos^2 \tau + ((1+2 \mathcal{E})^2 \cosh \chi \sin \tau + 2 \mathcal{E} (\mathcal{E} \cos \phi + \sin \phi) \sinh \chi)^2\,; \label{chiDiez}
\eea
Here we use the notation $\mathcal{E} = \coth \frac{\pi R}{2}$.

\section{Geodesics in the BTZ geometry \label{BTZgeodesics}}

We start from the BTZ coordinate patch of the AdS$_3$ spacetime: 
\be
ds^2 = - (r^2-R^2) d t^2 + \frac{d r^2}{r^2-R^2}+r^2 d\varphi^2\,, 
\ee
where $t \in [0,\ +\infty)$, $r \in (R,\ +\infty)$, and for now $\varphi \in \RR$. The geodesics in this geometry are described by the following formulae \cite{Bal2012}:
\bea
r(\lambda)^2 &=& \Gamma_-^2 + (\Gamma_+^2 - \Gamma_-^2) \cosh^2(\lambda-\lambda_0) \label{r(l)}\,,\\
\varphi(\lambda) &=& \varphi_0 + \frac{1}{R} \arctanh \left( \frac{\Gamma_-}{\Gamma_+} \tanh(\lambda - \lambda_0) \right)\,,\label{phi(l)}\\
t(\lambda) &=& t_0 + \frac{1}{R} \arctanh \left( \sqrt{\frac{R^2-\Gamma_-^2}{\Gamma_+^2-R^2}} \tanh(\lambda - \lambda_0)\right)\,;\label{t(l)}
\eea
The constants $\Gamma_\pm$. $R$ are subject to the restriction 
\be
0 < \Gamma_-^2 < R^2 < \Gamma_+^2\,;
\ee
One can express the variables as a function of $r\in[\Gamma_{+},\infty[$:
\begin{eqnarray}
\varphi_{\pm}(r)&=&\varphi_{0}\pm\frac{1}{R}\text{arctanh}\left(\frac{\Gamma_{-}}{\Gamma_{+}}\sqrt{\frac{r^{2}-\Gamma_{+}^{2}}{r^{2}-\Gamma_{-}^{2}}}\right), \label{eq:xpm} \\
t_{\pm}(r)&=&t_{0}\pm\frac{1}{R}\text{arctanh}\left(\sqrt{\frac{R^{2}-\Gamma_{-}^{2}}{\Gamma_{+}^{2}-R^{2}}}\sqrt{\frac{r^{2}-\Gamma_{+}^{2}}{r^{2}-\Gamma_{-}^{2}}}\right), \label{eq:ypm}\\
\lambda_{\pm}(r)&=&
\lambda_{0}\pm\ln\left(\frac{\sqrt{r^{2}-\Gamma_{-}^{2}}+\sqrt{r^{2}-\Gamma_{+}^{2}}}{\sqrt{\Gamma_{+}^{2}-\Gamma_{-}^{2}}}\right).\label{eq:lambdapm}
\end{eqnarray}
We are interested in geodesics with endpoints on the boundary. From the equations (\ref{phi(l)}) and (\ref{t(l)}), the separation between two boundary endpoints $(\varphi_{a}, t_{a}, r_{a}=\infty)$ and $(\varphi_{b}, t_{b}, r_b =\infty)$ is given by
\begin{eqnarray}
\Delta \varphi &=& \varphi_2 - \varphi_1 =  \varphi(\lambda\rightarrow\infty)-\varphi(\lambda\rightarrow-\infty)=\frac{2}{R}\text{arctanh}\left(\frac{\Gamma_{-}}{\Gamma_{+}}\right), \label{deltaT}\\
\Delta t &=&t_2 - t_1 = t(\lambda\rightarrow\infty)-t(\lambda\rightarrow-\infty)=\frac{2}{R}\text{arctanh}\left(\sqrt{\frac{R^{2}-\Gamma_{-}^{2}}{\Gamma_{+}^{2}-R^{2}}}\right),.\label{deltaPhi}
\end{eqnarray}
In this paper we use BTZ coordinates in the spacetime with BTZ identification generated by the holonomy (\ref{btz-hol}). In the BTZ coordinate patch it periodizes the angular coordinate: $\varphi \sim \varphi + 2\pi n$, where $n \in \ZZ$. We choose the fundamental domain such that the worldline of the first particle is in the middle of the domain, and the worldline of the second particle is at the periodically identified boundary, i. e. we set $\varphi \in [-\pi, \pi)$. Depending on the value of the difference of $\Delta \varphi = \varphi_2 - \varphi_1$, a geodesic in the BTZ patch of the pure AdS$_3$ space without topological defects is pulled back to either a direct ($|\Delta \varphi| < \pi$) geodesic, or a geodesic which winds around the horizon (($|\Delta \varphi| > \pi$) on the topological quotient. The above formulas for the geodesics hold when pulled back to the BTZ coordinates, but with addition of the term $2\pi n$ to the $\Delta \varphi$. 

With that in mind, let us now focus on geodesics in BTZ coordinates, taking into account all windings. The equations (\ref{deltaT}-\ref{deltaPhi}) can be used to express the integration constants $\Gamma_\pm$ through the coordinates of the endpoints.
\bea
&& \Gamma_+ = R \sqrt{\frac{1+\tanh^2 \frac{R \Delta t}{2}}{\tanh^2 \frac{R \Delta t}{2} + \tanh^2 \frac{R (\Delta \varphi + 2\pi n)}{2}}}\,, \label{Gamma+}\\
&& \Gamma_- = \Gamma_+ \tanh \frac{R (\Delta \varphi + 2\pi n)}{2} \label{Gamma-}\,,
\eea
From the equation (\ref{r(l)}) it is clear that the $\Gamma_+$ has the meaning of the closest distance along the radial coordinate to the horizon, into which the geodesic with given endpoints can approach, i. e. the depth of the geodesic in the bulk. The formula (\ref{Gamma+}) shows that winding geodesics always approach to the horizon closer than direct geodesics. This fact is key to our understanding of dynamics of the pole structure of non-equilibrium correlators, as explained in the main text. 

The (regularized) length of a geodesic can be found to be
\begin{eqnarray}
\Delta\lambda&=&\lambda_{+}(r\rightarrow r_{0})-\lambda_{-}(r\rightarrow r_{0})= 2\ln\left(\frac{\sqrt{r_{0}^{2}-\Gamma_{-}^{2}}+\sqrt{r_{0}^{2}-\Gamma_{+}^{2}}}{\sqrt{\Gamma_{+}^{2}-\Gamma_{-}^{2}}}\right) 
\end{eqnarray}
in terms of a regularized AdS boundary at $r =r_0$.
In the limit $r_{0}\rightarrow\infty$ this reduces to
\begin{equation} \label{eq:intermediate}
\Delta\lambda=2\ln\left(\frac{2r_{0}}{\sqrt{\Gamma_{+}^{2}-\Gamma_{-}^{2}}}\right).
\end{equation}
In terms of coordinates of the endpoints, this formula leads to the expression
\be
\mathcal{L}_{\text{ren}} (a,\ b) = \log\left(2(\cosh[R(\varphi_b - \varphi_a + 2\pi n)] - \cosh[R(t_b - t_a)])\right) +2 \log \left(\frac{r_0}{R}\right)\,, \label{geodesicLength}
\ee
where $r_0$ is the near-boundary cut-off.
From this formula, it is clear that the length of \textit{any} winding geodesic is always larger than the length of \textit{any} direct geodesic. Therefore, the winding geodesics in BTZ geometry will always give subleading contributions to holographic correlation functions for any placement of endpoints. 

\section{Lorentzian correlators in thermal equilibrium} \label{Lorentzian}

Consider real-time two-point correlation functions of a scalar operator $\mathcal{O}$ of scaling dimension $\Delta$ in holographic $(1+1)$d CFT at finite temperature $T > \frac{R}{2\pi}$ above the Hawking-Page threshold, so that $R > 1$. In this case the holographic dual bulk spacetime is the static BTZ black hole. The Wightman two-point correlator than reads \cite{Keski98,Skenderis08}:
\be
\langle \mathcal{O}_\Delta (t, \varphi) \mathcal{O}_\Delta (0,0) \rangle_T = \sum_{n = -\infty}^{+\infty} \left(\frac{1}{2(\cosh [R (\varphi + 2 \pi n)] - \cosh [R (t - i \epsilon)]}\right)^\Delta\,; \label{GWeq}
\ee
Let us assume for definiteness that $t > 0$. Then in analogy to the zero-temperature vacuum case of CFT on a cylinder \cite{OS,AAT,AKT}, this expression can be rewritten as following sum over images:
 \bea
 \langle \mathcal{O}(t, \varphi)\mathcal{O}(0,0)\rangle_T =
\sum_{n = -\infty}^{+\infty} \left(\frac{1}{2(\cosh [R (\varphi + 2 \pi n)] - \cosh R t)}\right)^\Delta \times \Phi_{W}^\Delta (R; t, \varphi; n)\,\label{GWeqUnfolded};
\eea
where the phase factor is introduced
\be
\Phi_{W}^\Delta (R; t, \varphi; n) =  \exp\left(-i \pi\Delta \theta(\cosh R t  + \cosh[R (\varphi+2\pi n)])\right)\,. \label{PhaseW}
\ee
If initial points are spacelike-separated, then all phase factors in this expression equal to $1$, and thus the Wightman correlator for spacelike-separated points can be interpreted as sum over all geodesics in the BTZ black hole geometry with metric (\ref{schw}). The $n$ is the winding number. The leading contribution to the sum is given by the direct geodesic with $n = 0$, and winding geodesics give exponentially suppressed contributions to the correlator. If points are timelike-separated, then the contribution with $n = 0$ is interpreted as contribution from the reflection geodesic times the exponential factor $\e^{-i \pi \Delta}$. The same goes for contribution from image points.

One can rewrite other real-time two-point functions in similar manner \cite{AAT,AKT,Bal2012}. The difference is in the phase factors, which can be recovered using common QFT definitions of Green's functions through Wightman correlators. For example, for the retarded Green's function one obtains
\be
\Phi_{ret}^\Delta (R; t, \varphi; n) = -2i\sin[ \pi \Delta \,{\mbox{sign}}(t)] \theta(t)\theta (\cosh R t  - \cosh[R ( \varphi+2\pi n)])\,; \label{PhaseRet}
\ee
and for the Feynman propagator one obtains 
\bea
\Phi_{F}^\Delta (R; t, \varphi; n) =&& \left[\theta(t) \e^{-i\,\pi\,\Delta \,\text{sgn}(t)}+\theta(-t)\e^{i\pi\,\Delta \,\text{sgn}(t)}] \right] \theta(\cosh R t - \cosh [R (\varphi + 2 \pi n)])\nn\\
&& + \theta(-\cosh R t + \cosh [R (\varphi + 2 \pi n)])\label{PhaseF}\,.
\eea

\end{document}